\documentclass{pasj00}

\def\Umlaut#1{\"{#1}}

\def\commenta{$^*$}
\def\commentb{$^\dagger$}

\def\submitted{submitted}
\def\inpress{in press}
\def\astroph#1{ (astro-ph/#1)}

\DeclareAbbreviation\AAHam{Astron. Abh. Hamburg. Sternw.}
\DeclareAbbreviation\AARv{Astron. Astrophys. Rev.}
\DeclareAbbreviation\an{Astron. Nachr.}
\DeclareAbbreviation\AcA{Acta Astron.}
\DeclareAbbreviation\Afz{Astrofizika}
\DeclareAbbreviation\AnTok{Tokyo Astron. Obs. Annals, Sec. Ser.}
\DeclareAbbreviation\Ap{Astrophysics}
\DeclareAbbreviation\ARep{Astron. Rep.}
\DeclareAbbreviation\ATel{Astronomer's Telegram}
\DeclareAbbreviation\ATsir{Astron. Tsirk.}
\DeclareAbbreviation\AcApS{Acta Astrophys. Sinica}
\DeclareAbbreviation\AstL{Astron. Letters}
\DeclareAbbreviation\BaltA{Baltic Astron.}
\DeclareAbbreviation\BASI{Bull. Astron. Soc. India}
\DeclareAbbreviation\BeSN{Be Star Newsletter}
\DeclareAbbreviation\GCN{GCN}
\DeclareAbbreviation\ibvs{Inf. Bull. Variable Stars}
\DeclareAbbreviation\JAD{J. Astron. Data}
\DeclareAbbreviation\JAVSO{J. American Assoc. Variable Star Obs.}
\DeclareAbbreviation\JBAA{J. British Astron. Assoc.}
\DeclareAbbreviation\LowOB{Lowell Obs. Bull.}
\DeclareAbbreviation\MitVS{Mitteil. Ver\"{a}nderl. Sterne}
\DeclareAbbreviation\MmSAI{Mem. Soc. Astron. Ita.}
\DeclareAbbreviation\Msngr{Messenger}
\DeclareAbbreviation\NewA{New Astron.}
\DeclareAbbreviation\NewAR{New Astron. Rev.}
\DeclareAbbreviation\OAP{Odessa Astron. Publ.}
\DeclareAbbreviation\Obs{Observatory}
\DeclareAbbreviation\PASA{Publ. Astron. Soc. Australia}
\DeclareAbbreviation\PAZh{Pis'ma AZh}
\DeclareAbbreviation\PhR{Phys. Rep.}
\DeclareAbbreviation\PVSS{Publ. Variable Stars Sect. R. Astron. Soc. New Zealand}
\DeclareAbbreviation\PZ{Perem. Zvezdy}
\DeclareAbbreviation\PZP{Perem. Zvezdy Pril.}
\DeclareAbbreviation\QJRAS{QJRAS}
\DeclareAbbreviation\RMxAA{Rev. Mexicana Astron. Astrof.}
\DeclareAbbreviation\RvMA{Reviews of Modern Astron.}
\DeclareAbbreviation\Sci{Science}
\DeclareAbbreviation\SvA{Soviet Astronomy}
\DeclareAbbreviation\SvAL{Soviet Astronomy Letters}
\DeclareAbbreviation\VeSon{Ver\"{o}ff. Sternw. Sonneberg}
\DeclareAbbreviation\VSOLJBul{VSOLJ Variable Star Bull.}
\DeclareAbbreviation\yCat{VizieR Online Data Catalog}
\DeclareAbbreviation\ZA{Z. Astrophys.}

\def\ASPConf#1#2{ASP Conf. Ser. #1, #2}
\def\IAUColloq#1#2{IAU Colloq. #1, #2}

\def\PublisherCambridge{Cambridge: Cambridge University Press}
\def\PublisherKluwer{Dordrecht: Kluwer Academic Publishers}
\def\PublisherASP{San Francisco: ASP}
\def\PublisherReidel{Dordrecht: D. Reidel Publishing Company}
\def\PublisherSpringer{Berlin: Springer-Verlag}
\def\PublisherUAP{Tokyo: Universal Academy Press}

\begin{document}
\SetRunningHead{T. Kato et al.}{Variable Star Network}

\Received{}
\Accepted{}

\title{Variable Star Network: World Center for Transient Object Astronomy
       and Variable Stars}

\author{Taichi \textsc{Kato}, Makoto \textsc{Uemura}, Ryoko \textsc{Ishioka}}
\affil{Department of Astronomy, Kyoto University,
       Sakyo-ku, Kyoto 606-8502}
\email{tkato@kusastro.kyoto-u.ac.jp, ishioka@kusastro.kyoto-u.ac.jp, uemura@kusastro.kyoto-u.ac.jp}

\author{Daisaku \textsc{Nogami}}
\affil{Hida Observatory, Kyoto University, Kamitakara, Gifu 506-1314}
\email{nogami@kwasan.kyoto-u.ac.jp}

\author{Chatief \textsc{Kunjaya}}
\affil{Institute of Technology, Bandung, Indonesia}
\email{kunjaya@as.itb.ac.id}

\author{Hajime \textsc{Baba}}
\affil{Center for Planning and Information Systems, The Institute of
       Space and Astronautical Science (ISAS), \\
       Sagamihara, Kanagawa 229-8510}
\email{baba@plain.isas.ac.jp}

\email{\rm{and}}

\author{Hitoshi \textsc{Yamaoka}}
\affil{Faculty of Science, Kyushu University, Fukuoka 810-8560}
\email{yamaoka@rc.kyushu-u.ac.jp}



\maketitle

\begin{abstract}
   Variable Star Network (VSNET) is a global professional-amateur network
of researchers in variable stars and related objects, particularly in
transient objects, such as cataclysmic variables, black hole binaries,
supernovae and gamma-ray bursts.  The VSNET has been playing a pioneering
role in establishing the field of {\it transient object astronomy},
by effectively incorporating modern advance in observational astronomy
and global electronic network, as well as collaborative progress
in theoretical astronomy and astronomical computing.
The VSNET is now one of the best-featured global networks in this field
of astronomy.  We review on the historical progress, design concept,
associated technology, and a wealth of scientific achievements powered
by the VSNET.
\end{abstract}

\section{Introduction}

\subsection{Transient Object Astronomy}

   With the progress of modern physics and astronomy, our outlook on
the universe is dramatically changing from the stationary universe
to the dynamic, ever changing, universe.  The dynamical phenomena in
the universe appear as variations at various time-scales ranging
from the cosmological evolutionary time scale to less than a milisecond.
These time-variations are becoming actually observable with the advent
of the modern observing equipment and technology.  Among them,
time-variations arising from extreme gravity as best exemplified by
black holes, and from degenerate objects, such as white dwarfs and
neutron stars, have been receiving extreme attention from various
fields of modern science, as the natural laboratory of general relativity
and quantum mechanics, which best represent the glorious success
of the ``century of physics".

   As can be easily expected from the extreme conditions, astronomical
phenomena under strong gravity or in degenerate conditions have
extremely short time-scales, and are known to be usually very
unpredictable.
These astronomical phenomena are now generally called
``transient phenomena", or referred to as ``transient objects".
The concept of {\it transient object astronomy} appeared very late in
the history of astronomy and now flourishing as a new modality of
astrophysical research.\footnote{
  As shown in Appendix \ref{sec:app:iaupos}, the VSNET is the earliest
  group which began using the term {\it transient objects} in the
  present context of astronomical significance.
}  This success greatly owed to the recent
great advancement of observing modalities, information technology and
computational astrophysics.  The Variable Star Network (VSNET),
the objective of this review, is one of the earliest and most successful
international groups that led to the modern success of transient object
astronomy.

\subsection{New Window to Transient Object Astronomy}
\label{sec:newwindow}

   In the research history of {\it transient object astronomy},
there were two major breakthroughs in the early 1990's.  The one is the
development of easy availability of CCDs and personal computers,
and the other is the advent of the internet.  These two breakthroughs
played a key role in establishing {\it transient object astronomy}
as one of the most popular contemporary astronomy topics.

   From the traditional viewpoint, CCDs were usually used as a
``faint-end" extension of the former photon detection methods, e.g.
photoelectric and photographic observations.  This naturally led to
a pursuit of observing fainter stars on long-exposure CCD images
(cf. \cite{how88faintCV1}; \cite{szk89faintCV2}; \cite{how90faintCV3};
\cite{how91faintCV4}).

   The founder of the VSNET was one of the first to break this
tradition, and was virtually the first person who systematically
turned modern CCD equipment to bright, transient objects,
such as classical novae and outbursting dwarf novae (the best examples
being \cite{kat91v838heriauc}; \cite{kat91v1251cygiauc};
\cite{kat91efpegiauc}, see the later sections for their scientific
achievements).  The traditional time-resolved observations
of classical novae and outbursting dwarf novae were almost restricted
to so-called target-of-opportunity (TOO) observations.
The best traditional examples include the 1978 outburst of WZ Sge
(\cite{pat78wzsgeiauc3311}; \cite{mat80wzsge}; \cite{pat81wzsge}),
and the 1986 outburst of SW UMa \citep{rob87swumaQPO}.
This kind of observations was usually
severely limited by the telescope time allocation, and many
important transient phenomena (e.g. the 1985 historical long outburst
of U Gem: \cite{can02ugem1985}) faded away without receiving sufficient
observational coverage.  Traditional proposals for telescope time were
also limited because of the transient and unpredictable nature of
these phenomena; there is no guarantee that there is a suitable
transient target at the time of allocated observation.  For this
reason, systematic observational research in these objects was
severely restricted to short-period, less unpredictable objects,
with an enormous effort of world-wide coordination
(e.g. VW Hyi: \cite{sch85DNoutburstvwhyi}; \cite{pri87vwhyimultiwavelength};
YZ Cnc: \cite{vanpar94suumayzcnc}).

   Timely circulation of alerts on transient objects or phenomena is
also crucially important, particularly for TOO-type observation.
Before the wide availablity of the internet, the typical way of
communicating such alerts was a phone call from an observer (usually
an amateur astronomer watching variable stars) to a variable star
organization, which was typically relayed (with some delay) to local
observers for confirmation.  The information, if it was recognized as
particularly important, was then distributed to world-wide observers
usually from the Central Bureau of Astronomical Telegrams (CBAT)
via telegrams, direct phone calls, or slow postcards.  It usually
took, even in best cases, a day or more before this crucial
information was relayed to the actual observer undertaking
a TOO observation.  The early stage of transient objects was usually
missed because of this delay.  For example, the detection of the
1986 historical outburst of SW UMa was relayed via an astronomical
telegram only when the object reached a historical brightness of
$V\sim$9, although the outburst was initially reported $\sim$1.5 below
the peak brightness.  There had been very few early stage
observations (i.e. within a day of the event detection) of transient
objects before the 1990's.

\subsection{Early Public Electronic Communication}
\label{sec:earlyelec}

   This situation drastically changed with the public availability
of the internet.  In the early times (around 1990--1991), there
were only sporadic internet communications on observations, mainly
via personal e-mails and on public bulletin board systems.  This
strategy worked slightly better than in the past, the situation was
basically unchanged in that most of observers had to rely on
occasional communications or a slow access to news materials.

   From the necessity of publicly and electronically disseminating
urgent astronomical phenomena, there appeared e-mail exploders (mailing
lists).  The Scandinavian {\it varstars} list and the (mainly)
professional {\it novanet} by the Arizona State University team
played an early important role in publicly relaying information on
transient objects.\footnote{
Both networks do not exist at present.
}  The early-time progress of these electronic
communications is summarized in the {\it vsnet-history} list.\footnote{
$\langle$http://www.kusastro.kyoto-u.ac.jp/vsnet/Mail/vsnet-history/\\maillist.html$\rangle$.
}

\begin{figure*}
  \begin{center}
    \FigureFile(160mm,80mm){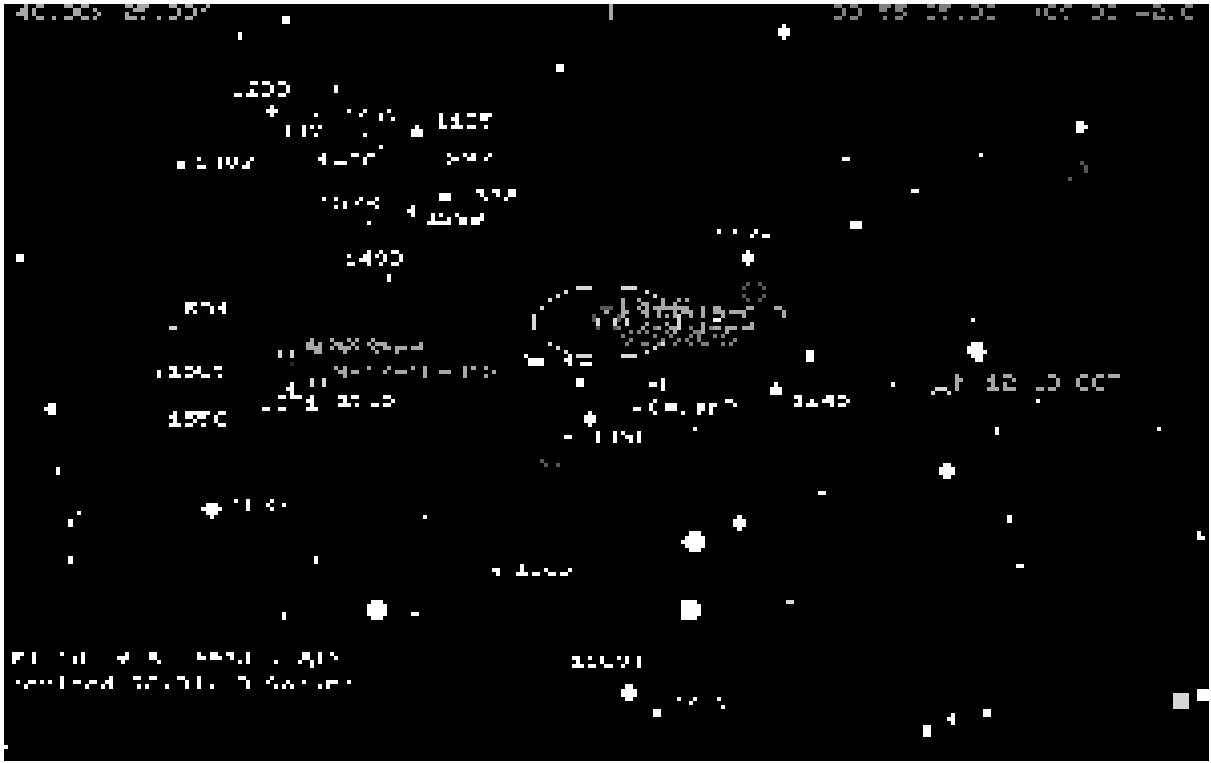}
  \end{center}
  \caption{Circulated electronic chart of SN 1993J (issued on 1993 April 13)
  showing $V$-band comparison stars for standardization.
  }
  \label{fig:sn93jcht}
\end{figure*}

\begin{figure*}
  \begin{center}
    \FigureFile(160mm,80mm){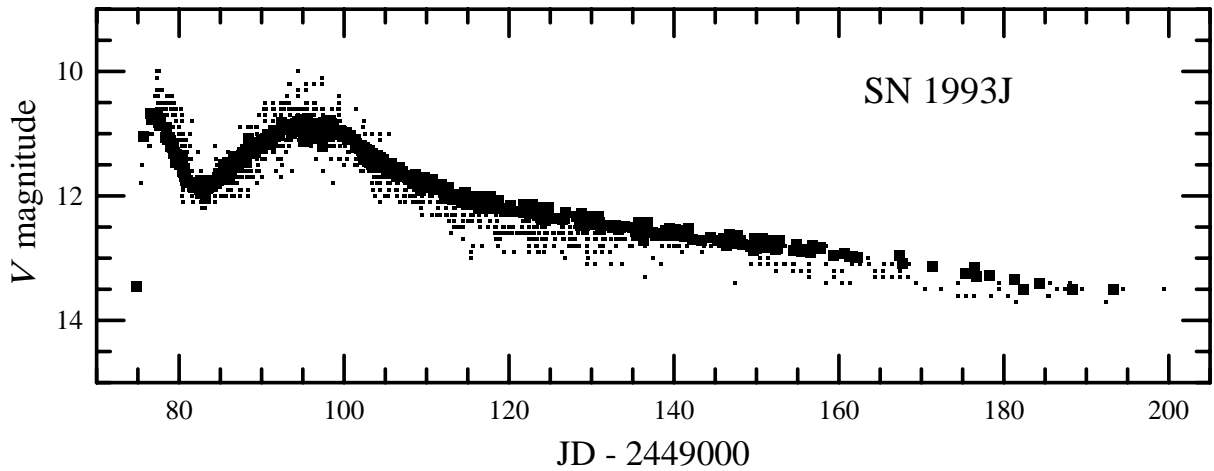}
  \end{center}
  \caption{Light curve of SN 1993J, drawn from the ``SN 1993J photometry
  update" (see text).  The Large and small dots represent (nearly)
  $V$-band and visual observations.
  }
  \label{fig:sn93j}
\end{figure*}

   The scientific role of wide-availability of these e-mail exploders
was recognized upon the appearance of SN 1993J in M 81
(\cite{rip93sn1993jiauc5731}; \cite{pea93sn1993j}).
This supernova showed an unusual early-time light curve and a spectral
transition from a type-II to type-Ib supernova
(\cite{nom93sn1993jnat}; \cite{swa93sn1993j};
\cite{fil93sn1993jletter}; \cite{pod93sn1993j}).
In communicating nightly rapid changes and distributing
most up-to-date observation
strategies, the e-mail exploders played a more crucial role than ever.
Another advantage of e-mail exploders as a {\it standardization tool}
of observations became evident (figure \ref{fig:sn93jcht}).
Early-time non-standard observations
were quickly corrected using the updated photometric comparison stars,
and questionable observations were examined real-time to clarify the
cause.  This led to a huge world compilation of SN 1993J photometry
updates (see figure \ref{fig:sn93j}) contributed by a number of
volunteers, including the VSNET founder.\footnote{
The final version of the ``photometry update" is publicly available at
$\langle$ftp://vsnet.kusastro.kyoto-u.ac.jp/pub/vsnet/SNe/\\sn1993j/sn.mag$\rangle$
}  This high-quality, uniform compilation of real-time
observations greatly contributed to real-time theoretical modeling of
this object (e.g. \cite{nom93sn1993jnat}), spectroscopy
(e.g. \cite{bar93sn1993j}; \cite{tan93sn1993j}; \cite{clo95sn1993j};
\cite{tra93sn1993j}) and photometry (e.g. \cite{vandri93sn1993j};
\cite{whe93sn1993j}).
We published our own results in \citet{oky93sn1993j}.  We also contributed
to a number of International Astronomical Union Circulars (IAUCs)
(\cite{kat93sn1993jiauc5747}; \cite{zim93sn1993jiauc5750};
\cite{kin93sn1993jiauc5755}; \cite{fil93sn1993jiauc5760};
\cite{tra93sn1993jiauc5780}; \cite{hu93sn1993jiauc5783}).
The complete history of this SN 1993J story can be also seen in the
{\it vsnet-history} archive.

\subsection{Opening of the Electronic Era of Transient Object Astronomy}
\label{sec:openingera}

\begin{figure*}
  \begin{center}
    \FigureFile(160mm,80mm){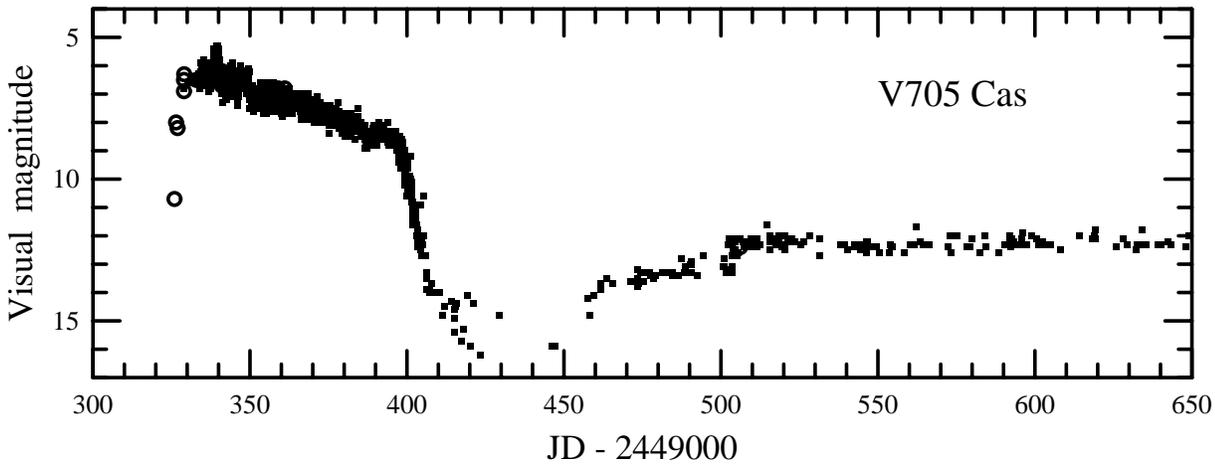}
  \end{center}
  \caption{Light curve of V705 Cas (Nova Cas 1993), covering the rise
  and the ``great fade'', and rebrightening phases, drawn from
  the observations circulated through the early development of the VSNET.
  The dots and open circles represent visual and photographic/CCD
  observations, respectively.
  }
  \label{fig:v705}
\end{figure*}

   Upon the recognition of the importance of e-mail exploders on the
occasion of SN 1993J, more systematic efforts were taken to standardize
the communication and data reporting method.  In relation to reporting
observations, we started widely disseminating observations of regular
variable star observations, mainly submitted to the Variable Star
Observers League in Japan (VSOLJ),\footnote{
$\langle$http://www.kusastro.kyoto-u.ac.jp/vsnet/VSOLJ/vsolj.html$\rangle$
and see
$\langle$http://vsolj.cetus-net.org/$\rangle$ for the VSOLJ
Variable Star Bulletin page.
} and those personally reported to us.
People started recognizing the scientific importance of widely
disseminating regular observations, which can be readily reflected on
scheduling new observations.  New findings based on widely reported
observations (e.g. superhump detection of a dwarf nova) were also
relayed real-time, which worked as a positive feedback to original
observers.  The prototype of VSNET-type e-mail exploders was thus
established in 1993.  The next major astronomical event at this stage
of the history was the discovery of Nova Cas 1993 (V705 Cas).
This nova showed considerable degree of early-time fluctuations,
as well as a later dust-forming episode.  During all of the stages
of evolution of the nova explosion, the data circulating strategy
established at the time of SN 1993J played an impressive role:
the comprehensive compilation of V705 Cas by Yasuto Takenaka (see
figure \ref{fig:v705}) was cited in a Nature paper
\citep{sho94v705casdust} as best authenticated
optical record of this nova.  The nova was later
even symbolically called {\it an electronic nova} \citep{pep95v705cas},
representing the opening of new electronic era of transient object
astronomy.

\subsection{Establishment of VSNET}

   The information of these transient objects and regular variable star
observations was initially relayed manually, or relayed on existing
less specified e-mail exploder systems.  In 1994, our own
e-mail exploder system (VSNET) started working.  This service smoothly
took over the past manual e-mail announcement systems, and immediately
received wide attention both from amateur and professional communities.
The establishment of the VSNET thus became the ``prototype" of world-wide
amateur-professional collaborations based on public e-mail communication.
This initiative later led to a flourishing VSNET Collaboration
(section \ref{sec:vsnetcollab}).  The early history was reviewed by
D. Nogami et al. (1997) in ``Electronic Publishing, Now and the Future'',
Joint Discussion 12 of the 23rd IAU General Assembly.  Considering the
historical significance in the advent of {\it transient object astronomy}
and the current unavailability of this document in a solid publication,
we reproduce the presented contents in Appendix \ref{sec:app:iaupos}
(in order to preserve the original contents, we only corrected minor
typographical errors).
The VSNET mailing list system now has more than 1300 subscribers from
more than 50 countries all over the world.

\section{VSNET as a Modern-Day ``Electronic Variable Star Organization''}

\subsection{Historical Development of the Electronic Reporting System}
\label{sec:rephist}

   During the very initial stage of the development of the VSNET,
we simply relayed observations to those who (potentially) need the data.
However, it soon became evident, from the experiences with SN 1993J and
V705 Cas (subsection \ref{sec:earlyelec}), that there is a need for
a newly designed reporting system adapted for electronic data exchanges.
Since we already had sufficient experience with relaying VSOLJ reports
to the world-wide variable star observers, it was a natural solution
to extend the VSOLJ format to an international version.  This changes
were minimal, by introducing Universal Time (UT)-based system and the
extension of coding system of observers.  The details of the reporting
system is described in Appendix \ref{sec:app:report}.

   By globally collecting data, we soon recognized the necessity for
setting up a dedicated e-mail list for reporting observations,
{\it vsnet-obs}.\footnote{
$\langle$http://www.kusastro.kyoto-u.ac.jp/vsnet/Mail/vsnet-obs/\\maillist.html$\rangle$.  Because of the large number of articles, the online archive is
subdivided; see
$\langle$http://www.kusastro.kyoto-u.ac.jp/vsnet/Mail/index.html$\rangle$.
for the complete message archive.
} The alert list {\it vsnet-alert}\footnote{
$\langle$http://www.kusastro.kyoto-u.ac.jp/vsnet/Mail/vsnet-alert/\\maillist.html$\rangle$.
} was prepared at the same time, which has been one of the most renowned
and reliable sources of noteworthy phenomena of variable stars and
transient objects, and the messages in {\it vsnet-alert} have been
frequently cited in the professional literature as the primary source
of information (e.g. \cite{DownesCVatlas2}; \cite{DownesCVatlas3};
\cite{RitterCV6}; \cite{RitterCV7}; \cite{kub03CVCat}).

\subsection{Standardization}
\label{sec:standard}

   The VSNET played an important role in standardization of variable
star observing.  There are several steps in standardization:
(1) standardization of reporting format (see subsection \ref{sec:rephist}),
(2) standardization of comparison stars, and other minor steps.
In standardizing comparison stars, the VSNET group also took an initiative
in the history of transient object astronomy.  The earliest example
includes SN 1993J and classical novae, for which reliable CCD-based (or
sometimes photoelectric) comparison star sequences were determined and
distributed through the VSNET lists.  Before this standardizating efforts
were taken, the nova researchers had to cope with often unreliable early
reports in IAUCs, which were often based on various sources of comparison
stars with notoriously diverse photometric quality.  With the advent of
the VSNET, the modern-day nova observations have now become as reliable
as those of other variable stars with well-established comparison star
sequences.  The VSNET group has been paying attention to the quality of
the original discovery reports, and have issued several magnitude updates
superseding the IAUC announcements.  The same effort has been taken for
supernova photometry, although the faintness and the large number of
the target objects have made it a more difficult task than in classical
novae.

\begin{figure*}
  \begin{center}
    \FigureFile(140mm,210mm){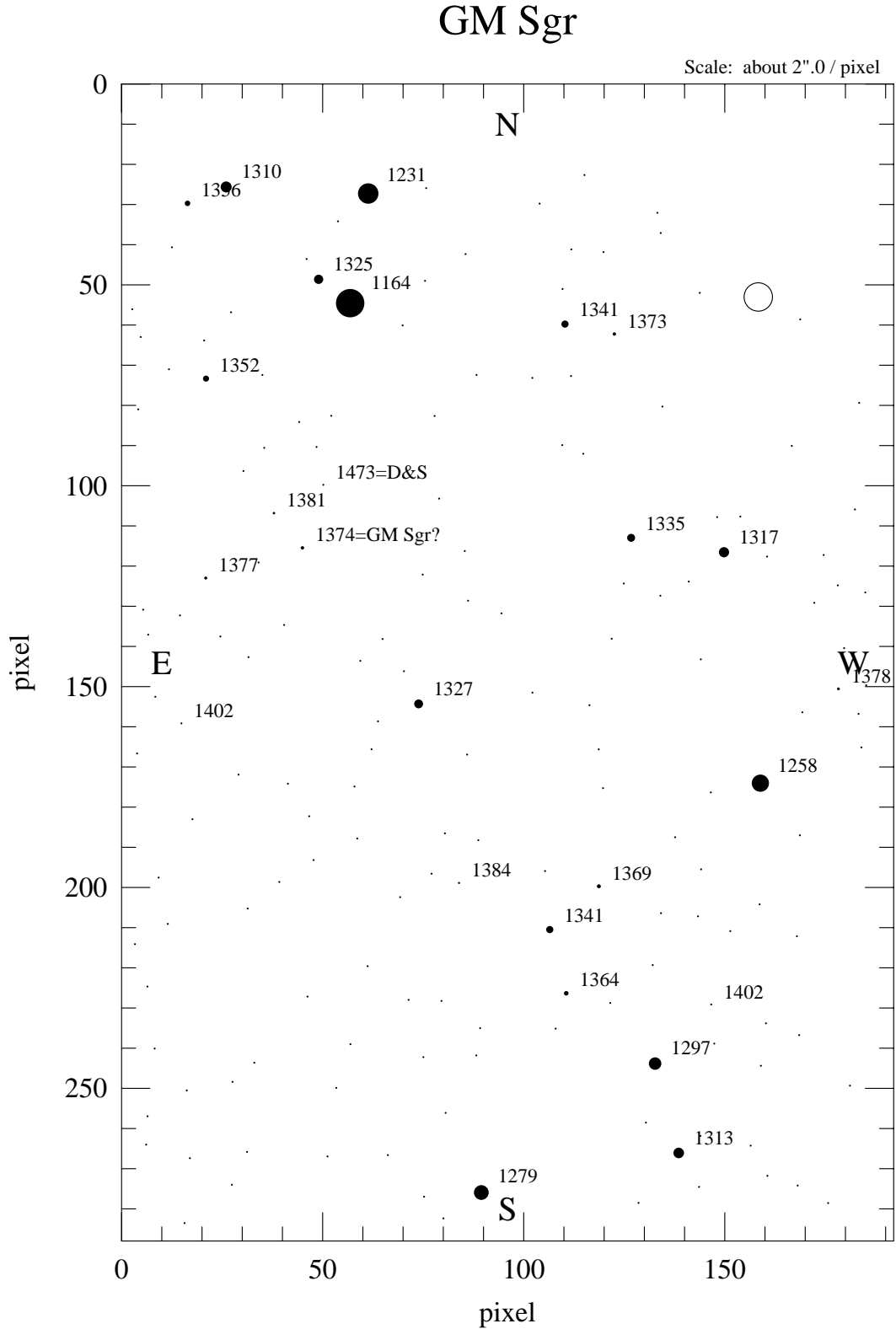}
  \end{center}
  \caption{Example of VSNET chart issued in 1996.  The object is GM Sgr
  (currently renamed as V4641 Sgr; note that this chart provided the
  corrected identification against the previous publication).
  The availablity of a chart with the correctly identified object
  enabled the discovery of the giant outburst in 1999
  (see subsection \ref{sec:xray:fxt}).
  }
  \label{fig:gmchart}
\end{figure*}

   In observation of (non-transient) ordinary variable stars, the VSNET
took the initiative to standardize the comparison star magnitudes
to the modern $V$ system, from various old systems including
the traditional Harvard visual photometric system.
The effort was initially taken to revise the
faint-end magnitudes for cataclysmic variables (CVs) and peculiar
variables using the CCD camera at Ouda Station, Kyoto University
\citep{Ouda}.  These results were continuously released as ``VSNET charts"
through the VSNET lists (see figure \ref{fig:gmchart}).
Several independent contributed efforts,
by Rafael Barbera (Grup d'Estudis Astronomics)\footnote{
$\langle$http://www.astrogea.org/$\rangle$.
}, and Fraser Farrell (Astronomy Society of South Australia),\footnote{
$\langle$http://www.assa.org.au/$\rangle$.
}
as well as those by the VSNET administrator team, were made
to write software packages to graphically display this VSNET chart
format.  Brian Skiff has been continuously contributing to the VSNET
in photoelectrically standardizing sequences for selected variable stars;
this initiative was globally taken over with the CCD works, notably by
Arne Henden and Bruce Sumner.\footnote{
$\langle$ftp://ftp.nofs.navy.mil/pub/outgoing/aah/sequence/sumner/$\rangle$.
}  These standardized sequences and charts
have own lists {\it vsnet-sequence} and {\it vsnet-chart}.\footnote{
$\langle$http://www.kusastro.kyoto-u.ac.jp/vsnet/Mail/\\vsnet-sequence/maillist.html$\rangle$ and
$\langle$http://www.kusastro.kyoto-u.ac.jp/vsnet/Mail/vsnet-chart/maillist.html$\rangle$.
}

   For the bright end, we were the group to first extensively use
Hipparcos and Tycho magnitudes at the earliest epoch (1997) of the public
release of these catalogs.  We immediately made public variable star
charts based on Hipparcos and Tycho magnitudes.\footnote{
$\langle$ftp://vsnet.kusastro.kyoto-u.ac.jp/pub/vsnet/charts/HIPTYC/$\rangle$
}
Since then, this adoption of the standard
$V$-band system (selected for $B-V$ colors) for visual photometry
has been gradually becoming the global standard.  Since the public
release of Tycho-2 catalogue, this standard was extended to a slightly
fainter magnitude.\footnote{
$\langle$ftp://vsnet.kusastro.kyoto-u.ac.jp/pub/vsnet/charts/\\Tycho-2/$\rangle$
}

   For poorly observed faint objects, we also took an initiative
(vsnet-chat 700, in 1998)\footnote{
$\langle$http://www.kusastro.kyoto-u.ac.jp/vsnet/Mail/vsnet-chat/\\msg00700.html$\rangle$
} to calibrate USNO catalog magnitudes since the early release of
the USNO A1.0 catalog \citep{USNOA10}.
This calibration has been widely used, when an alternative sequence is
not readily available.

\subsection{Collaboration with the VSOLJ}
\label{sec:vsoljcolab}

   From the very beginning of its history, the VSNET has been in
continuous collaboration with the VSOLJ.
The activity includes hosting the VSOLJ
alert and discussion mailing lists, distributing the VSOLJ reports and
prediction of variable stars, and hosting the public VSOLJ database and
light curves (see subsection \ref{sec:publc}).  A part of standardization
schemes (subsection \ref{sec:standard}) has been developed in collaboration
with the VSOLJ administrator group (see also Appendix \ref{sec:app:report})
and through a discussion with world-wide variable star leaders at the
International VSOLJ Meeting held in conjunction with the 23rd IAU General
Assembly.

   The other pioneering aspect of the VSNET--VSOLJ collaboration
was the introduction of CCDs for CV photometry.  This work was mainly
done in collaboration with Makoto Iida, who took the world initiative
to monitor faint CVs with a commercial CCD.  This collaboration led to
fruitful scientific results (NY Ser: \cite{iid95nyser};
DV Dra: \cite{iid95dvdra}; BC UMa, see
$\langle$http://www.kusastro.kyoto-u.ac.jp/vsnet/DNe/bcuma.html$\rangle$).
These successful results were demonstrated in a number of international
conferences, including the Padova CV Conference in 1995 and the Keele
CV Conference in 1997, and the results became gradually digested by the
professional community.  This pioneering amateur-professional
collaboration in transient objects and CVs finally led to the most
successful VSNET Collaboration (section \ref{sec:vsnetcollab}), and the
strategy was taken over, albeit with a lesser degree of the original
``flavor" of amateur-professional relations and publicity policy,
by a number of following world-wide groups with the similar strategies.

\subsection{Collaboration with the Southern Hemisphere Observers}

   Since the VSNET service initially started with
the coordinating role
among mostly northern hemisphere observers, there were initially few
reports from the southern observers.  With the outstanding activity
of the VSNET, there arose a number of requests from professional
astronomers who were planning for TOO observations of southern dwarf
novae.  This situation has been gradually and progressively improved
by increasing contributions from the southern observers, particularly
by the members of Royal Astronomical Society of
New Zealand (RASNZ).\footnote{
$\langle$http://www.rasnz.org.nz/$\rangle$.
} By now, these contributions, notably by Rod Stubbings, have enabled
a number of rare outburst detections and early circulation of these
phenomena.  Together with the collaboration with southern CCD observers,
the scientific achievements of transient object astronomy in the
southern hemisphere is explosively growing (e.g. microquasar V4641 Sgr:
\cite{stu99v4641sgriauc}; \cite{uem02v4641sgrletter};
\cite{uem02v4641sgr},
southern SU UMa-type dwarf novae: \cite{kat03v877arakktelpucma};
\cite{kat03bfara}; \cite{kat03nsv10934mmscoabnorcal86}), which will be
reviewed in later sections.

   Most recently, we started collaborating with the Nagoya University
team for identifying X-ray transients (see subsection \ref{sec:sci:xraynova})
with the Simultaneous-3color InfraRed Imager for Unbiased Survey
(SIRIUS) camera installed at the Infrared Survey Facility (IRSF)\footnote{
$\langle$http://www.saao.ac.za/facilities/irsf/irsf.html$\rangle$.
}
situated at the South African Astronomical Observatory, Southerland,
South Africa.
Several scientific achievements have already been issued
(\cite{kat03xtej1720iauc}; \cite{bab03igr17464iauc}; \cite{nag03j1720}).
A work on V359 Cen in collaboration with the Microlensing Observations
in Astrophysics (MOA) project\footnote{
$\langle$http://moa.scitec.auckland.ac.nz/$\rangle$.
} team has been also published \citep{kat02v359cen}.

\section{World-Wide Web and Anonymous FTP Service}

\subsection{Historical Development}

   From the very beginning of the VSNET, all messages posted to VSNET
public mailing lists are opened to the public.  A message can be relayed
(unless the sender otherwise specifies the usage) to third-party
members, or can be posted to a different mailing list or a public news
service.  In the earliest times, these messages were archived (they were
sometimes made public) at individual receivers.

   From the necessity of publicly providing standardized charts and
related materials (see subsection \ref{sec:standard}), we initially
used private anonymous ftp service operated at the Department of
Astronomy, Kyoto University.
This was replaced by the official VSNET anonymous ftp
service in 1995 July.  Contributed programs, notably VSNCHT written
by Rafael Barbera, to graphically display VSNET-format charts were
made public from the start of the service.  Magnitude summaries of
selected objects such as new novae, our own and contributed CCD
images of outbursting dwarf novae were soon made available through
the anonymous ftp service.

   We set up the official VSNET World-Wide Web (WWW) service\footnote{
$\langle$http://www.kusastro.kyoto-u.ac.jp/vsnet/$\rangle$.
} in 1995 June.  The web pages have been continuously updated,
particularly in announcing newly discovered transient object phenomena.
Even during these updates, we have paid all our efforts to preserve
the original URLs for future reference; almost all pages that existed
in the past can be tracked with the original URLs even now.  The other
design of the VSNET WWW system was in the combined usage of the WWW
and ftp services.  In the earliest times, not all internet users were
able to use window-based browsers.  We have thus set up two ways
(WWW and anonymous ftp) of access to the desired data.  With this
feature, a user is able to get necessary images or programs even
without a browser or a fast internet connection.  This feature became,
however, less important with the wide availablity of the window-based
browsers.  The ftp service soon included a complete archive of the
VSNET mailing list messages.

   In 1997 April, we started public WWW service of all archival
messages posted to the VSNET lists.  The WWW archive is automatically
updated by using the MHonArc system.\footnote{
$\langle$http://www.mhonarc.org/$\rangle$.
}  We once implemented the Namazu-based\footnote{
  The Namazu full-text WWW-based search engine was developed by one of the
  VSNET administrators, Hajime Baba. This is one of the best used systems
  among search engines for non-commercial, personal or academic uses.
  See $\langle$http://www.namazu.org/$\rangle$ for the Namazu project.
}
full-text search engine on the VSNET WWW service, but this
was discontinued because publicly available search engines now have
equivalent functions.

\subsection{Public Light Curve Archive}
\label{sec:publc}

   With the development of the VSNET WWW service, we started public
light curve archive service in 1996 August.\footnote{
$\langle$http://www.kusastro.kyoto-u.ac.jp/vsnet/LCs/index.html$\rangle$.
}  These light curves are drawn and regularly updated from {\it vsnet-obs}
reports, which were incorporated into the VSNET database (Appendix
\ref{sec:app:report}).  At present, the regular updates of these (static)
light curves archive are performed by a Java$^{\rm TM}$-based light curve
generator engine wrapping the Linux-based VSNET database system.

   The VSNET light curve archive also hosts light curves drawn from
the VSOLJ and Association Fran\c{c}aise des Observateurs d'Etoiles
Variables (AFOEV)\footnote{
$\langle$http://cdsweb.u-strasbg.fr/afoev/$\rangle$.
}
 public database, by courtesy of the respective
organizations.  These light curves can be easily reached from the
VSNET top page, as well as from the AFOEV website.

   We also implemented individual variable star pages\footnote{
$\langle$http://www.kusastro.kyoto-u.ac.jp/vsnet/gcvs/index.html$\rangle$.
} generated from the GCVS electronic edition.  These pages provide
handy links to the light curves (VSNET, VSOLJ, AFOEV) and links to
the relevant pages and charts on the VSNET.

\subsection{Data Search Engine and Light Curve Generator}
\label{sec:datasearch}

   Soon after the establishment of the data reporting system, we also
implemented a Common Gateway Interface (CGI)-based data search engine
in 1996 June.\footnote{
$\langle$http://www.kusastro.kyoto-u.ac.jp/vsnet/etc/searchobs.html$\rangle$.
}  This service returns the observations of a specified variable star
selected from the VSNET public database, which is coherently updated
from the regular variable star reports to the VSNET.  This type of
interactive variable star data browser was the first one in the world,
and became a prototype of the subsequent similar services.  This
interactive data search engine has been widely used and frequently
referred in the professional literature.  A number of WWW services,
including the well-known ``Latest Supernovae" page\footnote{
$\langle$http://www.rochesterastronomy.org/snimages/$\rangle$.
} by David Bishop, provides links to most up-to-date compilations of
variable star reports by directly referring to the VSNET data search
engine.  This WWW-based data search engine provides the modern-day
extension of the early-time magnitude summaries
(subsection \ref{sec:openingera}).

   In 1998 March, this VSNET data search engine furthermore started
providing a Java$^{\rm TM}$ applet-based interactive light curve
interface,\footnote{
$\langle$http://www.kusastro.kyoto-u.ac.jp/vsnet/etc/drawobs.html$\rangle$.
} with which a user can freely browse the data (both VSNET and
VSOLJ observations) with a graphical user interface (GUI).
This was one of the earliest Java$^{\rm TM}$ applications in public
astronomical service.

\section{VSNET Collaboration}
\label{sec:vsnetcollab}

\subsection{VSNET as World Center of Time-Series Observations of CVs}
\label{sub:cvcenter}

   As described in section \ref{sec:vsoljcolab}, the VSNET played a
pioneering and essential role in establishing a new modality of
internet-based amateur-professional collaboration.  This direction
was one of the main aims of the VSNET from the very start of its
internet presence.

   Among various kinds of variable stars, CVs are the ``canonical''
class of transient objects, as briefly reviewed in subsection
\ref{sec:newwindow}.  With the common interest in CVs and related
systems, the VSNET amateur-professional collaboration originally
mainly focused on CVs, especially on unpredictable outbursts of
dwarf novae.  The actual collaborative studies were done on the
existing VSNET list, most frequently on {\it vsnet-alert} and
{\it vsnet-obs}.  In most cases, real-time reports of visual detections
of outbursts in dwarf novae, usually after some verification process
involving the VSNET administrator team, triggered the actual observing
campaigns.  This process was usually performed within several hours and
a day of the detection.  This prompt reaction to event triggers of
transient objects later enabled an efficient reaction to gamma-ray burst
(GRB) triggers (section \ref{sec:sci:grb}).
The unique feature of the VSNET
campaigns on transient objects is that they are a collaborative effort
between visual observers and CCD observers.  The VSNET is historically
the first organization that realized the high productivity
involving traditional visual variable star observations, although
this importance in transient object astronomy had long been stressed
and had been a dream among researchers.

   The early-time successful cooperative works include:
recurrent nova V3890 Sgr (\cite{kat90v3890sgriauc}, phone call),
detection of superhumps in AQ Eri (\cite{kat91aqeri}, phone call and e-mail),
detection of superhumps in V1251 Cyg (\cite{kat91v1251cygiauc};
\cite{kat95v1251cyg}, phone call and e-mail),
WX Cet in 1991 (\cite{kat95wxcet} independent detection and e-mail),
EF Peg (\cite{kat91efpegiauc}; \cite{kat02efpeg}, phone call and e-mail),
HV Vir in 1992 (\cite{kat01hvvir}, e-mail),
SW UMa in 1992 (\cite{kat92swumasuperQPO}, phone call).
Since then, most information is relayed by e-mail and e-mail exploder
systems (most of the works were conducted upon response to real-time
outburst detection alerts reported through the VSNET):
V344 Lyr (\cite{kat93v344lyr}), HY Lup = Nova Lup 1993
(\cite{kat93novalup}), AK Cnc (\cite{kat94akcnc}; \cite{men96akcnc}),
T Leo (\cite{kun94tleo}),
KV And (\cite{kat94kvand}), AY Lyr (\cite{nog94aylyr}),
CY UMa (\cite{kat95cyuma}), KV And in 1994 (\cite{kat95kvand}),
FO And (\cite{kat95foand}), TT Boo (\cite{kat95ttboo}),
discovery of ER UMa (\cite{kat95eruma}), which will be described
in subsection \ref{sec:sci:dwarfnova}, PU Per (\cite{kat95puper}),
HS Vir (\cite{kat95hsvir}), GO Com (\cite{kat95gocom}),
DH Aql (\cite{nog95dhaql}), V1159 Ori (\cite{nog95v1159ori}),
HV Aur (\cite{nog95hvaur}), V725 Aql (\cite{nog95v725aql}, own detection),
RZ LMi (\cite{nog95rzlmi}), V1028 Cyg in 1995 (\cite{bab00v1028cyg}).
This continuous stream of scientific reports based on the VSNET
amateur-professional cooperations brought a great impact on the
community \citep{lil96vwhyi}.

   During this early-epoch collaborative works through the VSNET,
several noteworthy rare (typically once in a decade to several decades)
phenomena occurred and were studied in unprecedented detail
through the VSNET:
AL Com in 1995 (\cite{mat95alcomiauc}; \cite{nog95alcomiauc}: the outburst
was detected by an AAVSO member, soon relayed to the VSNET, which enabled
the detection of ``early superhumps", which are observed only for
several nights after the start of the outburst.  See \cite{kat96alcom};
\cite{nog97alcom} for our summaries of this event).\footnote{
The full details of the phenomenon, made public in the Keele CV Conference
in 1996 \citep{kat96alcomproc}, can be seen at
$\langle$http://www.kusastro.kyoto-u.ac.jp/vsnet/DNe/alcom1.html$\rangle$.
The star was nominated as the ``star of the year" in this conference.
}  This information on the VSNET further enabled extensive follow-up
studies by different groups (\cite{pat95alcomiauc}; \cite{pat96alcom};
\cite{pyc95alcomiauc}; \cite{pyc95alcom}; \cite{how96alcom};
\cite{szk96alcomIUE}).

   The 1996--1997 outburst of EG Cnc (\cite{sch96egcnciauc}) was
a great surprise.  This outburst was detected by Patrick Schmeer,
a VSNET member, and immediately relayed through the VSNET alert system.
Together with the early detection of early superhumps
(\cite{mat96egcnciauc}; \cite{mat98egcnc}), the object showed
an unexpected sequence of post-superoutburst rebrightenings
(\cite{kat97egcnc}).  This unexpected phenomenon was discovered by
ourselves and the collaborative efforts through the VSNET.\footnote{
$\langle$http://www.kusastro.kyoto-u.ac.jp/vsnet/DNe/egenc.html$\rangle$.
It was our greatest pleasure that a number of speakers in the Wyoming
CV Conference in 1997 presented this VSNET webpage as representing the
most unusual activity of a dwarf nova.
}  This detection of the spectacular rebrightening phenomenon, which
most clearly illustrated the power of real-time information exchange,
also led to a number of observational and theoretical papers
(\cite{osa97egcnc}; \cite{osa01egcnc}; \cite{pat98egcnc};
\cite{szk00TOADs}; \cite{szk02egcnchvvirHST}).  This impressive
phenomenon exhibited to the public, through the VSNET,
the ``real-time science" in the making.

\begin{figure*}[t]
  \begin{center}
    \FigureFile(160mm,80mm){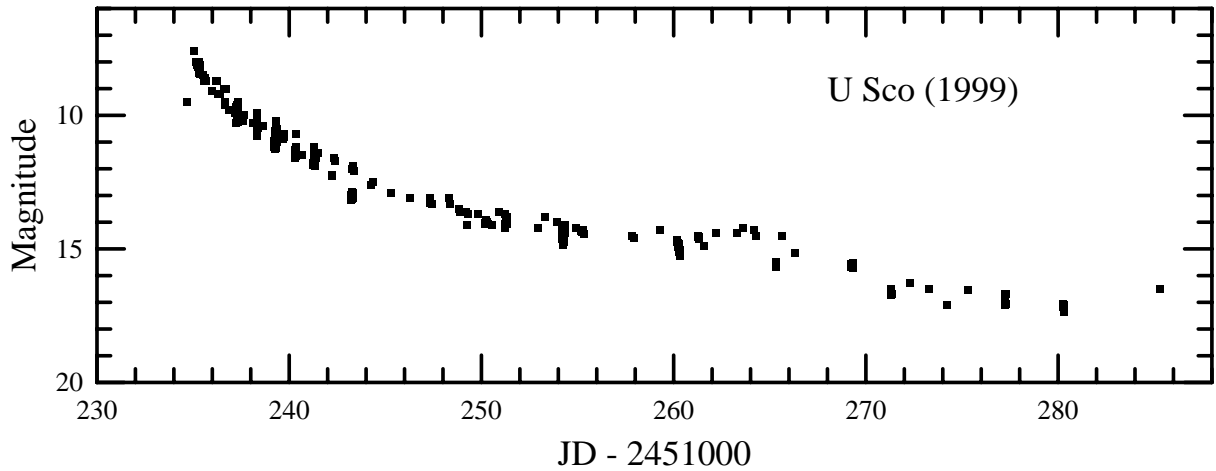}
  \end{center}
  \caption{Light curve of the 1999 outburst of the recurrent nova U Sco
  from VSNET observations.}
  \label{fig:usco}
\end{figure*}

   The 1999 outburst of the recurrent nova U Sco (\cite{sch99uscoiauc};
figure \ref{fig:usco})
again illustrated the ability of the international alert network system
provided by the VSNET.\footnote{
$\langle$http://www.kusastro.kyoto-u.ac.jp/vsnet/Novae/usco.html$\rangle$.
}  The outburst detection was made by Patrick Schmeer, and the alert was
immediately disseminated through the VSNET.  With this quick notification,
a very stringent upper limit became immediately available, which was
obtained only less than four hours before the outburst detection.  The
early news enabled an American observer to catch the real optical
maximum ($m_{\rm v}$ = 7.6), which was more than one magnitude brighter
than had been supposed for this recurrent nova.  It is very likely
the delays in delivering information were partly responsible for the
underestimate of the maximum magnitudes in past outbursts.
Almost all important findings near the maximum light (within one day
of detection) were obtained before the relevant IAUC was issued.
The VSNET collaborative study on this outburst additionally led to the
first-ever detection of eclipses during the outburst
(\cite{mat03usco}; \cite{hac00uscoburst}; \cite{hac00uscoqui}).
This outburst again produced a rich scientific outcome from various
researchers (\cite{kah99uscoSSS}; \cite{lep99usco}; \cite{mun99usco};
\cite{kiy99usco}; \cite{ike00cicamnsgr98usco}; \cite{anu00usco};
\cite{eva01uscoIR}; \cite{iij02uscospec}): when compared to the
results from previous outbursts (1979: \cite{bar79usco}; \cite{war79usco},
1987: \cite{ros88usco}; \cite{sek88usco}), the 1999 result could be
even referred to as the ``victory in the electronic era" of transient
object astronomy.

\begin{figure*}[t]
  \begin{center}
    \FigureFile(170mm,90mm){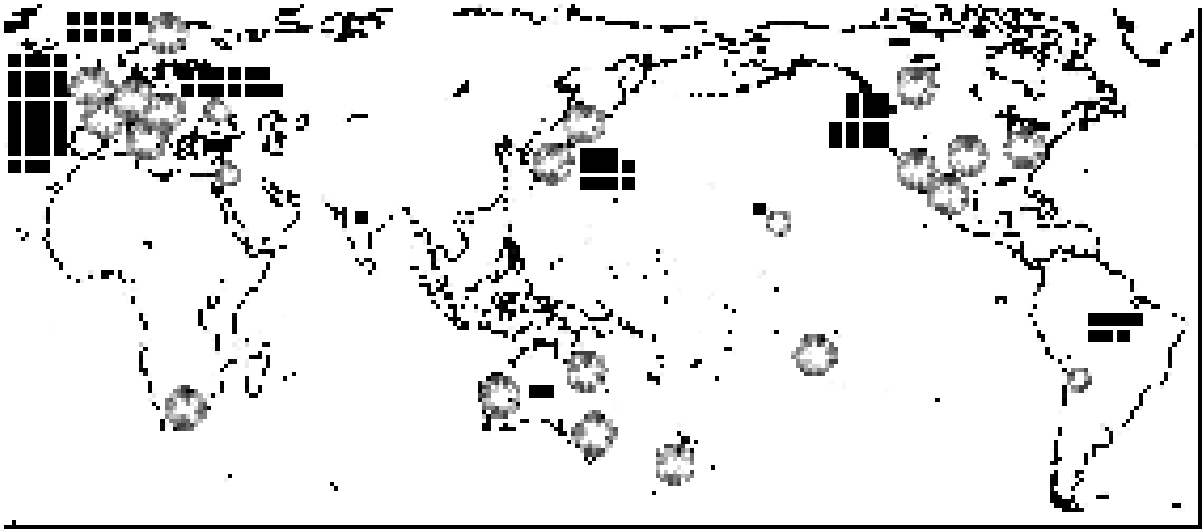}
  \end{center}
  \caption{Global distribution of observers of VSNET Collaboration.
  The large and small ``burst'' marks represent regular and occasional
  participants to CCD photometric campaigns.
  The number of filled squares relative numbers of represent (primarily)
  visual or photographic observations from various regions
  of the earth.
  }
  \label{fig:obsmap}
\end{figure*}

   After experiences with these, and other spectacular transient
phenomena (which will be reviewed in section \ref{sec:science}), we
set up new lists ({\it vsnet-campaign}\footnote{
$\langle$http://www.kusastro.kyoto-u.ac.jp/vsnet/Mail/\\vsnet-campaign/maillist.html$\rangle$.
} and other sublists), which primarily deal with campaigns
on selected targets.  With the establishment of these dedicated lists,
we started referring to our amateur-professional world-wide
collaboration group as VSNET Collaboration.  This (not too
tightly bound) group has become harboring subsequent intensive studies
undertaken in the VSNET.  Figure \ref{fig:obsmap} shows the global
distribution of contributors.

\subsection{VSNET Campaign Lists and Summaries}

   The VSNET campaign list and sublists were thus progressively
established since 2000, encapsulating a wide range of transient
astronomical phenomena.  The VSNET campaign lists are subdivided
into categories based on object classes (e.g. {\it vsnet-campaign-dn}
for dwarf novae, {\it vsnet-campaign-xray} for X-ray binaries).
Most recently, following the discovery of very unusual objects or
phenomena of public interest, we sometimes set up a separate list
focused on single objects (e.g. {\it vsnet-campaign-v4641sgr}\footnote{
$\langle$http://www.kusastro.kyoto-u.ac.jp/vsnet/Mail/\\vsnet-campaign-v4641sgr/maillist.html$\rangle$.
} for the microquasar V4641 Sgr (subsection \ref{sec:sci:xraynova}),
{\it vsnet-campaign-v838mon}\footnote{
$\langle$http://www.kusastro.kyoto-u.ac.jp/vsnet/Mail/\\vsnet-campaign-v838mon/maillist.html$\rangle$.
} for the most unusual stellar explosion (V838 Mon) with an astounding
light echo, {\it vsnet-campaign-sn2002ap}\footnote{
$\langle$http://www.kusastro.kyoto-u.ac.jp/vsnet/Mail/\\vsnet-campaign-sn2002ap/maillist.html$\rangle$.
} for the nearest hypernova SN 2002ap (see subsection \ref{sec:sci:SN}).
These new features enabled
interested theoreticians to share real-time information of these most
unusual objects.  The number of VSNET campaign individual lists is 36
(2003 August).

   Summaries of the activities of the VSNET Collaboration have been
compiled by Makoto Uemura and issued on a weekly basis (VSNET campaign
news).\footnote{
$\langle$http://www.kusastro.kyoto-u.ac.jp/vsnet/Mail/\\vsnet-campaign-news/maillist.html$\rangle$.
}  This information has been summarized as a yearly review
of the activity of the VSNET Collaboration.\footnote{
$\langle$http://www.kusastro.kyoto-u.ac.jp/vsnet/Summary/$\rangle$.
}  These VSNET campaign news and summaries comprise an important
(authorized) part of the nearly real-time contribution from the VSNET
to the other organizations (section \ref{sec:otheralert}), and are
sometimes cited in themselves as a convenient record of the activities of
particular objects.

   Nowadays the VSNET campaigns are more or less continuously
undertaken.  In order to inform the current targets of interest,
we have recently set up a notification list {\it vsnet-campaign-target}
to campaign contributors.

\subsection{Contribution to Other World-Wide Alert Systems}
\label{sec:otheralert}

   From the very start of the AstroAlert system\footnote{
$\langle$http://skyandtelescope.com/observing/proamcollab/astroalert/$\rangle$.
}
maintained by the Sky Publishing Co.,
the VSNET has been nominated as one of the authorized
information providers.
We now regularly contribute ``News from VSNET" to this alert system,
primarily notifying the transient astronomical phenomena of current
interest.  We also occasionally issue alerts on particularly urgent
phenomena and nova discoveries.  The news have been relayed to
the AAVSO, the {\it astro-l} mailing list, and are widely distributed as
the primary source of information on transient astronomical phenomena.

\section{Variable Star Related Activities}

   In addition to main scientific achievements (section \ref{sec:science}),
the VSNET group has been engaged in various activities in the field of
variable stars and variable star-related matters.  Here we show our
representative activities.

\subsection{Variable Star Identifications}
\label{sec:vsid}

   The VSNET group has been historically engaged in providing updated
information to the General Catalogues of Variable Stars (GCVS:
\cite{GCVS}; \cite{sam00GCVSbaltic}; \cite{sam02GCVSelectronic1};
\cite{sam03GCVSelectronic2}; \cite{NSVSibvs})
team,\footnote{
$\langle$http://www.sai.msu.su/groups/cluster/gcvs/gcvs/$\rangle$.
}
and provided a large amount of variable
star identifications and suggestions, mainly based on examination of
the historical literature with the modern technology.
The present most relevant lists are {\it vsnet-id} and
{\it vsnet-gcvs}\footnote{
$\langle$http://www.kusastro.kyoto-u.ac.jp/vsnet/Mail/vsnet-id/\\maillist.html$\rangle$ and
$\langle$http://www.kusastro.kyoto-u.ac.jp/vsnet/\\Mail/vsnet-gcvs/maillist.html$\rangle$.
}, which deal with variable star identifications and the GCVS revision
project, respectively.  The identifications reported to {\it vsnet-id}
are also relayed to the SIMBAD\footnote{
$\langle$http://simbad.u-strasbg.fr/Simbad$\rangle$
} office.  There is a series of solid papers
on systematic variable star identifications
(\cite{kat99var1}; \cite{kat99var2}; \cite{kat99var3}; \cite{kat99var4};
\cite{kat99var5}; \cite{kat99var6}; \cite{kat99var7}; \cite{kat99var8};
\cite{kat01v391sctv2435sgr}).
There have also been collaborations with the
Multitudinous Image-based Sky-survey and Accumulative Observations
(MISAO) project,\footnote{
$\langle$http://www.aerith.net/misao/$\rangle$.
}
which used their own CCD images to identify variable objects near
the cataloged positions
(\cite{yos99nsv25425}; \cite{yos00v854oph}; \cite{yos00v949sgr};
\cite{MISAOastrometry2}; \cite{MISAONSV}).  These collaborative
works were conducted also in collaboration with the VSOLJ.

   In identifying variable stars, our own chart-plotting system
(Appendix \ref{sec:app:chart}) has played an important role.

\subsection{New Variable Stars}
\label{sec:newvar}

   The VSNET has been providing the role of a world center of reporting
newly discovered variable stars.  There is a dedicated mailing list
{\it vsnet-newvar}.\footnote{
$\langle$http://www.kusastro.kyoto-u.ac.jp/vsnet/Mail/\\vsnet-newvar/maillist.html$\rangle$.
}  The reported new variable stars are usually checked by the VSNET
administrator group for identification with other astronomical sources
(e.g. X-ray sources, infrared sources, and emission-line stars).  These
identification processes are also conducted on the public mailing list.
These reports on new variable stars on {\it vsnet-newvar} are now
regarded as the primary source of original information by the GCVS
team (e.g. \cite{NameList75}; \cite{NameList76}; \cite{NameList77}).

   A preliminary list (newvar.cat) of newly reported, or objects with
some peculiarity (which can be candidate variable stars) has been updated,
and made public\footnote{
$\langle$ftp://vsnet.kusastro.kyoto-u.ac.jp/pub/vsnet/others/newvar.cat$\rangle$.
} as a reference for those who are looking for new variable stars and
variable star identifications.  Examination of the properties of
poorly known variable stars are also our regular works
(many results have been reported through {\it vsnet-gcvs} lists, as well
as through some solid publications such as \cite{kat03v1899sgr};
\cite{kat03uwaqr}; \cite{kat03zzaqr}, which were based on the ASAS-3
public data; see subsection \ref{sec:novaconfirm}).

\subsection{Confirmation of Novae}
\label{sec:novaconfirm}

   Confirmation of reported nova candidates is another regular work
conducted by the VSNET administrator team, as well as scientific work
in novae (subsection \ref{sec:sci:nova}).  This work takes advantage
of the established VSNET software and catalog systems in variable star
identifications (subsection \ref{sec:vsid}) and in confirming new
variable stars (subsection \ref{sec:newvar}).  The VSNET system in
screening and confirming newly reported nova candidates is one of the
most reliable and efficient among all the existing nova confirmation
systems.  In particular, the VSNET team has even succeeded in
recognizing novae from otherwise dismissed new variable star reports
(e.g. V1548 Aql: \cite{uem01v1548aqliauc}; \cite{kat01v1548aql}).
The recent discovery of V463 Sct (\cite{has00v463sct};
\cite{uem00v463sct}; \cite{kat02v463sct}) is another example of the
reliability of the VSNET nova identification system.\footnote{
See
$\langle$http://www.kusastro.kyoto-u.ac.jp/vsnet/\\Novae/hadv46.html$\rangle$
for the full story.
} The reverse case in that the VSNET played an important role in
disqualifying a suspected nova.  The most recent example was with
V4006 Sgr, which was originally reported as a probable nova in IAUC
\citep{nak03v4006sgriauc8113}.  This object was soon identified with
a known variable star through the public discussion in the VSNET.\footnote{
$\langle$http://www.kusastro.kyoto-u.ac.jp/vsnet/Novae/v4006sgr.html$\rangle$
}  The initial discovery announcement was subsequently readily corrected
within a day of the original announcement \citep{gre03v4006sgriauc8115}.

   Most recently, we refer to the public photometric real-time database
provided by the ASAS-3 team \citep{ASAS3}\footnote{
$\langle$http://www.astrouw.edu.pl/$^{\sim}$gp/asas/asas\_asas3.html$\rangle$
} for confirming southern nova suspects.
The most recent successful example includes V2573 Oph = Nova Oph 2003
(\cite{tak03v2573ophiauc8166}).
This application of ASAS-3 public data is also true for confirming
new variable stars (e.g. V2552 Oph = Had~V98: \cite{kat03hadv98}) and
better characterizing already known variable stars
(e.g. \cite{kat03v504cen}).  The individual variable star pages on
VSNET website provides convenient links to the ASAS-3 pages
(see subsection \ref{sec:publc}).

   Upon recognition of new novae, the VSNET team makes an announcement
to nova researchers and the variable star community to enable early-time
confirmation and follow-up observations.  The VSNET is now recognized as
one of the most powerful media for disseminating such a kind of alerts
to the open community, and is relied on by many professional
nova researchers.

   For this purpose, VSNET takes an ``open policy'' of any nova (and
supernova, variable star etc.) discovery announcements,
i.e. such announcements will be immediately released and made public.
The full reasoning of the policy and the actual recommended reporting
procedure is described in
$\langle$http://www.kusastro.kyoto-u.ac.jp/vsnet/etc/discovery.html$\rangle$
and the links from this page.  The VSNET has a dedicated list for
announcing on new novae {\it vsnet-discovery-nova},\footnote{
$\langle$http://www.kusastro.kyoto-u.ac.jp/vsnet/Mail/\\vsnet-discovery-nova/maillist.html$\rangle$.
} although this kind of information is usually also
posted to {\it vsnet-alert}
from the convention.  In any case, readers are strongly recommended to
refer to the above page before attempting to make an actual nova
discovery report.

\subsection{Confirmation of Supernovae}
\label{sec:conf:SN}

  As well as confirmation of nova candidates, confirmation of suspected 
supernovae is also undertaken from the very beginning of the VSNET.  
The first example was SN 1995D in NGC 2962 \citep{nak95sn1995Ddiscovery}, 
which was discovered by Reiki Kushida.  The object was confirmed on the 
same night in Japan by three individual observers
(vsnet-alert 30, 31, 32, 1995 February)\footnote{
$\langle$http://www.kusastro.kyoto-u.ac.jp/vsnet/Mail/vsnet-alert/\\msg00030.html$\rangle$ and so on.}, and the discovery announcement led the
multicolor photometry starting soon after the discovery
\citep{sad96sn1995D}.

  Spectroscopic confirmation of several supernovae has also been 
reported on the VSNET lists.  The earliest one was for SN 1995al
(vsnet-alert 266, 
1995 November),\footnote{
$\langle$http://www.kusastro.kyoto-u.ac.jp/vsnet/Mail/vsnet-alert/\\msg00266.html$\rangle$.
}
which was distributed well earlier than the formal publication
in IAUC \citep{aya95sn1995alSpec}.  Such a quick distribution
of spectral type of supernova made it possible to schedule larger
telescopes or space-borne instruments to observe it.  The nearest
type-Ic hypernova SN 2002ap (vsnet-alert 7120, 2002 January)\footnote{
$\langle$http://www.kusastro.kyoto-u.ac.jp/vsnet/Mail/alert7000/\\msg00120.html$\rangle$.
}
was the most successful recent example (\cite{kin02sn2002apSpecIAUC}, 
\cite{kin02sn2002apSpecApJ}), which was followed by
many instruments including the Subaru telescope \citep{kaw02sn2002apspecpol}
or the XMM-Newton \citep{sor02sn2002apXMM}.
Plentiful information, including prediscovery observations
\citep{yam02sn2002appredisc}, has been posted to the 
specially made sublist {\it vsnet-campaign-sn2002ap}.

   These contributions to early-time observation are summarized
in subsection \ref{sec:sci:SN}, as well as the scientific results drawn
from them.

\subsection{Hosting Announcements of External Organizations}

   The VSNET has been playing a role providing an electronic 
medium for announcements made by other variable star-related organizations.
The examples include the early announcements of AAVSO Newsletters
(vsnet-alert 14, 1995 January), Alert Notices (vsnet-alert 58, 1995
March), News Flashes (vsnet-alert 344, 1996 February),
website announcement (vsnet-alert 178, 1995 August),
ALEXIS\footnote{
$\langle$http://nis-www.lanl.gov/$^{\sim}$asocops/alxhome/$\rangle$.
} transients (vsnet-alert 321, 1996 February),
International TOAD Watch website\footnote{
$\langle$http://www-astro.physics.ox.ac.uk/$^{\sim}$erik/itw/$\rangle$.
} (vsnet-alert 439, 1996 June),
website for Hungarian Astronomical Association --
Variable Star Section\footnote{
$\langle$http://vcssz.mcse.hu/$\rangle$.
} (HAA/VSS, vsnet 756, 1996 September),
website for Center for Backyard Astrophysics\footnote{
$\langle$http://cba.phys.columbia.edu/$\rangle$.
} (CBA, vsnet 872, 1996 November),
website for British Astronomical Association,
Variable Star Section\footnote{
$\langle$http://www.britastro.com/vss/$\rangle$.
} (BAAVSS, vsnet 893, 1996 November),
website for Group of Amateur Astronomers, Czechia,
(GAMA, vsnet-alert 596, 1996 November),
IAU Commission 42, Bibliography on Close Binaries\footnote{
$\langle$http://a400.sternwarte.uni-erlangen.de/ftp/bcb/$\rangle$.
} (BCB, vsnet 981, 1997 January), and numerous announcements on
call for observations, electronic publications, and international
conferences.\footnote{
We now have a dedicated list for conference announcements:
$\langle$http://www.kusastro.kyoto-u.ac.jp/vsnet/Mail/\\vsnet-conference/maillist.html$\rangle$.
}  The VSNET website and mailing lists thus have been
{\it a world center of variable star-related announcements}.

\section{Scientific Achievements of VSNET}
\label{sec:science}

\subsection{Dwarf Novae}
\label{sec:sci:dwarfnova}

\begin{figure*}
  \begin{center}
    \FigureFile(160mm,220mm){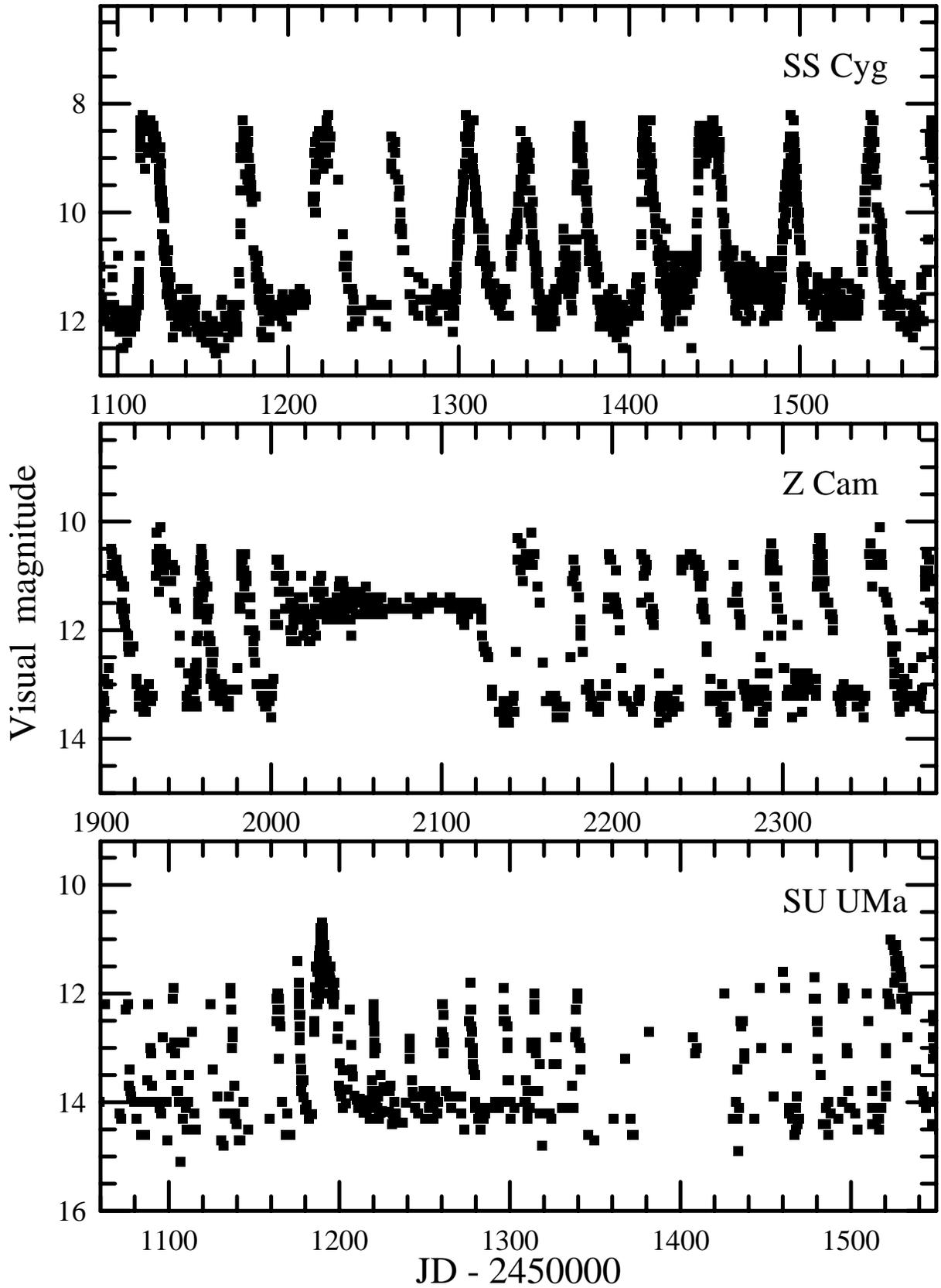}
  \end{center}
  \caption{Long-term light curve of three representative classes
  of dwarf novae (SS Cyg, Z Cam, and SU UMa) drawn from VSNET
  observations.  Portions of the light curves show variation in
  outburst cycle lengths and quiescent magnitude (SS Cyg), a standstill
  (Z Cam), and two superoutbursts (SU UMa).
  }
  \label{fig:dn}
\end{figure*}

   Dwarf novae (cf. \cite{osa96review}) have historically been the
best-studied class of objects by the VSNET Collaboration and its precedent
mailing list-based cooperation (hereafter referred to the VSNET
Collaboration throughout this section).  Figure \ref{fig:dn} shows
a comparison of long-term light curves of three representative classes
of dwarf novae (cf. \cite{osa96review}) drawn from the VSNET
observations.\footnote{
  These data of various kinds of variable stars are freely available
  (from the URL in subsection \ref{sec:datasearch}), or upon request to the
  VSNET administrator (vsnet-adm@kusastro.kyoto-u.ac.jp).  There is no
  specially requested authorization for the usage of these data in
  scientific publications other than the usual scientific manner of
  acknowledgement (please refer to the message from the data search engine
  for a recommended form of reference).
  The data usually cover observations since 1995, and are particularly
  convenient for making correlation analysis with other modern observations
  (e.g. spectroscopy and multiwavelength observations).
}

   The early work (up to 1996) almost immediately doubled the number
of SU UMa-type dwarf novae, by the detection of superhumps
(\cite{vog80suumastars}; \cite{war85suuma}; \cite{war95suuma}) during
their superoutbursts.  In typical SU UMa-type dwarf novae, superoutbursts
only occur approximately once per year, and last for $\sim$10 d.
The intervals of superoutbursts (supercycles) are not constant, making
it difficult to plan a scheduled observation.  The observations of
SU UMa-type dwarf novae in the past were more or less TOO-type
observations (subsection \ref{sec:newwindow}).  We first prepared a
comprehensive list of candidates of SU UMa-type dwarf novae, selected
by various criteria from the literature and from our own observations.
This strategy was in some way similar to the other projects such as:

\vskip 2mm
\begin{itemize}
\item High-galactic latitude CV search:
\citet{how90highgalCV}; \citet{szk92CVspec}; \citet{dri94highgalCV};
\citet{aug95highgalCV}
\item Its descendant concept of ``Tremendous Outburst Amplitude
Dwarf Novae" (TOADs): \citet{how95TOAD}; \citet{how95TOADproc};
\citet{szk95TOAD}; \citet{how00TOAD}
\item Recurrent Object Programme:\footnote{
$\langle$http://www.theastronomer.org/recurrent.html$\rangle$.
} e.g. \citet{poy96alcom}, and the
related extensive work by \citet{VanmunsterOutburstActivity}
\end{itemize}
\vskip 2mm

   Although all of these other projects or works basically relied on
catalog selections (e.g. \cite{how95TOAD}) or were based on the (usually
poorly known) past activities.  Our strategy was different in that:
(1) our selection was based on comprehensive and extensive search
through all available literature (this work was conducted in collaboration
with the VSOLJ), (2) we had examined the Palomar Observatory Sky Survey
(POSS) plates and our own systematic CCD survey to get unique
identifications and more reliable outburst amplitudes of the objects,
(3) we applied a theoretical background (e.g. \cite{osa85SHexcess};
\cite{osa89suuma}; \cite{ich94cycle}) for selecting the objects,
and (4) we had amateur collaborators to share our interest in monitoring
faint dwarf novae with a CCD (subsection \ref{sec:vsoljcolab}).
The combination of these factors brought an unprecedented success in
discovering new SU UMa-type dwarf novae.  Since the early success
story was already described in subsection \ref{sub:cvcenter}, we mainly
focus on the recent contributions of the VSNET Collaboration to this
field.

\subsubsection{Discovery of ER UMa Stars}
\label{sec:sci:eruma}

   In the canonical picture of SU UMa-type dwarf novae, the mass-transfer
(driven by angular momentum loss from the binary) in these systems
was believed to be the result of gravitational wave radiation (GWR:
cf. \cite{pac67CVGWR}; \cite{fau71GWR}; \cite{tut79GWR}; \cite{taa80GWR};
\cite{pac81CVGWR}),
since the fully convective secondary at this ($\leq$2 hr) orbital
period ($P_{\rm orb}$) is generally considered improbable to sustain
magnetic braking, which plays a major role in longer period
($P_{\rm orb}\geq$3 hr) CVs
(\cite{rit76CVevolution}; \cite{rit80CVevolution}; \cite{why80CVevolution};
\cite{pac81CVevolution}; \cite{rap82CVevolution}; \cite{rap83CVevolution};
\cite{spr83periodgap}; \cite{pac83CVecolution};
\cite{ver84periodgap}; \cite{pat84CVevolution};
\cite{rit85CVperiodgap}).
Since the angular momentum loss by GWR is a unique function of
component masses and orbital separation, and the masses of the primary
white dwarfs and the main-sequence secondary stars only have a small
degree of diversity, the SU UMa-type stars were historically considered
as an ``one parameter system", i.e. the basic properties are determined
by a single parameter, $P_{\rm orb}$.

   This concept so widely prevailed that statistics and classifications
were usually given following this concept (\cite{war95suuma};
\cite{vog93suuma}), and that numerical simulations implicitly assumed
this canonical picture \citep{osa89suuma}.  In observations, a search
for new SU UMa-type dwarf novae was usually restricted to dwarf novae
with long recurrence times (the best example being the Recurrent Object
Programme).  Our discovery, however, of the ER UMa stars completely
changed this picture.

   ER UMa (=PG~0943+521) was originally classified as a novalike object
\citep{gre82PGsurveyCV} selected from its ultraviolet excess
(Palomar-Green Survey: \cite{PGsurvey}, see also \cite{rin93PGCV}).
This object was, together with other PG CVs selected from
\citet{gre82PGsurveyCV}, regularly monitored by the VSOLJ members for
potential activities.  In 1992, Makoto Iida (VSOLJ) noticed that
this object shows dwarf nova-like outbursts (\cite{iid94eruma},
see {\it vsnet-history} archive for full details in the public reports
circulated in 1992--1993).  Because of the presence of long-lasting
states of intermediate brightness (which later turned out to be
superoutbursts), this star was originally considered to be a
Z Cam-type dwarf nova, which is characterized by the presense of
standstills in addition to dwarf nova-type outbursts
(see e.g. \cite{hel01book}, Sect. 5.4; see also \cite{war74vysclzcam};
\cite{mey83zcam}; \cite{opp98zcam}).

   In early 1994, following the detection of a bright outburst by
Gary Poyner, our CCD observations revealed the presence of superhumps.
Combined with the visual observations electronically reported, this
object was finally identified as an SU UMa-type dwarf nova with an
unexpectedly short (43 d) supercycle ($T_{\rm s}$).  The shortest
$T_{\rm s}$ before the discovery of ER UMa was 134 d of YZ Cnc
(\cite{szk84AAVSO};
\cite{vanpar94suumayzcnc}).  It later turned out through a discussion
at the Padova CV Conference in 1994\footnote{See also
  \citet{hon95erumarzlmiv1159oriv446her}; the object was originally
  presented in the conference as the best observational proof for
  the mass-transfer burst model of dwarf nova outbursts.  Our detection
  of superhumps was presented in a later talk by \citet{osa95DNproc},
  which invoked a hot discussion, against the original topics to be
  presented by this talk, regarding the true nature of ER UMa.
  At the time of this conference, the tone of the audience was even
  skeptical about the presence of an SU UMa-type dwarf nova with such
  a short supercycle.
} that this object was independently studied at least by two groups:
by the RoboScope (\cite{hon92CCDphotometry}) team \citep{rob95eruma}
and by \citet{mis95PGCV}.  Among all groups, the VSNET team was the
first to unambiguously identify the nature of this object
\citep{kat95eruma}.  Spectroscopic \citep{tho97erumav1159ori}
as well as photometric \citep{kat95erumaqui} identifications
of the orbital periods later confirmed this identification.

   Once this discovery was announced, new members of the group of
ER UMa stars were immediately identified through impetuous real-time
competitions through the VSNET public lists:
V1159 Ori (\cite{nog95v1159ori}; \cite{rob95eruma})
and RZ LMi (\cite{rob95eruma}; \cite{nog95rzlmi}).
It later turned out that the supercycle of V1159 Ori was independently
recognized by \citet{pat95v1159ori}, but was only interpreted within
the classical framework.  The supercycle of RZ LMi is exceptionally
short (19 d), which is the shortest record of supercycles up to now
[see also \citet{kat99erumareview} for the basic observational review,
and \citet{kat02v503cyg}; \citet{kat03erumaSH} for recent discussions of
ER UMa stars].  Later additions to these ER UMa stars include
DI UMa \citep{kat96diuma}, an RZ LMi-like system, and IX Dra
\citep{ish01ixdra}, both of which were discovered by the VSNET
Collaboration.

   From the standpoint of the disk-instability model, these discoveries
immediately led to theoretical interpretations
(\cite{osa95eruma}; \cite{osa95rzlmi}), and significantly contributed
to the ``unified theory of dwarf novae" (\cite{osa96DNproc};
\cite{osa96review}).
These theoretical calculations indicate an unexpectedly (from the
canonical picture based on the GWR-driven angular momentum loss)
high mass-transfer rate (ER UMa: \cite{osa95eruma}).  The extremely
short $T_{\rm s}$ of RZ LMi requires an additional (still poorly understood)
mechanism \citep{osa95rzlmi}.  This theoretical effort recently
led to ramifications of ideas including the effect of extremely low
mass ratio ($q$ = $M_2$/$M_1$) \citep{hel01eruma} or the effect
of irradiation \citep{bua02suumamodel}, which need to be investigated
by future work.  It has been speculated \citep{hel01eruma} that these
mechanisms are partly responsible for the manifestation of the
unusual properties of still the still poorly understood
WZ Sge-type dwarf novae,
which will be discussed later (subsection \ref{sec:sci:wzsge}).

   The required high mass-transfer is still a mystery.  Although there
have been a number of suggestions, including the long-term effect
of nova explosions
(\cite{kat95eruma}; \cite{sch98novaCVevolution}; \cite{pat98evolution};
\cite{kol01mdotvar}), which is, in some sense, a modern extension of
the ``old" discussion of the ``nova hibernation" scenario
(\cite{Hibernation}; \cite{pri86hibernation}; \cite{liv87hibernation};
\cite{liv88novaoutbursthibernation}; \cite{due92novahibernation}; 
\cite{nay92Hibernation}; \cite{muk95hibernation}),
irradiation-induced mass-transfer feedback
(\cite{mcc98masstransfercycle};
\cite{mcc98CVirradiationevolution}), none of them has succeeded in
explaining the required high mass-transfer rates in ER UMa.  Recent
observations (\cite{kat01v1159ori}; \cite{fri99diuma}) suggest that
the high mass-transfer rates in ER UMa stars are less likely the result
from a secular evolutionary effect, but may be more related to the
activity cycle in the secondary star
(\cite{bia88CVcycle}; \cite{bia90CVcycle}; \cite{ak01CVcycle}).
The origin of the unusually high activity of ER UMa stars is still
an open question.

   From the observational side, there has been a systematic search,
mainly conducted by the VSNET Collaboration and relevant
results communicated to the public in real time, to find
intermediate systems between ER UMa stars and classical SU UMa-type
dwarf novae.  Several interesting objects were recognized
as a result:
NY Ser = PG~1510+234 (SU UMa-type dwarf nova in the period gap:
\cite{nog98nyser}),
HS Vir = PG~1341$-$079 (SU UMa-type dwarf nova with very short outburst
recurrence times: \cite{kat95hsvir}; \cite{kat98hsvir};
\cite{men99hsvir}; \cite{kat01hsvir}; \cite{men01wxcettumencuvelhsvir}),
SX LMi (low amplitude SU UMa-type dwarf nova: \cite{nog97sxlmi};
\cite{wag98sxlmi}; \cite{kat01sxlmi}),
CI UMa (SU UMa-type dwarf nova with irregular supercycle behavior:
\cite{nog97ciuma}),
V1504 Cyg (SU UMa-type dwarf nova with a short $T_{\rm s}$,
\cite{nog97v1504cyg}),
V503 Cyg (unusual SU UMa-type dwarf nova with unusual outburst behavior,
$T_{\rm s}\sim$89 d: \cite{kat02v503cyg}; \cite{har95v503cyg}),
SS UMi (normal SU UMa-type dwarf nova with a short $T_{\rm s}\sim$84.7 d:
\cite{kat98ssumi}; \cite{kat00ssumi}),
BF Ara (normal SU UMa-type dwarf nova with the shortest
$T_{\rm s}\sim$84.3 d: \cite{kat01bfara}; \cite{kat03bfara}),
V344 Lyr (large-amplitude SU UMa-type dwarf nova with a short supercycle:
\cite{kat93v344lyr}; \cite{kat02v344lyr}),
MN Dra = Var73 Dra (SU UMa-type dwarf nova with $T_{\rm s}\sim$60 d:
\cite{nog03var73dra}).

   The evolution of superhumps in ER UMa stars is also known to be
unusual.  \citet{kat96erumaSH} reported the early presence of
large-amplitude superhumps, which later turned out to be part of
an unexpected early phase reversal of superhumps \citep{kat03erumaSH}.
This phenomenon resembles the so-called
late superhumps (\cite{hae79lateSH}; \cite{vog83lateSH};
\cite{vanderwoe88lateSH}; \cite{hes92lateSH}) seen in late stages of
superoutbursts in SU UMa stars, but the striking difference is
that they appear in the early stage of the superoutburst.
The origin of this phase reversal is not yet understood.
\citet{gao99erumaSH} reported some peculiar features in superhumps of
ER UMa.  \citet{mas97SXTeruma} suggested a possible link between
ER UMa stars X-ray transients by comparing the evolution of their
superhumps.

   To summarize, the discovery of ER UMa stars brought a revolutionary
turning point of dwarf nova studies: the SU UMa-type dwarf novae
are no longer ``one-parameter systems" (see e.g. \cite{war98CVreviewWyo};
\cite{nog98CVevolution}).  Research on short-$P_{\rm orb}$ CVs
(mainly SU UMa-type dwarf novae) is continuously broadening,
and is now becoming one of ``the mainstreams" of CV research.

\subsubsection{WZ Sge-Type Dwarf Novae}
\label{sec:sci:wzsge}

\begin{figure*}
  \begin{center}
    \FigureFile(150mm,100mm){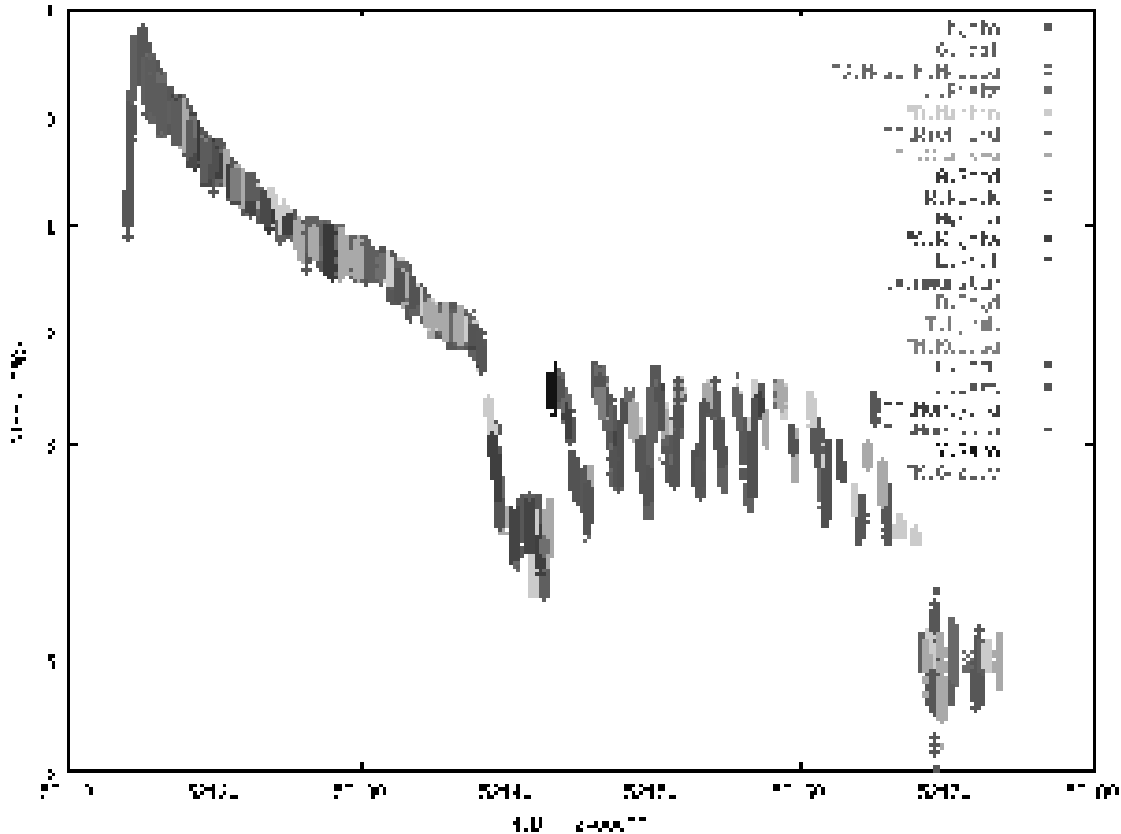}
  \end{center}
  \caption{CCD light curve of the 2001 outburst of WZ Sge by the
  VSNET Collaboration.}
  \label{fig:wzlc}
\end{figure*}

   WZ Sge-type dwarf novae are a peculiar subtype of SU UMa-type dwarf
novae.\footnote{
  Historically, the concept of WZ Sge-type dwarf novae referred to
  large-amplitude, rarely outbursting dwarf novae (cf. \cite{bai79wzsge};
  \cite{dow81wzsge}; \cite{pat81wzsge}; \cite{odo91wzsge}).  This
  extension of this standpoint led to the concept of Tremendous
  Outburst Amplitude Dwarf Novae (TOADs), mainly proposed by
  \citet{how95TOAD}.  WZ Sge-type dwarf novae were sometimes even referred
  to as extreme TOADs (cf. \cite{how03wzsgeIR}).  The modern viewpoint,
  however, is that the outburst amplitude is one of the general properties of
  WZ Sge-type dwarf novae, rather than the definition.  WZ Sge-type
  dwarf novae are now better recognized as an extreme population of
  SU UMa-type dwarf novae (\cite{kat01hvvir}, see also the discussion
  in relation to the ``TOAD" concept: \cite{kat03hodel}).  We therefore
  primarily use the term WZ Sge-type dwarf novae rather than TOADs.
}  The properties of WZ Sge-type dwarf novae include: (1) long
($\sim$ 10 yr or more) outburst recurrence time, (2) large ($\sim$ 8 mag)
outburst amplitude, (3) very few, or sometimes no, occurrence of
(isolated) normal outbursts, (4) presence of ``early superhumps",
which are modulations having periods very close to the orbital periods,
during the earliest stage of superoutbursts, and (5) frequent occurrence
of post-superoutburst rebrightening (see \cite{kat01hvvir} and
\cite{kat02wzsgeESH} for modern observational reviews).

   These properties are difficult to explain, even with the recent
progress of the disk-instability theory, and the WZ Sge-type dwarf
novae have been continuously providing challenging problems to both
theoreticians and observers.  One of the main difficulties reside in
their extremely long recurrence times.  If one assumes the standard
disk-instability model, the recurrence time is limited by the diffusion
time in quiescence (\cite{ich94cycle}; \cite{osa96review}).  In order
to avoid a thermal instability resulting from this diffusion, one needs
to assume an extremely low viscosity parameter in quiescence
($\alpha_C < 0.00005$, \cite{sma93wzsge}; $\alpha_C < 0.003$,
\cite{osa95wzsge}).  Although the origin of such a low viscosity is
becoming positively resolved by considering a very cold disk with
a low electronic conductivity \citep{mey98wzsge}, there still exist
a number of arguments to avoid an extremely low quiescent viscosity.
For example, \citet{las95wzsge}, \citet{war96wzsge} assumed
evaporation/truncation of the inner disk to prevent thermal instability
to occur.  \citet{ham97wzsgemodel} and \citet{bua02suumamodel} presented
slight modifications of these ideas.  The models, however, are expected
to show much shorter outburst lengths than in the low-$\alpha_C$ model
(\cite{osa95wzsge}).  Some models thereby assume an enhanced
mass-transfer during a superoutburst (as originating from
\cite{pat81wzsge} and extended by \cite{pat02wzsge}), there is,
however, no concrete observational evidence supporting
this supposed enhanced
mass-transfer \citep{osa03DNoutburst}.  A careful analysis of the
observations of the 2001 outburst of WZ Sge also supports this lack of
mass-transfer (R. Ishioka et al. in preparation).  Some authors
ascribed the existence of a brown-dwarf secondary star to the origin
of the required low $\alpha_C$ (\cite{mey99diskviscosity}).
This possibility, together with the expectations from the theoretical
viewpoint (e.g. \cite{pac71CVminimumperiod}), led to a wide interest
to search for brown dwarfs in WZ Sge-type dwarf novae
(\cite{how97periodminimum}; \cite{pol98TOAD}; \cite{cia98CVIR};
\cite{vantee99v592her}; \cite{pat01SH}; \cite{how01llandeferi};
\cite{men01j1050}; \cite{kat02v592her}; \cite{men02CVBD}; \cite{lit03CVBD}).
Although earlier reports tended to suggest the presence of a brown dwarf,
the evidence is less clear from more recent detailed studies
(\cite{ste01wzsgesecondary}; \cite{kat02v592her}; \cite{men02CVBD};
\cite{how02llandefpegHST}; \cite{lit03CVBD}).
Most recently, the observational lack of CVs having
brown dwarf secondaries is even becoming a serious problem
(\cite{bar03CVminimumperiod}).  In all aspects of disk-instability
problems, origin of disk viscosity, and late-stage evolution of
compact binaries, WZ Sge-type dwarf novae continue to be key objects.

   Since the outbursts of WZ Sge-type dwarf novae are quite rare,
these objects have best illustrated the ability and the achievement
of the VSNET as the real-time network.  Before the VSNET was established,
most (presumable) WZ Sge-type outbursts were only poorly studied
(e.g. AL Com: \cite{ber64alcom}, UZ Boo: \cite{bai79wzsge},
PQ And: \cite{mca88pqandiauc}; \cite{hur88pqandiauc1}; \cite{hur88pqandiauc2};
\cite{hur88pqandiauc3}, GW Lib: \cite{maz83gwlibiauc},
other systems: \cite{ric92wzsgedip}).  The establishment of the VSNET
has been one of the greatest steps toward understanding the WZ Sge-type
dwarf novae.  In particular, without the collaboration with the VSOLJ
and without the VSNET alerts, the early part of the 2001 outburst of
WZ Sge would have still remained a mystery.

   The earliest work on WZ Sge-type dwarf novae was on HV Vir in 1992.
Although this object was originally recorded as a classical nova in 1929
(\cite{sch31hvvir}; \cite{due84hvvir}), we started monitoring for a
future potential outburst as a candidate WZ Sge-type dwarf nova
\citep{kat01hvvir}.  This outburst was discovered by Patrick Schmeer
\citep{sch92hvviriauc}, and was immediately relayed through the alert
network.  The VSNET team was the first to record periodic modulations
(early superhumps) in the light curve.\footnote{
This was immediately reported to the CBAT (see {\it vsnet-history} archive),
but was not published for an unknown reason.  See \cite{kat01hvvir}
more details.
}  The final result on this superoutburst was published as
\citet{kat01hvvir}, which has been chosen as a best modern reference on
new WZ Sge-type stars in \citet{Astrophysics2002}.  After $\sim$10 yr,
the object again went into superoutburst, which was observed in detail
by the VSNET Collaboration \citep{ish03hvvir}.

   The next object was UZ Boo in 1994 (\cite{iid94uzbooiauc};
\cite{sze94uzboo}).  The outburst detection was also immediately relayed
through the VSNET alert network.
Due to the short visibility, only a preliminary
superhump period of 0.0619 d was obtained \citep{kat01hvvir}, although
the post-superoutburst rebrightenings reported through the VSNET have
raised a possible link between WZ Sge-type dwarf novae and soft X-ray
transients (SXTs, or X-ray novae, subsection \ref{sec:sci:xraynova})
\citep{kuu96TOAD}.  This suggestion has been more substantiated by
\citet{uem00j1118pasj}, \citet{kuu00wzsgeSXT}, \citet{kuu01j1118};
this relation is becoming one of the major contemporary topics in
SXT outbursts.

   In 1995 and 1996, outstanding WZ Sge-type outbursts of AL Com and
EG Cnc occurred.  These outburst detections were immediately relayed
through the VSNET, and produced a wealth of scientific results as
already introduced in subsection \ref{sec:vsnetcollab}.  AL Com again
underwent an superoutburst in 2001.  A timely outburst announcement by
Steve Kerr on VSNET enabled early-time observations when the object
was still rising.  The detection of growing early superhumps was
already reported \citep{ish02wzsgeletter}, which was actually the first
detection of growing early superhumps before the 2001 spectacular
outburst of WZ Sge itself.

   In late 2000, another spectacular outburst of RZ Leo occurred,
which had long been suspected as a candidate WZ Sge-type dwarf nova
(\cite{mcn85rzleoiauc}; \cite{mat85rzleoiauc}; \cite{cri85rzleoiauc};
\cite{ric85rzleo}).  The VSNET Collaboration succeeded in detecting
both early superhumps and ordinary superhumps (\cite{ish00rzleoiauc};
\cite{ish01rzleo}), giving credence to the WZ Sge-type nature of
this dwarf nova.

   The unexpected outburst of the prototype WZ Sge in 2001 was one of
the greatest astronomical phenomena in recent years.  It was
detected by a Japanese amateur observer, Tomohito Ohshima, was
immediately relayed to the VSNET, enabling early coverage
(\cite{ish01wzsgeiauc7669}; \cite{ish02wzsgeproc}; \cite{kuu02wzsge}),
which was one of the greatest achievements of the VSNET as an alert
network (see figure \ref{fig:wzlc}).
This outburst produced a burst of scientific results
from a number of researchers, both ground-based and satellite-borne:
detection of early superhumps \citep{kat01wzsgeiauc7672}, growth of
ordinary superhumps (\cite{kat01wzsgeiauc7678}; \cite{kat01wzsgeiauc7680}),
spectroscopic detection of spiral patterns (\cite{ste01wzsgeiauc7675};
\cite{bab01wzsgeiauc7678}) whose details have been published in
\citet{bab02wzsgeletter}, ``real-time" modeling of the outburst
\citep{can01wzsge}, Chandra observation \citep{whe01wzsgeiauc7677},
HST observation \citep{kni02wzsgeHSToscillation}, far ultraviolet
spectroscopy \citep{lon03wzsgeFUSE}, infrared spectroscopy
\citep{how03wzsgeIR}, extensive photometry (\cite{ish02wzsgeletter};
\cite{pat02wzsge}; R. Ishioka et al. in preparation), and theoretical
modeling of the superhump light curve \citep{osa03wzsgetomography}.

   Other WZ Sge-type dwarf novae observed and reported by the
VSNET Collaboration include: UW Tri (\cite{kat01uwtri}),
LL And (\cite{kat03lland};
data cited in \cite{how94lland}, \cite{how96lland}),
V2176 Cyg (\cite{nov01v2176cyg}),
CG CMa (\cite{kat99cgcma}), V592 Her (\cite{kat02v592her}).
Among them, V2176 Cyg showed a ``dip" phenomenon,
noted for first time since the 1995 outburst of AL Com (\cite{nog97alcom}),
and V592 Her was confirmed to be a dwarf nova with an exceptionally
large outburst amplitude.

   The VSNET Collaboration has first systematically demonstrated that
all well-observed WZ Sge-type dwarf novae show ``early superhumps"
during the earliest stage of their superoutbursts (which may be the best
defining characteristic of WZ Sge-type dwarf novae: \cite{kat02wzsgeESH}).
These early superhumps are usually double-wave (sometimes more complex)
variations, which have periods extremely close to the orbital periods
(e.g. \cite{ish02wzsgeletter}).  Although there is a historical
(\cite{pat81wzsge}) and a modern version (\cite{pat02wzsge})
interpretation, that the phenomenon results from an enhanced mass-transfer,
it is now understood as the result of some sort of
resonance on the disk (2:1 resonance: \cite{osa02wzsgehump} or
vertical 2:1 resonance: \cite{kat02wzsgeESH}).

   The identification of RZ Leo as a WZ Sge-type dwarf nova
(\cite{ish01rzleo}), is a surprise from this point of view,
since the superhump period (0.078529 d) of
RZ Leo is anomalously long compared to the canonical picture of
WZ Sge-type dwarf novae.  This identification indicates that neither
a brown-dwarf secondary (\cite{mey99diskviscosity}) nor an extreme
mass-ratio, enabling 2:1 resonance (\cite{osa02wzsgehump}), may be
a necessary condition for the WZ Sge-type outburst phenomenon.
This implication is presently under discussion for
WZ Sge-type dwarf novae (cf. \cite{kat02wzsgeESH}; \cite{osa02wzsgehump}).
As concerns early superhumps, \citet{kat97tleo} detected a smooth
transition from the orbital to superhump period in a more usual
SU UMa-type dwarf nova, T Leo.  This phenomenon may be somehow related
to the evolution of early superhumps \citep{kat96alcom}.

   Several large-amplitude SU UMa-type dwarf novae share some common
properties with WZ Sge-type dwarf novae, particularly in the lengthening
of the superhump period.  The periods of the superhumps are usually
not constant, but show a significant period derivative ($P_{\rm dot}$).
The $P_{\rm dot}$ are usually negative in classical SU UMa-type dwarf
novae (\cite{war85suuma}; \cite{pat93vyaqr}; for a recent progress, see
\cite{kat03v877arakktelpucma}).  This negative $P_{\rm dot}$ is usually
considered to be a result of a decrease in the angular velocity of
precession of a shrinking eccentric disk \citep{osa85SHexcess}.
The decrease may also be a result of inward propagation of the
eccentricity wave (cf. \cite{lub91SHa}; \cite{lub91SHb}; \cite{lub92SH}).
A small number of SU UMa-type dwarf novae, notably many WZ Sge-type
dwarf novae, are known to show positive $P_{\rm dot}$.
This effect was first clearly detected in
V1028 Cyg \citep{bab00v1028cyg}, although this phenomenon first appeared
in solid publications on SW UMa (\cite{sem97swuma}; \cite{nog98swuma}),
V485 Cen \citep{ole97v485cen}, and on the WZ Sge-type star
AL Com (\cite{pyc95alcom}; \cite{how96alcom}; \cite{nog97alcom}).
The SU UMa-type dwarf novae newly
identified by the VSNET Collaboration as having positive $P_{\rm dot}$
include:
HV Vir (WZ Sge-type star, \cite{kat01hvvir}),
WX Cet \citep{kat01wxcet},
EG Cnc (WZ Sge-type star, \cite{kat97egcnc}),
and XZ Eri \citep{uem03xzeri}.  The true origin of this phenomenon
is not yet well understood (\cite{bab00v1028cyg}; \cite{kat01hvvir};
\cite{ish03hvvir}; \cite{kat03v877arakktelpucma}; \cite{kat03bfara}).

   Post-superoutburst rebrightenings are also a renowned feature of
the WZ Sge-type dwarf novae (\cite{kat97egcnc}; \cite{pat98egcnc};
\cite{ish03hvvir}), for which \citet{osa97egcnc}, and \citet{osa01egcnc}
presented an interpretation based on the slow viscosity decay in
the early post-superoutburst state.  This mechanism would require
a mass reservoir in the outer disk \citep{osa03DNoutburst}, whose
original observational implication was proposed by \citet{kat98super}.
The existence, or non-existence, of post-superoutburst rebrightenings
have been systematically studied by the VSNET Collaboration in almost
all SU UMa-type dwarf nova.  Recent examples include GO Com
\citep{kat95gocom}, WX Cet (\cite{kat95wxcet}; \cite{kat01wxcet}),
V1028 Cyg \citep{bab00v1028cyg}, V725 Aql \citep{uem01v725aql},
and in the unusual system (see subsection \ref{sec:sci:ultrashort})
EI Psc \citep{uem02j2329}.
The statistics is presented in \citet{kat98super}.

   A missing link between WZ Sge-type dwarf novae and ordinary
SU UMa-type dwarf novae has been sought.  CT Hya (\cite{nog96cthya};
\cite{kat99cthya}) was suggested to be one such system, though
a more recent statistical analysis suggests a more rigid segregation
between WZ Sge-type dwarf novae and ordinary SU UMa-type dwarf novae
\citep{kat03hodel}.  Several systems possibly related to WZ Sge-type
dwarf novae have been also studied by the VSNET Collaboration:
CC Scl = RX J2315.5$-$3049 (\cite{ish01j2315}).

   Long $P_{\rm orb}$ SU UMa-type dwarf novae with rare outbursts
have also been systematically studied by the VSNET Collaboration.
The objects include: EF Peg (\cite{kat91efpegiauc}; \cite{kat02efpeg})
and V725 Aql (\cite{nog95v725aql}; \cite{uem01v725aql}).  The origin
of the necessary low mass-transfer rate in such long $P_{\rm orb}$
systems is still a problem.

\subsubsection{Ultra-Short Period Systems and Helium Cataclysmic Variables}
\label{sec:sci:ultrashort}

   From the standard evolutionary scenario of compact binaries, there
should be a ``period minimum" (e.g. \cite{pac71CVminimumperiod}) at
which the mass-losing secondary star becomes degenerate and the binary
period starts to lengthen.  This period is observationally determined
to be $\sim$1.3 hr, which is about 10\% longer than the theoretical
predictions (\cite{pac71CVminimumperiod}; \cite{kol99CVperiodminimum};
\cite{kin01minPorb}; \cite{kin02CVperiodminimum}).
This discrepancy has not been yet resolved, although several attempts
have been made (e.g. \cite{ren02CVminimum}; \cite{bar03CVminimumperiod})
to reconcile theory with observation.

   There exist, however, hydrogen-rich systems with periods well below
this theoretical minimum period.  The ``classical" object is V485 Cen
(\cite{aug93v485cen}; \cite{aug96v485cen}; \cite{ole97v485cen});
the faintness of this object, however, prevented
a detailed observational study.
In the past few years, the VSNET team found a hydrogen-rich,
nearby bright system (EI Psc = 1RXS J232953.9+062814) having a short
period comparable to V485 Cen (\cite{uem01j2329iauc};
\cite{uem02j2329letter}; \cite{uem02j2329}).  Both the radial-velocity study
(\cite{tho02j2329}) and the superhump period analysis (\cite{uem02j2329};
\cite{ski02j2329}) independently confirmed that the secondary star of
this system is more massive than what is expected for this orbital
period, suggesting that the mass donor is an evolved core with a thin
hydrogen envelope.  From this finding, \citet{uem02j2329letter} suggested,
following the evolutionary calculations by \citet{pod03amcvn},
that both EI Psc and V485 Cen can be ancestors of helium CVs (or AM CVn
stars, \cite{war95amcvn}; \cite{sol95amcvnreview}) consisting of a
white dwarf and a mass-losing helium white dwarf.  If this interpretation
is confirmed, this object would become the first direct observational
evidence that helium CVs are descendants of a certain class of CVs
with hydrogen-rich appearance (cf. \cite{web84IBWDrcrbSNIa};
\cite{tut96IBWD}; \cite{nel01amcvnpopulationsynthesis};
\cite{yun02LMXBamcvnevolution}).  This object, given its proximity and
relatively heavy component masses, is also considered to be an excellent
candidate for next generation experiments of directly detecting
gravitational wave radiation
(\cite{hil00amcvnGR}; \cite{str02v407vuldecay}).

   In clarifying the nature of EI Psc, proper motion studies played
an independent important role in identifying the object as a nearby
object.  This finding was soon confirmed by later researchers
(\cite{big01j2329iauc}), and the same technique has been applied to
different sorts of objects by the
VSNET Collaboration (RX J2309.8+2135: \cite{kat02j2309},
V379 Peg: \cite{kat02v379peg}, CW Mon: \cite{kat03cwmon}).
This application of astrometry soon became the global standard in
studying CVs and related systems (\cite{kat01ftcam}, see also
recent entries of a survey of proper motions in
Downes et al. online CV catalog\footnote{
$\langle$http://icarus.stsci.edu/$^{\sim}$downes/cvcat/$\rangle$.
}
).

   Helium CVs have been one of the best observed targets
by the VSNET Collaboration.  Among them, CR Boo has been identified
as the first ``helium ER UMa star" with a supercycle of 46.3 d
(cf. subsection \ref{sec:sci:eruma}) by \citet{kat00crboo}.
The VSNET team also joined a campaign to study superhumps in CR Boo
\citep{pat97crboo}.  \citet{kat00v803cen} identified a similar supercycle
in the helium CV, V803 Cen.  \citet{kat01v803cen} further identified
standstills in V803 Cen, which were initially suggested for
CR Boo in \citet{kat00crboo}.  \citet{kat01crboo} later detected
a transition of CR Boo to a state of a short supercycle (14.7 d),
which they called ``the second supercycle".  This phenomenon is
still difficult to explain.  \citet{kat03v803cen} studied
V803 Cen for its long-term behavior and its 2003 June superoutburst,
and revealed that the object (and probably also CR Boo) shows outburst
behavior similar to WZ Sge.  From these studies, both CR Boo and
V803 Cen have been well-established ``helium counterparts" to
hydrogen-rich SU UMa-type dwarf novae, in contrast to the traditional
VY Scl-type (variable mass-transfer rate from the secondary, see
subsection \ref{sec:sci:vyscl}) explanation
of high and low states in these systems (\cite{war95amcvn};
\cite{wou03sn2003aw}).  This interpretation is perfectly in line with
the dwarf nova-type interpretation \citep{tsu97amcvn}, although this
interpretation was not originally correctly applied to observation
(\cite{pat97crboo}; \cite{pat00v803cen}).

   Outburst detections of other helium dwarf novae (e.g. KL Dra = SN 1998di)
have been also announced through the VSNET, and provided necessary
fundamentals\footnote{
$\langle$http://www.kusastro.kyoto-u.ac.jp/vsnet/DNe/kldra.html$\rangle$.
} for detailed research (e.g. \cite{jha98kldraiauc6983};
\cite{sam00kldraiauc7382}; \cite{woo02kldra}).

\subsubsection{Eclipsing Dwarf Novae}
\label{sec:sci:dnecl}

   Since CVs are close binary systems, high-inclination systems show
eclipses.  The presence of eclipses in CVs historically provided most
crucial information about the geometry and fundamental physics
of the accretion disk or the accretion stream
(\cite{wal54dqher}; \cite{wal55dqher}; \cite{kra58dqher};
\cite{gre59dqher}; \cite{kra62ugem}; \cite{war71ugem})
and the clarification of the cause of outbursts
(\cite{osa74DNmodel}; \cite{sma76ugem}; \cite{sma84ugemdiskradius};
\cite{krz85oycarsuper}).

   In recent years, the eclipse mapping technique (\cite{EclipseMapping})
has been used to geometrically resolve the accretion disk by numerically
modeling the eclipse light curve (and sometimes line variations) of CVs
(\cite{hor85rwtri}; \cite{mar87mapping}; \cite{woo89oycar};
\cite{bap91v4140sgr}; \cite{bap93eclipsemapping}).
This and analogous methods have also been used to study the
time-evolution of the accretion disk during dwarf nova outbursts
(\cite{hor85zcha}; \cite{woo89ippeg}; \cite{rut92oycar};
\cite{wol93ippeg}; \cite{bob97ippeg}),
discuss the presence of a spiral pattern
(which may be theoretically predicted spiral shocks:
\cite{saw86spiralshock}; \cite{spr87ADspiralshock};
\cite{mat90ADspiralshock}; \cite{mat00CVspiralshock}) in dwarf novae
(\cite{bap01exdra}; \cite{bap00exdraletter}; \cite{bap00ippeg}),
for mapping of the superoutbursting disks,
(\cite{bru96oycareclipsemapping}), to spectrally resolve
the accretion disk (\cite{rut93uxuma}; \cite{rut94uxuma}),
and to directly obtain physical
parameters of the accretion disk
(\cite{vri99physicalparametermapping}; \cite{vri02v2051ophmapping}).

   The VSNET Collaboration played an important role in studying
eclipsing CVs, especially eclipsing dwarf novae.  The initial efforts
were made to follow the eclipses of the rising phase of an IP Peg outburst
\citep{nog96ippeg}.  The VSNET alert lists provided an initiative
role of systematic studies when the northern new eclipsing dwarf nova
EX Dra (= HS 1804+6753) was discovered \citep{nog96ippeg}.  This action
was soon extended to observe rare outbursts of an eclipsing SU UMa-type
dwarf nova, DV UMa \citep{nog01dvuma}.  The 1995 outburst of HT Cas
was also spectacular.  This is a well-known eclipsing dwarf nova
(cf. \cite{pat81DNOhtcas}; \cite{zha86htcas}; \cite{woo95htcasXray};
\cite{muk97htcas}).  In 1995, the VSNET team received a request
for an optical ground-based campaign coordinated with
the Hubble Space Telescope (HST).
The observation by the VSNET team succeeded in correcting the eclipse
ephemeris, which is readily reflected on the HST observing schedule,
when the object suddenly jumped into an outburst!  Thanks to this
coincidence, we were able to obtain eclipse information only two days
prior to the outburst maximum, which precluded the enhancement of the
hot spot as expected from the mass-transfer burst \citep{bab99htcas}.
The eclipses during this outburst was also followed by another group
\citep{ioa99htcas}, who observed this object in response to this
outburst detection.  The results of eclipse mapping, together with later
outburst observations, have been recently reported \citep{bab03htcas}.

   IR Com (= S 10932), a system very similar to HT Cas, has also been
extensively studied by the VSNET team.  In particular, we detected the
1996 January outburst, and succeeded in taking the earliest eclipse
observations \citep{kat02ircom},  The true nature of this
object had remained unclear before (\cite{ric95ircom}; \cite{ric97ircom}).
The real-time circulation of this outburst detection and eclipse
information enabled a third-party follow-up observations
\citep{ric96ircomiauc}.

   In most recent years, the VSNET Collaboration discovered a deeply
eclipsing bright SU UMa-type dwarf nova (IY UMa = Tmz~V85)
in the northern hemisphere (\cite{uem00iyumaiauc}; \cite{uem00iyuma};
\cite{sam00iyumaiauc7353}).
For the first time in history, this observation yielded the simultaneous
discovery of superhumps and eclipses.  This system, the only bright
normal SU UMa-type dwarf novae suitably situated for northern telescopes,
has been proposed as the best candidate object for next generation
detectors on huge telescopes (cf. \cite{moo01iyuma}; \cite{ste03iyumaSTJ};
\cite{ray03STJspec}).  This and the subsequent
outbursts were followed by a number of teams, resulting in
rich physical insights (\cite{pat00iyuma}; \cite{wu01iyuma};
\cite{sta01iyuma}; \cite{rol01iyuma}; \cite{rol02iyuma};
\cite{sma03iyuma}).  With strong emission lines of He\textsc{ii} and
C\textsc{iii}/N\textsc{iii} in outburst (\cite{wu01iyuma}; such objects
are known to be quite rare: \cite{mor02DNspectralatlas}),
this object is a good candidate for spatially resolving a superoutbursting
disk by the emission-line eclipse mapping method \citep{mak02ADmapping},
as well as with classical Doppler tomography of the velocity field
\citep{DopplerTomography}.  We also succeeded in identifying the
supercycle \citep{kat01v2051ophiyuma}.

   The other outstanding object is DV UMa, which was observed during the
entire stage of the 1999 December superoutburst following
the outburst report by Timo Kinnunen.\footnote{
$\langle$http://www.kusastro.kyoto-u.ac.jp/vsnet/DNe/dvuma9912.html$\rangle$.
}  This observation first fully covered the early evolution of eclipses
in this rarely outbursting system.

   In 2002 February, a collaborative effort on GY Cnc = RX~J0909.8+1849
led to the discovery of the eclipsing nature of this dwarf nova
(\cite{gan00gycnc}; \cite{kat00gycnc}; \cite{sha00gycnc};
\cite{tho00gycnc}).  Eclipse observations during the 2001 November
outburst revealed the noticeable absence of the hot spot during the
late stage of an outburst \citep{kat02gycnc}.  This observation suggested
that GY Cnc may be the first long-$P_{\rm orb}$ object sharing common
properties with HT Cas and IR Com.

   Recent detailed outburst (or superoutburst) observations of eclipsing
dwarf novae include: XZ Eri (\cite{uem03xzeri}) and OU Vir (R. Ishioka
et al. in preparation).  Both stars show prominent superhumps as
well as eclipses.  XZ Eri is the first eclipsing SU UMa-type dwarf nova
with a positive period derivative \citep{uem03xzeri}.  OU Vir is another
object continuously receiving world-wide attention (\cite{van00ouvir};
\cite{mas02ouvir}), for which we succeeded in determining the first
reliable orbital and superhump periods.  V2051 Oph is another eclipsing
CV, which had been thought to be a low-field polar (\cite{war83v2051oph};
\cite{wen84v2051oph}; \cite{war87v2051oph}) for which
the VSNET Collaboration first provided
unambiguous clarification of its SU UMa-type nature by securely detecting
superhumps and supercycles (\cite{kiy1998v2051oph};
\cite{kat01v2051ophiyuma}).  With the help by this clear identification,
this object has also been receiving special attention both with ground-based
and satellite-borne observations
(\cite{bap98v2051ophHST}; \cite{bru00htcasv2051ophippeguxumaflickering};
\cite{ste01v2051oph}; \cite{vri02v2051ophmapping})

   The recovery and clarification of the nature of the ``lost" dwarf nova
V893 Sco is another noteworthy achievement by the VSNET Collaboration.
This object had long been lost, when Katsumi Haseda (VSOLJ) reported an
outbursting object (at a nominally different position from that of the
originally reported V893 Sco) to the VSNET.  After careful research of
the discovery material, as a part of confirmatory process of a new
variable star (subsection \ref{sec:newvar}), this newly reported object
was eventually identified with the lost V893 Sco \citep{kat98v893sco}.
This was only the beginning of the story; the object soon turned out
to be the brightest, and presumably one of the nearest, eclipsing dwarf
novae below the period gap (\cite{tho99v893sco}; \cite{mat00v893sco};
\cite{bru00v893sco}).  This object has been extensively studied since
its recovery (\cite{mas01v893sco}; \cite{kat02v893sco}).

   CW Mon (\cite{kat03cwmon}) shows grazing eclipses during certain
stages of outbursts.  Together with the transient appearance of pulsed
signals, the presence of a premaximum halt in the outburst light curve
and relatively strong X-ray radiation, this object has been suspected
to be an intermediate polar (see also subsection \ref{sec:sci:IP}).

\subsubsection{Super-Quasi-Periodic Oscillations}
\label{sec:sci:QPO}

   Quasi-periodic oscillations (QPOs) are short-period, quasi-periodic
oscillations widely observed in accreting binary systems including
CVs (cf. \cite{war95book}).  QPOs in CVs are usually subdivided into
two classes.  One is dwarf nova oscillations (DNOs) observed during
outbursts of dwarf novae.  DNOs have short periods (usually 19--29 s)
and long coherence times (\cite{rob73DNO}; \cite{szk76DNO};
\cite{pat81DNOhtcas}; \cite{hil80DNO}).  The other is QPOs, which have
longer (40 to several hundred seconds) and shorter coherence times
(usually less than $\sim$10 wave numbers).  We discovered a potentially
new class of QPOs (super-QPOs) during the 1992 superoutburst of
SW UMa \citep{kat92swumasuperQPO}.  These super-QPOs have long
(several hundred seconds) periods and long coherent times (more than
several tens of wave numbers).  In some cases \citep{kat92swumasuperQPO},
the amplitude can be quite large (up to 0.2 mag).  The most outstanding
feature of super-QPOs is that they are observed only during certain
stages of SU UMa-type superoutbursts.  In SW UMa (1992) and EF Peg
(\cite{kat02efpeg}), the super-QPOs were observed during the growing
stage of superhumps.  A similar, but less striking, probable appearance
of super-QPOs was also recorded by the VSNET Collaboration during the
early stage of an superoutburst of NSV 10934
(\cite{kat03nsv10934mmscoabnorcal86}).

   During the 2000 superoutburst of SW UMa, similar super-QPOs temporarily
appeared during the decay phase of a superoutburst
(vsnet-alert 4331).\footnote{
$\langle$http://www.kusastro.kyoto-u.ac.jp/vsnet/DNe/swuma00.html$\rangle$.
}  These observations suggest that the appearance of super-QPOs is
closely related to the growth and decay of superhumps, or related to
the existence of heating/cooling waves (\citet{yam95DNoscillation}
proposed that some sort of QPOs can be an excitation of trapped
oscillations around the discontinuity of physical parameters).

   Different interpretations have also been suggested.  For example,
\citet{war02DNO} suggested that the super-QPOs may be a result of
interaction between the weak magnetism of the white dwarf and some
kind of wave in the inner accretion disk.  Although this explanation
would be compatible with the suggested presence of a weak magnetic field
in SW UMa (\cite{sha86swumaXray}; \cite{rob87swumaQPO};
\cite{szk88swumaEXOSATIUE}; \cite{ros94v426ophswumav348pup}), there
would be a need for a different mechanism to explain why these super-QPOs
only appear only during temporary stages of superoutbursts.

   In EF Peg \citep{kat02efpeg}, a rapid decrease in the periods of
super-QPOs was recorded.  From this finding, \citet{kat02efpeg}
suggested a rapid removal of angular momentum from an orbiting blob
in the accretion disk, via a reasonable viscosity in a turbulent disk.
The origin of super-QPOs is still an open question, but their prominent
profile is expected to provide crucial information about the origin
of QPOs in CVs \citep{war02DNO}.

\subsubsection{Low States in Dwarf Novae}
\label{sec:sci:dnlow}

   Although some nova-like CVs (VY Scl-type stars, subsection
\ref{sec:sci:vyscl}) are best known to
show ``low states", during which mass-transfer from the secondary is
reduced (\cite{war95book}; \cite{rob81mvlyr}), this phenomenon
has not been clearly confirmed to occur in dwarf novae.  Although there
have been claims for ``low states" (HT Cas: \cite{zha86htcas};
\cite{woo95htcasXray}; \cite{rob96htcas}, IR Com: \cite{ric95ircom};
\cite{kat02ircom}, WW Cet: \cite{rin96wwcet}; BZ UMa:
\cite{kal86bzumaiauc}), it is not evident whether or not these phenomena
directly reflect a reduced mass-transfer from the secondary, since
the state change in the disk (especially the viscosity parameter) would
reproduce similar phenomena.  Extensive studies on selected well-observed
dwarf novae (e.g. \cite{hon98zcam}; \cite{opp98zcam})
have found no evidence for the long-term variation of mass-transfer.

   During the extensive work by the VSNET Collaboration, we discovered
that the Z Cam star RX And underwent a deep fading in 1996 September
(vsnet-obs 3750).\footnote{
$\langle$http://www.kusastro.kyoto-u.ac.jp/vsnet/Mail/obs3000/\\msg00750.html$\rangle$.
}  This fading (and the lack of outbursts) lasted until 1997 January,
which yielded the first unambiguous detection of a temporarily reduced
mass-transfer in dwarf novae.  This phenomenon was thoroughly studied by
\citet{kat02rxand}.  Careful research on the historical light curve
indicated that similar phenomena were sporadically observed in RX And
(\cite{kat02rxand}; \cite{sch02rxand}), but had been overlooked mainly
because of the confusion of the true quiescent identification, and the
lack of real-time circulation of information.  The VSNET Collaboration
succeeded in detecting such a phenomenon through real-time, regular
monitoring of light curves of dwarf novae, and through a prompt reaction
when an anomalous change was observed (the deep quiescence of RX And
was originally confirmed by our own CCD observation).  This is another
aspect how effectively the VSNET alert network worked, besides other
outburst-type transient events.  This detection, announced world-wide,
thereby led to a prompt HST observation \citep{sio01rxand}, which
revealed the presence of a hot white dwarf.  \citet{kat03rxandsuuma}
further reported the detection of short fading episodes in RX And
and SU UMa, which may be a result of temporary reduction of mass-transfer
rate.

\subsubsection{Other Dwarf Nova Works}

   Other dwarf nova-related works, not covered by the above
subsections, by the VSNET Collaboration include:

\vskip 2mm
\begin{enumerate}

\item[(a)]
Dwarf novae in the period gap: GX Cas, V419 Lyr
(\cite{nog98gxcasv419lyr}) (see subsection \ref{sec:sci:eruma}
for NY Ser and MN Dra)

\item[(b)]
Time-variation in more usual SU UMa-type dwarf novae and candidates:
AW Gen (\cite{kat96awgem}),
RZ Sge (\cite{kat96rzsge}),
V1113 Cyg (\cite{kat96v1113cyg}; \cite{kat01v1113cyg}),
CC Cnc (\cite{kat97cccnc}; \cite{kat02cccnc}),
VZ Pyx (\cite{kat97vzpyx}),
CY UMa (\cite{kat97cyuma}; \cite{kat99cyuma}),
PU Per (\cite{kat99puper}),
AQ Eri (\cite{kat99aqeri}; \cite{kat01aqeri}),
CT Hya (\cite{kat99cthya}),
V364 Peg (\cite{kat99v364peg}),
QW Ser (\cite{kat99qwser}; \cite{nog03qwser}),
BZ UMa (\cite{kat99bzuma}),
CI Gem (\cite{kat99cigem}),
TY Vul (\cite{kat99tyvul}),
KV Dra (\cite{nog00kvdra}),
V844 Her (\cite{kat00v844her}),
QY Per (\cite{kat00qyperiauc}),
TY Psc (\cite{kun01typsc}; \cite{kun03typsc}),
V630 Cyg (\cite{nog01v630cyg}),
V369 Peg (\cite{kat01v369peg}),
UV Gem, FS And, AS Psc (\cite{kat01uvgemfsandaspsc}),
RX Cha (\cite{kat01rxcha}),
YZ Cnc (\cite{kat01yzcnc}),
IR Gem (\cite{kat01irgem}),
FT Cam (\cite{kat01ftcam}; \cite{kat02ftcam}),
GZ Cnc, NSV 10934 (\cite{kat01gzcnc}; \cite{kat02gzcncnsv10934}),
DM Dra (\cite{kat02dmdra}),
SU UMa (\cite{kat02suuma}),
DM Lyr (\cite{nog03dmlyr})

\item[(c)]
Time-variation in SS Cyg-type dwarf novae:
V1008 Her (\cite{kat99v1008her}),
V1101 Aql (\cite{kat99v1101aql}),
DK Cas (\cite{kat01dkcas}),
HH Cnc = Tmz~V36 (\cite{kat01tmzv36}),
IS Del (\cite{kat01isdel}),
IZ And (\cite{kat01izand}),
DX And (\cite{kat01dxand}),
AH Eri (\cite{kat01aheri}),
CG Dra (\cite{kat01cgdra}),

\item[(d)]
Standstills of Z Cam-type dwarf novae:
VW Vul (\cite{kat99vwvul}),
AT Cnc (\cite{nog99atcnc}, Doppler tomography; \cite{kat01atcnc}),
Z Cam (\cite{kat01zcam}),
HL CMa (\cite{kat02hlcma}),
FX Cep (\cite{kat01fxcep}),
V363 Lyr (\cite{kat01v363lyr}),
EY Cyg (\cite{kat02eycyg}),
IW And (unusual Z Cam star: \cite{kat03iwand}),

\item[(e)]
Quiescent dwarf novae: UV Per (\cite{kat90uvper}),
GO Com (\cite{kat90gocom})

\item[(f)]
Classification:
BF Eri (\cite{kat99bferi}; \cite{kat00bferi}),
LX And (\cite{uem00lxand}; \cite{kat02lxand}),
HP And (with Subaru, \cite{nog03hpand})

\item[(g)]
Statistics and compilation:
\citet{nog97sxlmi}; \citet{kat03hodel}

\end{enumerate}
\vskip 2mm

   Most of the work was done with the help of the alert network and
the collaboration described in subsections \ref{sec:vsoljcolab} and
\ref{sub:cvcenter}.

\subsubsection{Other Global Contributions}

   As the VSNET has been mediating an enormous number of outburst alerts
of dwarf novae since its very early history, these alerts, as well as
long-term observations, contributed to world-wide dwarf nova studies by
different teams.  Since they are so numerous, we only list representative
ones:

\vskip 2mm
\begin{itemize}
\item V485 Cen: \citet{ole97v485ceniauc}; \citet{ole97v485cen}
\item PV Per: \citet{van97pvper}
\item TU Crt: \citet{men98tucrt}
\item KS UMa: \citet{wat98ksuma}; \citet{ole03ksuma}
\item KV Dra: \citet{van00kvdra}
\item WY Tri: \citet{van01wytri}
\item KX Aql: \citet{tap01kxaql}
\item XY Psc: \citet{hen01xypsc}
\item AH Her: \citet{spo01ahher}; \citet{spo02ahher}
\item CVs in the 2MASS survey: \citet{hoa02CV2MASS}
\item V844 Her: \citet{tho02gwlibv844herdiuma}
\item V2051 Oph: \citet{vri02v2051ophmapping}; \citet{vri03v2051oph}
\item QZ Ser: \citet{tho02qzser}, another peculiar dwarf nova discovered
by Katsumi Haseda (Had~V04), and announced in collaboration with the VSNET
(subsection \ref{sec:newvar})
\item IP Peg: \citet{bap02ippeg}
\item RX~J0944.5+0357: \citet{men02j0944}
\item V1504 Cyg: \citet{pav02v1504cyg}
\item GZ Cnc: \citet{tap03gzcnc}
\item V1141 Aql: \citet{ole03v1141aql}
\item QW Ser: \citet{ole03qwser}
\item FS Aur: \citet{tov03fsaur}
\item EM Cyg: \citet{spo03emcyg}
\item short-period dwarf novae: \citet{tho03kxaqlftcampucmav660herdmlyr}
\item faint CVs survey: \citet{mas03faintCV}
\item SU UMa stars: \citet{pat03suumas}
\end{itemize}
\vskip 2mm

\subsection{Novae and Recurrent Novae}
\label{sec:sci:nova}

\subsubsection{Classical Novae and Related Objects}
\label{sec:sci:CN}

   From the beginning of the VSNET, novae and recurrent novae have been
widely studied as one of the classical representatives of transient objects.
The earliest observations include the recurrent nova V3890 Sgr (1990),
whose exact identification was clarified by us \citep{kat90v3890sgriauc}.
The next advancement was with V838 Her (Nova Her 1991), whose eclipsing
nature and exact orbital period was clarified by our observation
(\cite{kat91v838heriauc}).  This is the first classical nova whose
evolution of eclipses was caught from the early decline stage of the
outburst (\cite{lei93v838her}; \cite{szk94v838herqzaur};
\cite{ing95v838her}).  The ``nova of the century"
\citep{sta94v1974cygskytel} V1974 Cyg (Nova Cyg 1992) was followed
with the advent of the e-mail alert list
(see {\it vsnet-history} messages).  This nova later turned out to be
a permanent superhumper system (\cite{sem94v1974cyg};
\cite{sem95v1974cyg}; \cite{ret97v1974cygSH}; \cite{ski97v1974cyg}).
The VSNET Collaboration later contributed to the international
observing campaigns of the superhumps (A. Retter et al., in preparation).

   The next major step was with V705 Cas (Nova Cas 1993), as introduced
in subsection \ref{sec:openingera}.  After this nova, the VSNET
has continuously provided public pages on individual novae,\footnote{
$\langle$http://www.kusastro.kyoto-u.ac.jp/vsnet/Novae/novae.html$\rangle$.
} which are referenced as a primary resource on recent novae.
V723 Cas (Nova Cas 1995) has been one of the best studied novae in the
VSNET history.  The object was discovered by Minoru Yamamoto,
whose report immediately triggered early follow-up observations
\citep{hir95v723casiauc}.  Several early reports discussed against the
classical nova-type classification (e.g. \cite{mun95v723casiauc6259}).
We were the first, with the enormous amount of information collected
by the VSNET, to predict that the object is a premaximum phase slow
nova resembling HR Del (vsnet 223).\footnote{
$\langle$http://www.kusastro.kyoto-u.ac.jp/vsnet/Mail/\\vsnet/msg00223.html$\rangle$.
}  This prediction was later confirmed by
a number of works (\cite{ohs96v723cas}; \cite{mun96v723cas};
\cite{iij96v723casiauc6365}; \cite{iij97v723casiauc6703};
\cite{iij98v723cas}).

   Although it was not a classical nova, V4334 Sgr (Sakurai's object)
in 1996 brought a major breakthrough in ``stellar evolution in real time"
(\cite{due96v4334sgriauc6328}; \cite{due96v4334sgr}).
This star is one of the best studied and discussed variable stars since
the late 1990's
(\cite{due97v4334sgr}; \cite{ark97v4334sgr}; \cite{asp97v4334sgr};
\cite{ark98v4334sgr}; \cite{eva98v4334sgrPN}; \cite{ark99v4334sgr};
\cite{kip99v4334sgrevolution};
\cite{ker99v4334sgr}; \cite{ker01v4334sgr};
\cite{asp99v4334sgr}; \cite{pav00v4334sgr};
\cite{due00v4334sgr}; \cite{pav01v4334sgr}; \cite{she01v4334sgr};
\cite{her01v4334sgr}; \cite{tat01v4334sgr};
\cite{due02v4334sgr}; \cite{cla02v4334sgr};
\cite{jef02v4334sgr}; \cite{kam02v4334sgrIR}; \cite{law02v4334sgr};
\cite{lyn02v4334sgrIR};
\cite{kau03v4334sgr}).
The VSNET not only relayed this breathtaking discovery of a final
helium flash object, but also promptly provided prediscovery observations
by Kesao Takamizawa (vsnet-alert 341)\footnote{
$\langle$http://www.kusastro.kyoto-u.ac.jp/vsnet/Mail/vsnet-alert/\\msg00341.html$\rangle$.
} (finally published by \cite{tak97v4334sgr}), which has been frequently
referenced and employed for theoretical modeling.
See subsection \ref{sec:sci:rcb} for its relation with R CrB-like stars
(cf. \cite{law03v4334sgrv605aqlfgsge}).

   In recent years, the VSNET Collaboration has been issuing scientific
results, as well as early announcements and identifications,
on individual novae:
V2487 Oph (Nova Oph 1998, recurrent nova candidate: \cite{nak98v2487oph};
\cite{hac02v2487ophproc}),
V4444 Sgr (Nova Sgr 1999, the possibility of this being a recurrent nova
has been discussed: \cite{kat04v4444sgr}),
V463 Sct (Nova Sct 2000, fast nova with an unusually prominent
premaximum halt: \cite{uem00v463sct}; \cite{kat02v463sct}),
V445 Pup (Nova Pup 2000, unusual nova with no indication of hydrogen
features: \cite{kat00v445pupiauc1}; \cite{kat00v445pupiauc2};
\cite{kat01v445pupiauc3}; \cite{nak01v445pup}; M. Uemura et al.
in preparation),
V1548 Aql (Nova Aql 2001, slow nova initially reported as a more usual
variable star: \cite{uem01v1548aqliauc}; \cite{kat01v1548aql}),
V1178 Sco (Nova Sco 2001, object originally confusedly reported to be
a novalike object, later turned out to be a genuine nova with
early stage oscillations: \cite{has01v1178sco}; \cite{kan01v1178sco};
\cite{kat01v1178sco}),
V2540 Oph (Nova Oph 2002, large-amplitude slow nova with strong
post-outburst oscillations: \cite{sek02v2540ophiauc};
\cite{kat02v2540oph}).  The VSNET Collaboration also joined multiwavelength
campaigns on novae with satellites
(e.g. V4743 Sgr: \cite{wag03v4743sgriauc8176}).
The VSNET has recently been relaying discovery, independent and prediscovery
detections of novae, as well as spectroscopic confirmation and early
photometric observations:
V2274 Cyg (Nova Cyg 2001:
           \cite{sat01v2274cygiauc7666}),
V4643 Sgr (Nova Sgr 2001:
           \cite{kat01v4643sgr}),
V4740 Sgr (Nova Sgr 2001 No. 3:
           \cite{wes01v4740sgriauc7708});
V4741 Sgr (Nova Sgr 2002:
           \cite{lil02v4741sgriauc7878};
           \cite{lil02v4741sgriauc7879}),
V4742 Sgr (Nova Sgr 2002 No. 2:
           \cite{yam02v4742sgriauc7972}),
V4743 Sgr (Nova Sgr 2002 No. 3:
           \cite{has02v4743sgriauc7975};
           \cite{wes02v4743sgriauc7975}),
V4744 Sgr (Nova Sgr 2002 No. 4:
           \cite{yam02v4744sgriauc8006}),
V4745 Sgr (Nova Sgr 2003:
           \cite{kat03v4745sgriauc8126};
           \cite{kat03v4745sgriauc8132}),
Possible Nova (2002) in NGC 205
           (\cite{yam02m110nova2002iauc8001}),
Possible Nova (2003) in Scutum:
           (\cite{nak03nsctiauc8190}).

   The discovery of a very unusual eruptive object (V838 Mon,
\cite{bro02v838moniauc7785}; \cite{wag02v838moniauc7785};
\cite{dellaval02v838moniauc7786})
was relayed through the VSNET alert system
during its early stage of eruption, enabling early-stage observations
(\cite{bed02v838moniauc7785}; \cite{bed02v838moniauc7790}).
The identification of the object with GSC, IRAS and 2MASS objects
was reported \citep{kat02v838moniauc7786}.  The early stage of this
object was most unusual, showing an M-type spectrum at outburst maximum.
On 2002 February 2, the object underwent a second major brightening,
which was quickly relayed via the VSNET alert system worldwide, and
the star became world-popular within a day.  The object subsequently
showed a prominent light echo when it faded (by the obscuration
by the forming dust).  There have been a number of works based on these
observations (\cite{kim02v838mon}; \cite{mun02v838mon};
\cite{gor02v838mon}; \cite{sob02v838mon}; \cite{bon03v838monnat};
\cite{cra03v838mon}; \cite{wis03v838mon}),
which more or less employed VSNET observations
and findings when discussing the peculiarity of this object.
The origin of this eruption is still a mystery.
Based on an idea proposed by \citet{sok03v838mon},
\citet{kat03ckvul} tried to explain the historical mysterious eruption
of CK Vul.

\subsubsection{Recurrent Novae and Supersoft X-Ray Sources}
\label{sec:sci:RN}

   The last two decades have dramatically changed our view of recurrent
novae.  The discovery of recurrent outbursts of V394 CrA in 1987
(\cite{lil87v394craiauc4428}; \cite{due88v394cra}; \cite{sek89v394cra}),
V745 Sco in 1989 (\cite{lil89v745scoiauc4820}; \cite{due89v745sco};
\cite{sek90v745sco}), V3890 Sgr in 1990 (\cite{jon90v3890sgriauc5004};
\cite{gon92v3890sgrIUE}; \cite{har93v745scov3890sgrIR}) and Nova LMC
1990 No. 2 (\cite{sek90novalmc19902}; \cite{sho91novalmc19902})
resulted in a dramatic increase of our knowledge in recurrent novae
(cf. \cite{sha89novareview}; \cite{anu92RN}; \cite{sek95RN}).

   Another epoch of recurrent nova discoveries arrived like a flurry
in the late 1990's and early 2000's.  All of these recurrent nova
outbursts were mediated via VSNET and followed in detail.  The initial
object in this series of recurrent nova outbursts was U Sco in 1999,
whose early history was described in detail in subsection
\ref{sub:cvcenter}.  This outburst of U Sco first enabled eclipse
observations in real time during outburst [the eclipsing nature of
U Sco was revealed only in 1990 \citep{sch90tpyxuscov394cra}; the
1999 outburst was the first outburst since this discovery; see also
\citet{kat99usco} for the retrospective detection of an eclipse
during the 1987 outburst].  This observation led to the first detection
of the period change in this system \citep{mat03usco}.  This finding
severely constrained the mass-transfer rate in quiescence
(\cite{hac00uscoburst}; \cite{hac00uscoqui}), which makes U Sco the
most promising candidate for an immediate precursor of a type-Ia
supernova \citep{hac00uscoburst}.

   The next object in this series was CI Aql.  This object had long been
suspected to be a dwarf nova based on the (apparently) small outburst
amplitude of the 1917 outburst \citep{due87novaatlas}.  Based on this
potential identification, the object had been monitored by amateur
variable star observers, notably by VSOLJ members and by members
of the Recurrent Object Programme (see subsection \ref{sec:sci:dwarfnova}).
The proposed quiescent counterpart, however, was found to be an eclipsing
binary which did not show CV characteristics \citep{men95ciaql}.
The same conclusion
had been reached with snapshot spectroscopy \citep{szk92CVspec},
showing no indication of hydrogen emission lines.  With this information,
almost all observers stopped monitoring for an outburst, although there
was the unexplained presence of a He\textsc{ii} emission line
\citep{gre96ciaql}.

   The news of a possible nova detection by Kesao Takamizawa on films taken
on 2000 April 28 arrived at the VSNET on April 29.  The reported position
was extremely close to that of CI Aql.  Minoru Yamamoto independently
detected this phenomenon, and reported it to be a brightening of CI Aql.
After careful examination of the identification, this possible nova is
identified as a recurrent outburst of CI Aql, 83 years after the 1917
discovery (\cite{tak00ciaql}; \cite{yam00ciaqliauc}).\footnote{
  The object was later found to be positively recorded one day earlier
  \citep{lil00ciaqliauc}.  An outburst in 1941--1942 was also later
  found in the plate archive \citep{sch01ciaqliauc7750}.
}  The nova-type nature of the outburst was soon clarified with
spectroscopy \citep{uem00ciaqliauc}, confirming that CI Aql is a new
recurrent nova.  Being already known as an eclipsing binary,
the evolution of the outburst light curve and the eclipse profile
were precisely
followed (\cite{mat01ciaql}; \cite{mat03ciaql}).  In particular,
a dip-like sudden fading in 2000 November was noted
\citep{mat00ciaqliauc}.  With these observational constraints,
\citet{hac01ciaql} succeeded in modeling the light curve, which
was further refined and extended (\cite{hac03SSS}; \cite{hac03j0513})
to explain the unique high/low transitions in the
supersoft X-ray source RX~J0513.9$-$6951 in the LMC.  Its galactic
counterpart, V Sge, has been very recently identified \citep{hac03vsge}.
These objects are now considered to be promising candidates for
precursors of type-Ia supernovae (e.g. \cite{hac99SNIaSSS};
\cite{hac03j0513}; for earlier and other suggestions, see e.g.
\cite{hac89SNIa}; \cite{hac96SNIa}; \cite{dellaval96RNSNIa};
\cite{hac99SNIasymbio}).  The discoveries and modern detailed
observations of U Sco and CI Aql outbursts thus provided firm
observational evidence for recurrent novae and supersoft X-ray sources
as immediate precursors of type-Ia supernovae.  Other representative
works (outside the VSNET) on CI Aql include
photometry \citep{sch00ciaql},
spectroscopy (\cite{kis01ciaql}; \cite{bur01ciaql}),
Chandra X-ray observation \citep{gre02ciaqlChandra},
modeling \citep{led03ciaqlmodel},
re-examination of the 1917 outburst \citep{wil00ciaql}.

   The next discovery of the series was on IM Nor (possible nova in 1920)
in 2002 January.  The outburst detection by William Liller
\citep{lil02imnoriauc7789} was quickly
relayed to the VSNET, and enabled early astrometric work to first firmly
identify the quiescent counterpart and its recurrent nova nature
(\cite{due02imnoriauc7799}; \cite{ret02imnoriauc7818}).
The light curve of IM Nor was published more than 50 years after
the 1920 outburst \citep{ell72imnor}.
There had been a suggestion of identification with the UHURU X-ray
source 2U 1536$-$52, which was later confirmed to be spurious
(\cite{wyc79imnor}, see also \cite{kat02imnoriauc7791}).
Although the outburst light curve in 1920 resembled that of a slow
recurrent nova T Pyx \citep{ell72imnor}, the unusually faint quiescence
inferred from \citet{wyc79imnor} had been a mystery.  The exact
identification with a new outburst solved this mystery, by the detection
of a considerable variation in quiescence \citep{kat02imnor}.
This suggestion was later confirmed by the detection of a short period
variation with eclipse-like fadings \citep{wou03imnor}.
\citet{kat02imnor} suggested, from the light curve and spectroscopic
appearance, that both CI Aql and IM Nor are members of
a new class of recurrent novae having intermediate properties between
classical novae and fast recurrent novae.

\subsection{Other Cataclysmic Variables}

\subsubsection{Intermediate Polars}
\label{sec:sci:IP}

   Intermediate polars (IPs), which are a class of magnetic CVs (MCVs)
having a magnetic white dwarf asynchronously rotating with the orbital
motion (sometimes referred to as DQ Her stars: for recent reviews,
see e.g. \cite{kin90IP}; \cite{pat94ipreview}; \cite{hel96IPreview};
\cite{buc00IPpower}; chapters 8 and 9 in \cite{hel01book}).
Although many ``classical" IPs are novalike systems with thermally
stable accretion disks, there are a number of IPs showing transient
outbursts.  Some of them look like dwarf novae (subsection
\ref{sec:sci:dwarfnova}), including well-known objects such as
GK Per (\cite{wat85gkperspin}; \cite{bia86gkper};
\cite{hut86aopscgkperpulsespec}; \cite{nor88gkperqui}; \cite{pat91gkper}),
DO Dra (sometimes called YY Dra, see \cite{NameList67}, \cite{kho88dodra}
for the official nomenclature, \cite{pat92dodra}; \cite{has97dodraHST}),
EX Hya (\cite{war81exhya}; \cite{jab85exhya}; \cite{hei87exhyaEinstein};
\cite{hel87exhyaspec}).  There has been a long-standing discussion
whether these IP outbursts originate from disk-instabilities or from
mass-transfer bursts (\cite{hel89exhya}; \cite{hel93tvcol};
\cite{hel00exhyaoutburst}).  The modern understanding is that at least
some of them are better understood as mass-transfer events (TV Col:
\cite{hel93tvcol}), while others can be understood as disk-instability
events (GK Per: \cite{kim92gkper}; \cite{yi92gkper}).

   Rapid circulation of outburst alerts are extremely important for
short-period systems, because these IP outbursts are usually very brief
(usually less than 1 d) and require
prompt follow-up observations
(e.g. TV Col: \cite{szk83tvcolflare}; \cite{szk84tvcolflare};
\cite{sch88tvcoloutburst}; \cite{hel93tvcol}; \cite{ret03tvcol},
EX Hya: \cite{bat86exhyaoutburst}; \cite{bat87exhyaiauc};
\cite{bon87exhya}; \cite{fuj87exhya};
\cite{hel89exhya}; \cite{rei90exhyaoutburst};
\cite{jon94exhyaiauc}; \cite{hel00exhyaoutburst},
DO Dra: \cite{wen83dodra}; \cite{wen83dodracycle};
\cite{hur85dodraiauc}; \cite{mcn85dodraiauc}; \cite{haz86dodrahistory};
\cite{dyc88dodraiauc}; \cite{kat88dodraiauc}).  Although these outburst
detections have historically been relayed via IAUCs, they were not
usually rapid enough to enable early-stage observations of these
outbursts.  The VSNET Collaboration has played a role in detecting,
and rapidly relaying these IP outbursts.  Since IP outbursts tend to
cluster (cf. \cite{bat86exhyaoutburst}, M. Uemura et al. in preparation),
rapid electronic announcements of these outbursts have dramatically
increased the chance of detailed follow-up observations, including
simultaneous observations with satellites.  The most remarkable recent
example includes X-ray/optical simultaneous observations of two outbursts
of DO Dra \citep{szk02dodra}, whose optical coverage was based on
observations by the VSNET Collaboration.

   The IP outbursts for which the VSNET Collaboration played
an important role of early notification include GK Per in 1996
(\cite{ish96gkperASCAiauc};
\cite{mor96gkperQPOmapping}; \cite{nog02gkper}), in 1999, whose occurrence
was notably predicted by a VSNET member, Tsutomu Watanabe
(vsnet-future 2),\footnote{
$\langle$http://www.kusastro.kyoto-u.ac.jp/vsnet/Mail/\\vsnet-future/msg00002.html$\rangle$.
} and real rise observed in detail by the VSNET Collaboration
(vsnet-alert 2652).\footnote{
$\langle$http://www.kusastro.kyoto-u.ac.jp/vsnet/Mail/alert2000/\\msg00652.html$\rangle$.
} These two outbursts were of special importance because detailed
modern time-resolved CCD observations set a stringent limit on the
expected occurrence of eclipses during outburst \citep{kim92gkper}.
These outbursts also provided an opportunity to study time-resolved
spectroscopy of the QPOs \citep{mor96gkperQPOmapping}, as well as
X-ray/optical detections of long-period QPOs
(\cite{mor99gkperQPO}; \cite{nog02gkper}).  Multicolor observations of
the 1996 outburst also provided observational constraints on the
outburst models \citep{nog02gkper}.  The 2002 outburst of GK Per
provided an opportunity to study magnetic accretion
\citep{bia03gkper}.  The long-term visual data, including the data
reported to the VSNET, have been used for analysis of outburst
properties \citep{sim02gkper}.

   In most recent years, the VSNET Collaboration has succeeded in
characterizing outburst properties of the unusual short-period intermediate
polar HT Cam \citep{tov98htcam}.  The outbursts are extremely brief
\citep{wat98htcam}, showing precipitous declines during the late part
of outburst.  From the time-difference of outburst maximum and the
maximum appearance of IP pulses during the 2001 outburst,
\citet{ish02htcam} concluded that
the outburst was triggered by a dwarf nova-type disk instability
phenomenon.  The existence of a precipitous later decline can be
explained by the truncation of the inner accretion disk.
The 2001 outburst of HT Cam was also studied by another group
\citep{kem02htcam}.  The outburst of HT Cam thus has been one of
milestones in the study of IP-dwarf nova relation.

   \citet{kat02v592cas} reported short-period,
nearly coherent, QPOs in V592 Cas, a nova-like star in the period
gap.  This object has been suggested to be a unique object in the
period gap showing both properties of superhumps and occasional IP-like,
nearly coherent, photometric oscillations.  Several other dwarf novae
have been suspected of IP-type signature, and were briefly discussed
in the relevant parts of subsection \ref{sec:sci:dwarfnova}.

\subsubsection{Polars}
\label{sec:sci:polar}

\begin{figure*}
  \begin{center}
    \FigureFile(160mm,80mm){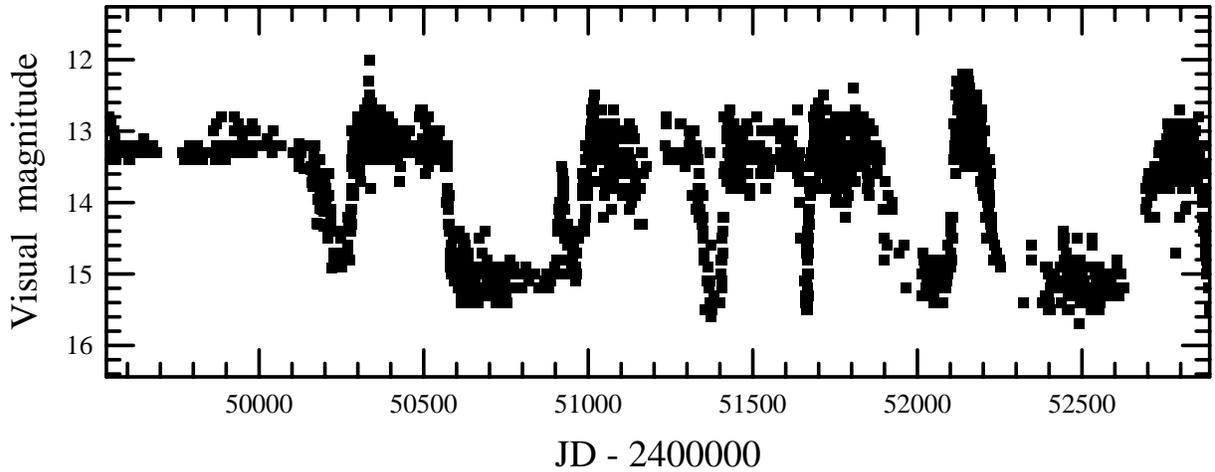}
  \end{center}
  \caption{Light curve of AM Her from VSNET observations.}
  \label{fig:amher}
\end{figure*}

   Bright polars (MCVs with synchronously rotating white dwarfs) have
been regularly monitored by the VSNET Collaboration.  Many polars,
notably AM Her (figure \ref{fig:amher}) and V834 Cen,
occasionally show low states, whose
occurrence has been notified through the VSNET alert system.
Long-term CCD monitoring of faint polars have been regularly
reported by Berto Monard.

   Coordinated multi-wavelength observations have been conducted on
several occasions.  The targets include AM Her, AR UMa, ST LMi, VV Pup,
AN UMa and EF Eri.  Some of these collaborative studies has been already
published as a solid paper \citep{hoa02vvpup}.

\subsubsection{VY Scl-Type Stars}
\label{sec:sci:vyscl}

\begin{figure*}
  \begin{center}
    \FigureFile(160mm,80mm){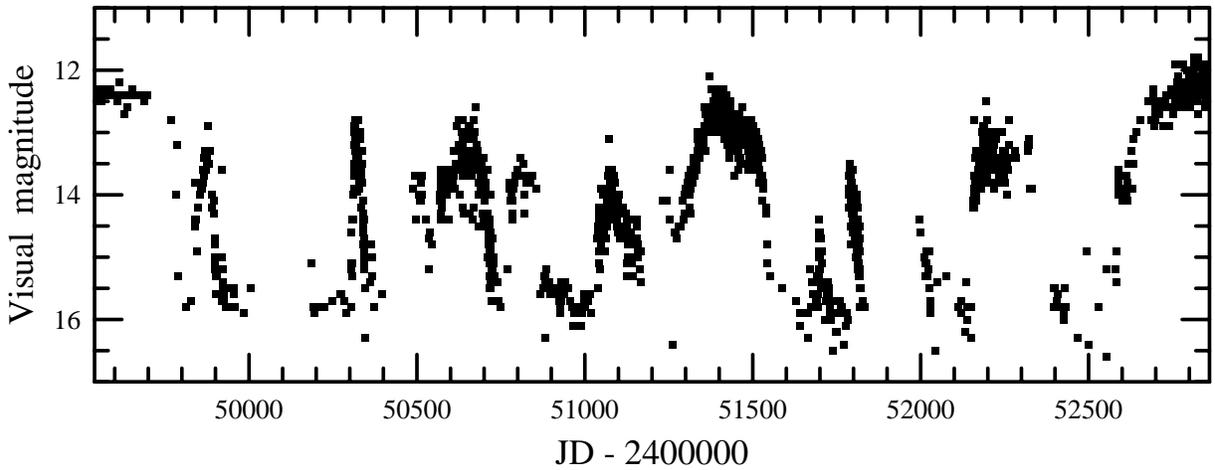}
  \end{center}
  \caption{Light curve of the VY Scl-type star MV Lyr from VSNET
  observations.}
  \label{fig:mv}
\end{figure*}

   VY Scl-type stars are novalike CVs with occasional low states,
or fading episodes (cf. \cite{war95book}; \cite{rob81mvlyr}; see
figure \ref{fig:mv} for the VSNET light curve of MV Lyr).
In some systems, these low states occur very infrequently
(the best example being TT Ari:
\cite{bor80ttariiauc}; \cite{fuh81ttari};
\cite{sha82ttariiauc}; \cite{hud84ttari}; \cite{sha85ttarilow};
\cite{hut85ttari}).\footnote{Although there was an interpretation of this
  object as a Z Cam-type dwarf nova permanently at standstill
  \citep{kra81ttari}, which was adopted by the GCVS team, it is generally
  considered as a VY Scl-type star with an occasional dramatic decrease
of mass-transfer.}  Such low states provide a
unique opportunity to study the white dwarf atmosphere or to directly
detect the secondary star (e.g. \cite{jam82ttariIRUV};
\cite{jam82ttariUV}).  From the viewpoint of disk-instability
model, the decreasing mass-transfer rate would produce a dwarf nova-type
disk instability if there is no special mechanism to suppress the
instability (\cite{hon94v794aql}; \cite{kin98DI}).  Observations,
however, tend to show smooth monotonous declines (\cite{hon94v794aql};
\cite{gre98vyscl}).  This makes a clear contrast to the ``low states"
in dwarf novae (cf. \ref{sec:sci:dnlow}).  There must be a mechanism
in VY Scl-type stars to somehow thermally stablizing the disk when
the mass-transfer is reduced (cf. \cite{lea99vyscl}).  Dense
observational coverage of VY Scl-type stars is therefore highly needed
immediately after the start of their declines.

   It is also known that VY Scl-type stars tend to show superhumps
(cf. \cite{pat99SH}).  Since the mass-ratios ($q$) of VY Scl-type stars
are not usually considered sufficiently small to enable excitation of the
3:1 resonance to produce superhumps (cf. \cite{hir90SHexcess};
\cite{woo00SH}; \cite{whi91SH}; \cite{mol92SHexcess}; \cite{mur98SH}),
there is apparently the need for an explanation of the cause of the
superhumps.  \citet{mur00SHvyscl} considered the variation in the
mass-transfer rate in VY Scl-type stars, and presented a working
hypothesis why VY Scl-type stars, with intermediate $q$, can show
superhumps.  This hypothesis also needs to be tested by more observations
of VY Scl-type stars during different brightness states.

   The VSNET Collaboration succeeded in early announcing a rare
fading of V751 Cyg, whose VY Scl-type nature was suspected more than
20 years ago \citep{rob74v751cyg} but had no comparable fading in
recent years.  The 1997 fading of V751 Cyg was originally reported to
the VSNET by Laszlo Szentasko, and its progress was followed in detail
by the VSNET Collaboration members, notably with CCD photometry at
Ouda Station.  This fading not only presented authentication of
V751 Cyg as a genuine VY Scl-type star, but also enabled X-ray
observations which led to the discovery of transient supersoft
X-ray emission (\cite{gre99v751cyg}; \cite{gre00v751cygvsge}).
This observation, suggesting a possible extension of luminous
supersoft X-ray sources (SSXS) toward a less-massive white dwarf,
led to a revolutionary change
in our view of VY Scl-type stars.  Other rare low states of VY Scl-type
stars announced through the VSNET Collaboration include
LQ Peg (=PG 2133+115) in 1999 (the second historical fading:
\cite{wat99lqpeg}; \cite{kat99lqpeg}; \cite{sch02lqpegproc}) and
BZ Cam in 1999 (the second historical fading:
\cite{wat00bzcam}; \cite{wat01bzcam}; \cite{kat01bzcam};
\cite{gre01bzcam}).  The 1999 fading of BZ Cam is notable in that
transient superhumps were detected during the fading \citep{kat01bzcam},
which may give support for the explanation by \citet{mur00SHvyscl}.

   The VSNET Collaboration also succeeded in presenting the first-ever
light curve of V504 Cen \citep{kat03v504cen}, which has been suspected
to be a VY Scl-type star from spectroscopy \citep{kil89v504cen}, but had
no solid photometric record qualifying the VY Scl-type nature.

   Well-known VY Scl-type stars, such as MV Lyr, have long been
best-observed targets by the VSNET Collaboration.  Some of these long-term
observations were employed to qualify the light curves of VY Scl-type
stars \citep{gre98vyscl}.  The well-known VY Scl-type star KR Aur was
intensively studied by the VSNET Collaboration \citep{kat02kraur},
which led to the detection of short-term variations having power-law
type temporal properties.

   The other notable object is V425 Cas, whose low state in 1998 was
announced by Timo Kinnunen through the VSNET, which led to our own
discovery of short-term (2.65 d), large-amplitude (up to 1.5 mag)
variations \citep{kat01v425cas}.  Such a type of variation had never
been never in any class of hydrogen-rich CVs, and \citet{kat01v425cas}
suggested that they are dwarf nova-type instabilities in a moderately
stabilized disk.  This discovery was introduced as the ``shortest period
dwarf nova" in Astrophysics in 2002 \citep{Astrophysics2002}.

\subsection{X-Ray Binaries and X-Ray Transients}
\label{sec:sci:xraynova}

   X-ray binaries are close binary systems which consist of a compact
object and a normal star.  The mass accretion from the normal
star onto the compact object generates strong X-ray emission.  A number
of X-ray binaries have been discovered as transients
(cf. \cite{LMXBcatalog}).  Their outbursts can be observed in all
wavelengths, hence simultaneous multi-wavelength observations have
played a key role to reveal the nature of X-ray binaries and X-ray
transients (e.g. \cite{oro97j1655}; \cite{hyn00j1118}).  Their outburst
cycle is generally longer than a year, and in some cases, longer than
decades.  Prompt observations of an early outburst phase are hence
important.  The VSNET has been providing information about X-ray
transients which enables prompt observations, not only for optical
observers, but also for X-ray, UV, IR, and radio observers.
Besides these recent studies, the earliest work by the authors include
infrared quiescent observation of V404 Cyg = GS 2023+338
\citep{kat90v404cyg}, which first revealed the existence of
the photometric period of 5.76 hr.  Another outstanding early result
was the discovery of superhumps and orbital variation in the
outbursting X-ray transient GRO J0422+32 = V518 Per
(\cite{kat92v518periauc}; \cite{kat93v518periauc}; \cite{kat95v518per}).
This observation has been one of the most comprehensive studies
of superhumps in ``classical" soft X-ray transients up to now
(cf. \cite{odo96BHXNSH}).

   Here we focus on the soft X-ray transients for which the VSNET
collaboration conducted intense world-wide campaigns.  Soft X-ray
transients are also called X-ray novae, whose common characteristics
were established in the mid-1990's (\cite{tan95BHXN}; \cite{che97BHXN}):
Their light curves are typically described with a Fast Rise and an
Exponential Decay (FRED).  The $e$-folding time is 30--40 d during the
decay phase.  In the FRED-type outburst, a reflare, or a secondary
maximum is observed $\sim 50$ d after the outburst maximum.  The
outbursts are considered to occur due to a sudden increase of mass
accretion rate in an accretion disk (\cite{min89BHXN};
\cite{hua89BHXN}).  Radial velocity studies of the secondary star have
revealed that a dozen of soft X-ray transients contains stellar-mass
black holes.  Soft X-ray transients thus provide an ideal laboratory of
the accretion physics onto the black hole, and they are called black
hole X-ray transients. 

   In the framework of classical soft X-ray transients, the optical and
X-ray emissions originate from the outer and the inner region of the
accretion disk, respectively.  Simultaneous optical and X-ray
observations therefore enable us to study the evolution of the accretion
disk and the mechanism of its activity.  The VSNET collaboration
enables us to obtain dense samples throughout outbursts.  With these
observations, we summarize the progress of studies of classical soft
X-ray transients in subsection \ref{sec:xray:cxt}. 

   On the other hand, several recent transients have exhibited
characteristics which are difficult to explain within the classical
framework.  The prompt observations by the VSNET collaboration has
played an important role, in particular, for the research on the
luminous fast transients (\cite{uem02v4641sgr}) and rapid optical
variations (\cite{uem00j1118pasj}; \cite{uem02v4641sgrletter};
\cite{uem04v4641sgr}).  We summarize our studies on these new classes of
activity in subsections \ref{sec:xray:fxt} and \ref{sec:xray:rapid}.

\subsubsection{Fast X-ray transients, V4641 Sgr and CI Cam}
\label{sec:xray:fxt}

   It has been proposed that the outburst of classical soft X-ray
transients is induced by thermal instability of the accretion disk
(\cite{min89BHXN}; \cite{hua89BHXN}).  This model can explain the large
$e$-folding time during their decay phase by considering a strong X-ray
irradiation which stabilizes the outer disk.  This standard picture
cannot, however, explain two atypical outbursts, that is, the fast
transients V4641 Sgr (\cite{hje00v4641sgr}; \cite{uem02v4641sgr}) and
CI Cam (\cite{bel99cicam}).  Their outburst duration was only a few days,
which is too short to be interpreted with the viscous decay of the
classical soft X-ray transients.

   On 1998 March 31, the All-Sky Monitor (ASM) of the
Rossi X-Ray Timing Explorer (RXTE) detected a new
X-ray transient named XTE J0421+560 (\cite{smi98cicamiauc}).
We were among the first to point out the presence of a supposed
symbiotic star CI Cam (\cite{ber95cicam}) within the error
(vsnet-alert 1621).\footnote{
See $\langle$http://www.kusastro.kyoto-u.ac.jp/vsnet/Mail/\\alert1000/msg00621.html$\rangle$.
}
This information was immediately relayed to X-ray and optical observers
through the VSNET, and the proposed identification was securely
confirmed with the discovery of an outbursting object at the location
of CI Cam (\cite{hje98cicamiauc6857}).  Contrary to
classical X-ray transients, this object started a rapid fading with
an $e$-folding time of $\sim 0.5$ d just after outburst maximum
(\cite{bel99cicam}).  Its X-ray spectrum can be described by an
absorbed power-law model with high-energy cutoff, which is atypical for
X-ray transients (\cite{bel99cicam}).
Optical observations reported
to the VSNET show that CI Cam brightened to 8.8 mag on April 3,\footnote{
See $\langle$http://www.kusastro.kyoto-u.ac.jp/vsnet/Xray/\\cicam.html$\rangle$.
}
and faded by $\sim 2$ mag within two days.  Spectroscopic observations
revealed that CI Cam is a B[e]--X-ray binary (\cite{bel99cicam}).  The
nature of the compact object is still unknown (\cite{bel99cicam};
\cite{orl00cicam}; \cite{hyn02cicam}; \cite{rob02cicam}).  The VSNET
Collaboration further obtained quiescent observations, which revealed
the presence of weak activity \citep{kat01cicam}.

   V4641 Sgr is a variable whose optical spectrum is that
of an A-type star (\cite{dow95CVspec};
note that the object was not correctly identified
in the literature until we published a reliable chart,
see subsection \ref{sec:standard}).
In 1999 August, Tsutomu Watanabe, a member of the
VSOLJ, noticed that the object entered an active state in the optical
range (\cite{wat99v4641sgr}).  This state was characterized by the
presence of large-amplitude variation having a possible periodicity
of 2.5 d \citep{kat99v4641sgr}.
On September 15, the state was terminated by a
short outburst reaching 8.8 mag independently detected by Rod Stubbings
and the Kyoto team, following Berto Monard's detection of a brightening
preceding this event (vsnet-alert 3475, 3477, 3478;
\cite{stu99v4641sgriauc}).\footnote{
See $\langle$http://www.kusastro.kyoto-u.ac.jp/vsnet/Xray/\\gmsgr.html$\rangle$
for the full story.
}  All these reports were immediately circulated through the VSNET alert
network; the resultant vigil eventually enabled the historical detection
of this perfectly unexpected giant outburst.

   Following our notification, a corresponding X-ray outburst was found
in the backlog recorded with the RXTE/ASM; the X-ray outburst was not
discovered ``real-time" even with X-ray all-sky monitor.
The X-ray outburst, after reaching an astonishing flux of 12 Crab at
the X-ray maximum, rapidly faded and returned to the
pre-outburst level within only 2 hours (\cite{smi99v4641sgriauc}). 
The duration of the optical outburst was also short ($\sim 5$ d)
(\cite{uem02v4641sgr}).  The VSNET collaboration team performed prompt
observation of this short X-ray outburst, and revealed that the optical
variation exhibited anti-correlation against the X-ray variation
(\cite{uem02v4641sgr}).  Spectroscopic observations revealed that
V4641 Sgr consists of a black hole and a late B-type star
(\cite{oro01v4641sgr}).  As reported below,
V4641 Sgr experienced active phases in the next couple
of years (\cite{uem02v4641sgrletter}; \cite{uem04v4641sgr};
\cite{rup03v4641sgriauc}), however, no comparable outburst to the
1999 September giant one has been yet observed.

   While their binary components are totally different, the short outbursts
of V4641 Sgr and CI Cam have several common characteristics: First,
spatially resolved radio jets were associated with the outburst in both
systems (\cite{hje98cicamiauc6872}; \cite{hje00v4641sgr}).  Second, the
peak luminosities reached the Eddington luminosity (\cite{rob02cicam};
\cite{rev02v4641sgr}).  Their high luminosity implies that
supercritical accretion occurred.  Since the matter falls with almost
free-fall time scale in the supercritical accretion, the problem of the
short duration of the outbursts can be reconciled.  The optical--X-ray
anti-correlation of V4641 Sgr may be understood with the scenario that the
optically-thick supercritical accretion flow absorbed the X-ray
emission and re-emitted the optical emission (\cite{rev02v4641sgr2};
\cite{uem02v4641sgr}).  It is, however, still unknown how the
supercritical accretion was induced.

\begin{figure*}
  \begin{center}
    \FigureFile(150mm,90mm){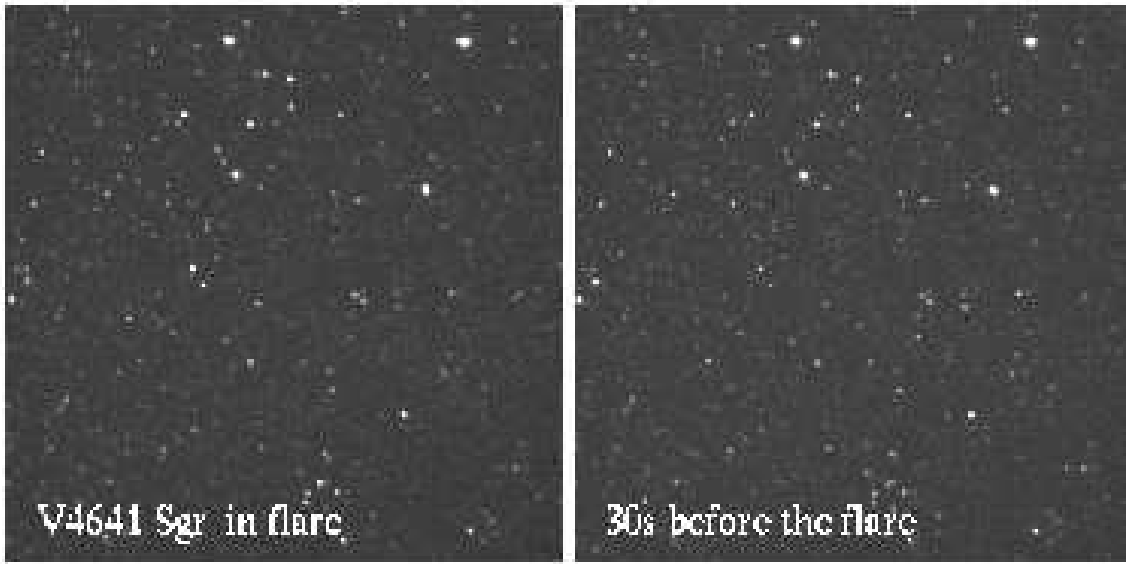}
  \end{center}
  \caption{``Optical flash'' from V4641 Sgr observed on 2002 July 7.}
  \label{fig:v4641flash}
\end{figure*}

\subsubsection{Short-term, non-thermal optical variations}
\label{sec:xray:rapid}

   In the classical picture, the optical emission is thermal emission from
the outer portion of the accretion disk where the temperature is
relatively low ($\sim 10^4$ K).  The observed time scale of optical
variations is hence long, such as superhumps (e.g. \cite{odo96BHXNSH})
and orbital period variations (e.g. \cite{kat95v518per}).  The black
hole binary system, GX 339-4 is, however, known to show rapid optical
variations of the time scale of seconds (\cite{mot82gx339};
\cite{mot83gx339}).  Such short time scale variations indicate that they
originate from an inner portion of the accretion flow.  They are
proposed to be cyclo-synchrotron emission from the inner region
(\cite{mot82gx339}), however, the mechanism to generate the emission and
variations is poorly known.  The VSNET collaboration team has recently
observed optical rapid variations of two sources, that is, XTE J1118+480
= KV UMa (\cite{uem00j1118pasj}) and V4641 Sgr
(\cite{uem02v4641sgrletter}; \cite{uem04v4641sgr}).

   XTE J1118+480 was discovered with the RXTE/ASM on 2000 March 29
(\cite{rem00j1118}).  We discovered an optical counterpart at 12.92
mag on March 30 (\cite{uem00j1118iauc}).  The optical light curve showed
so many fluctuations that \citet{uem00j1118pasj} suggested
the presence of optical rapid variations.
\cite{spr02j1118} detected optical short-term (a few
tens of seconds) variations which correlate with the X-ray variations.
These rapid optical variations are also proposed to be synchrotron
emission from the inner accretion flow (\cite{spr02j1118}).  Since XTE
J1118+480 remained at a low/hard state throughout the outburst, the
inner region is considered to be filled not by the standard disk, but by
the advection dominated accretion flow (ADAF: \cite{nar94ADAF}).  In the
ADAF region, the gas density is so low that the magnetic pressure is
dominant (\cite{min95lowstatedisk}).  The optical rapid variations are
probably generated at shock regions in such an inner region, which are
formed by magnetic reconnections or collisions of blobs in the
magnetically-dominated accretion flow (\cite{uem02v4641sgrletter}).

   After the luminous, short outburst in 1999 September, V4641 Sgr
experienced a new active phase in 2002 and 2003.  The VSNET
collaboration succeeded in detecting rapid optical variations during the
2002 active state (figure \ref{fig:v4641flash}),
whose detailed features and interpretation are reported
in \citet{uem04v4641sgr} in this volume.

\subsubsection{Classical soft X-ray transients}
\label{sec:xray:cxt}

   Superhumps, which have been originally studied in SU UMa-type dwarf
novae, are also observed in soft X-ray transients
(e.g. \cite{odo96BHXNSH}).  To observe the evolution of superhumps was
difficult in the case of soft X-ray transients because of their rare
outbursts, relatively long orbital period, and long durations of
outbursts made a dense sampling throughout outbursts difficult.
Our intense campaign of XTE J1118+480, however, first revealed
the evolution of superhumps in soft X-ray transients (\cite{uem02j1118}).
The superhump period was first 0.43\% longer than the orbital period,
and then decreased during the main outburst.
Superhumps in SU UMa-type dwarf
novae also exhibit this behavior, which can be interpreted by the
contraction of the elliptical accretion disk \citep{osa85SHexcess}
or the inward propagation of the eccentricity wave \citep{lub92SH}.

   It is proposed that the reflare is induced by the growing tidal
dissipation (\cite{ich94bxhn}; \cite{tru02SXTrebrightening}).  This
model was developed based on the fact that superhumps appeared only
after the reflare (\cite{odo96BHXNSH}).  On the other hand, superhumps
appeared even before the reflare in XTE J1118+480 (\cite{uem02j1118}).
The VSNET collaboration obtained a dense sample around the reflare of
XTE J1859+226, which is reported in \cite{uem04j1859} in this issue.  We
detected periodic variations, which may be superhumps, even before the
reflare of XTE J1859+226.  These observations are unfavorable for the
above scenario for the reflare.  It is also proposed that strong
X-ray irradiation onto the outer accretion disk may induce the reflare
(e.g. \cite{kin98SXTlightcurve}).

\subsubsection{High-Mass X-Ray Binaries}
\label{sec:xray:hmxb}

   SS 433 = V1343 Aql is an active high-mass X-ray binary (HMXB) with
relativistic jets, and
the nature of the binary system is still mysterious in many aspects.
Its optical magnitude is frequently monitored by VSNET observers,
and remarkable (brightening or fading) behavior is reported via
{\it vsnet-campaign-xray} when it occurs.

   In 1995, 1998, and 2000, simultaneous multi-wavelength observations
were organized by the ASCA team of RIKEN (N. Kawai et al., in preparation)
in order to determine an accurate ephemeris of the eclipse and
to compare light curves in the X-ray and optical wavelengths, which
yield a clue to understand the emitting regions in the binary system.
In these campaigns, VSNET played the role of a medium for exchanging
information on the optical parts, such as calls for optical
information, explanation of the background, and practical conditions
and notes for observation
(e.g. vsnet 103, for the 1995 campaign).\footnote{
$\langle$http://www.kusastro.kyoto-u.ac.jp/vsnet/Mail/vsnet/\\msg00103.html$\rangle$.
}

\subsection{Supernovae}
\label{sec:sci:SN}

  Our contributions to the spectroscopic type determination of supernovae
are summarized in table \ref{tab:SNspec}.  Confirmatory observations,
identifications and similar activities reported in IAUCs are also
listed in table \ref{tab:SNid}.  We also run the mailing list
{\it vsnet-campaign-sn},\footnote{
$\langle$http://www.kusastro.kyoto-u.ac.jp/vsnet/Mail/\\vsnet-campaign-sn/msg00785.html$\rangle$.
} which is now widely known as the most reliable and up-to-date
information of supernovae.
The ``Latest Supernovae" page (subsection \ref{sec:datasearch}) has been 
collaborating with it, which also provides USNO-A2.0 \citep{USNOA20}
based charts made
by Odd Trondal for almost every supernova brighter than 20th magnitude.
The charts gives blue and red magnitudes for photometric reference stars,
the latter ones have been widely used for ``CR" measurements
(see Appendix \ref{app:format}).

   Besides objects described in the subsections \ref{sec:earlyelec}
and \ref{sec:conf:SN}, here we mention here remarkable examples.

   SN 1997ef in UGC 4107 was discovered by Yasuo Sano, one of the most 
active participants of VSNET \citep{nak97sn1997efdisc}.  It was
originally announced as ``possible supernova", because its spectra was
quite unusual and could not convincingly be classified
as that of a supernova \citep{gar97sn1997efSpec}.
Further spectroscopic observations
revealed that it was likely an explosion with massive
(several M$_\odot$) ejecta.  The object consequently received
the designation of SN 1997ef (\cite{hu97sn1997efSpec},
\cite{fil97sn1997efSpec}).  \citet{fil97sn1997efSpec} and 
\citet{gar97sn1997efSpec2} suggested that it was likely be an explosion 
of a stripped very massive star, a suggestion supported by
theoretical modeling \citep{iwa00sn1997ef}.  It is the first 
example of the so-called ``type-Ic hypernova", a concept developed after
the suggested association SN 1998bw with GRB 980425
(\cite{gal98sn1998bw}, \cite{iwa98sn1998bw}), 
SN 2002ap (see subsection \ref{sec:conf:SN}), 
and the unambiguous identification of the supernova signature (SN 2003dh)
in GRB 030329 (see subsection \ref{sec:sci:grb}).

   SN 1997ei was discovered by Masakatsu Aoki
\citep{nak97sn1997eidisc} in NGC 3963.
The first spectroscopy reported in IAUC indicated that it is a type-Ia
supernova \citep{gar97sn1997eiSpec}.  Our spectroscopy showed some
peculiarity, then we reported that it could be a peculiar type-Ia supernova
\citep{aya97sn1997eiSpec}.  From later spectroscopic observations, it finally
turned out to be a type-Ic supernova (\cite{wan98sn1997eiSpec},
\cite{fil98sn1997eispec}).

   The SN 1998T case taught us the importance of identification.
It was produced in a pair of interacting galaxies,
and some catalogues of galaxies
gave discrepant designations for them.  The blobby nature of the host
galaxies also led to misidentifications of the supernova.  The discussion
on {\it vsnet-chat}, including the consultation of the NGC/IC project
\footnote{
$\langle$http://www.ngcic.com/$\rangle$.
}, led us to a correct identification of the galaxy, which was accompanied
by precise astrometry \citep{yam98sn1998Tid}.

   SN 1998bu in M 96 was discovered by Marko Villi.  It was the nearest
supernova since SN 1993J in M 81.  The first spectroscopy reported in
IAUC was a high-dispersion one, which could only determine the depth of the
interstellar absorption within our Galaxy and the host galaxy
\citep{mun98sn1998buSpec}.
Our spectrum revealed that it is a type-Ia supernova, and this report
was naturally 
posted to vsnet-alert 1785\footnote{
$\langle$http://www.kusastro.kyoto-u.ac.jp/vsnet/Mail/alert1000/\\msg00785.html$\rangle$.
}.  It was distributed earlier than the relevant IAUC
\citep{aya98sn1998buSpec},
which also included other spectroscopy \citep{mei98sn1998buSpec}.
In response to the type determination, the COMPTEL instrument
was pointed towards SN 1998bu in order to detect the
line $\gamma$-ray of $^{56}$Co decay, which had been detected
only from SN 1991T, the peculiar luminous type-Ia supernova.
Despite SN 1998bu is as close to us as SN 1991T,
the $\gamma$-ray lines was not
detected \citep{geo02sn1998buCGRO}, which may suggest
a diversity among type-Ia supernovae in the line $\gamma$-ray,
as well as in the light curves \citep{rie96snIalcvdiv}
and in the spectra \citep{nug95snIaspecdiv}.

   The first spectroscopic observation of SN 1998es indicated that
it is an intrinsically bright type-Ia supernova like SN 1991T
\citep{jha98sn1998esSpec}.  Our report confirmed it,
giving in addition the spectral evolution and the interstellar extinction
\citep{aya98sn1998esSpec}.

   SN 1999dn was a case similar to SN 1997ei.  Our spectrum suggested that
it was of type Ic with week He\textsc{i} lines \citep{aya99sn1999dnSpec}.
The same IAUC also contained two spectroscopic observations,
one of which \citep{tur99sn1999dnSpec} reached the same 
conclusion, but the other suggested it was of type Ia
\citep{qie99sn1999dnSpec}.
Later spectroscopy revealed that SN 1999dn is an intermediate event 
between type Ib and type Ic \citep{pas99sn1999dnSpec}.

   SN 2000ch was a very subluminous supernova.  It was originally
announced as a variable star in field of NGC 3432
\citep{KAI00sn2000chdisc}.
We noticed that the object can be seen on the DSS images since 1998
(vsnet-chat 2908),\footnote{
$\langle$http://www.kusastro.kyoto-u.ac.jp/vsnet/Mail/chat2000/\\msg00908.html$\rangle$.
} \citep{yam00sn2000ch}.  On the other hand, spectroscopy of this object 
\citep{wag00sn2000chSpec} suggested that it is located within
NGC 3432.  From these findings, \citet{fil00sn2000ch} assigned the supernova
designation of this object, resembling the ``type-V" SN 1961V.  The comment
of the discoverer (vsnet-chat 2944)\footnote{
$\langle$http://www.kusastro.kyoto-u.ac.jp/vsnet/Mail/chat2000/\\msg00944.html$\rangle$.
} finally supported this classification.

   SN 2001bf was yet another example similar to SN 1997ei and SN 1999dn.
Despite of its low signal-to-noise ratio, our spectrum clearly showed a deep
Si\textsc{ii} absorption feature, from which 
we estimated that it is a type-Ia supernova \citep{kaw01sn2001bfSpec}.
Another group suggested that it is of type Ic \citep{phi01sn2001bfSpec},
but the later spectral evolution revealed that it is indeed
a type-Ia supernova \citep{cho01sn2001bfSpec}.

   SN 2002ao was a slightly different case.  The first report on IAUC
quoted the resemblance with type-IIb supernovae (\cite{gal02sn2002aoSpec}, 
\cite{gal02sn2002aoSpec2}).  We estimated that it is of type Ic 
\citep{kin02sn2002aoSpec}, which is consistent with a later report 
\citep{fil02sn2002aoSpec} indicating the resemblance with SN Ic 1999cq.  
The rapid decline \citep{yam02sn2002aodecline} of SN 2002ao was another
common  feature with SN 1999cq \citep{mat00sn1999cqIcHe}.

   Other IAUC issues in relation with the activity of the VSNET
administrators include
IAUC 7033 (SN 1998eg: \cite{hur98sn1998egiauc7033}),
IAUC 8101 (SN 2003cg: \cite{hir03sn2003cgiauc8101}), and
IAUC 8171 (SN 2003gs: \cite{eva03sn2003gsiauc8171}).

\begin{table*}
\caption{Spectroscopy of supernovae.}\label{tab:SNspec}
\begin{center}
\begin{tabular}{ccccl}
\hline\hline
SN name   & Type & VSNET article\commenta & Relevant IAUC & Remark \\
\hline
1995al    & Ia   & alert 266 & 6256 \citep{aya95sn1995alSpec}
                  & see sec. \ref{sec:conf:SN}. \\
1997ei    & Ic   & alert 1399 & 6800 \citep{aya97sn1997eiSpec} & see text. \\
1998an    & Ia   & alert 1684 & 6878 \citep{aya98sn1998anSpec} & \\
1998aq    & Ia   & alert 1681 & 6878 \citep{aya98sn1998aqSpec} & \\
1998bu    & Ia   & alert 1785 & 6905 \citep{aya98sn1998buSpec} & see text. \\
1998es    & Ia,pec & alert 2394 & 7059 \citep{aya98sn1998esSpec} & see text. \\
1999bg    & II   & alert 2816 & 7137 \citep{aya99sn1999bgSpec} & \\
1999dn    & Ib/c & alert 3380 & 7244 \citep{aya99sn1999dnSpec} & see text. \\
1999gn    & II   & alert 3842 & 7336 \citep{aya99sn1999gnSpec} & \\
1999gq    & II   & alert 3862 & 7339 \citep{aya99sn1999gqSpec} & \\
2001bf    & Ia   & alert 5873 & 7625 \citep{kaw01sn2001bfSpec} & see text. \\
2001bg    & Ia   & alert 5871 & 7622 \citep{kaw01sn2001bgSpec} & \\
2001dp    & Ia   & alert 6299 & 7683 \citep{mig01sn2001dp} & \\
2002ao    & Ic   & camp-sn 339 & 7810 \citep{gal02sn2002aoSpec} & see text. \\
2002ap    & Ic,pec & alert 7120 & 7811 \citep{kin02sn2002apSpecIAUC}
                    & see sec. \ref{sec:conf:SN}. \\
2002bj    & II(n?) & camp-sn 363 & 7844 \citep{kin02sn2002bjSpec} & luminous\\
2002bo    & Ia   & alert 7241 & 7848 \citep{kin02sn2002boSpec} & \\
2002bu    & IIn  & alert 7259 & 7864 \citep{aya02sn2002buSpec} & \\
2002fk    & Ia   & alert 7516 & 7976 \citep{aya02sn2002fkSpec} & \\
2003J     & II   & camp-sn 534 & 8048 \citep{aya03sn2003Spec} & \\
2003K     & Ia   & camp-sn 534 & 8048 \citep{aya03sn2003Spec} & \\
\hline
 \multicolumn{5}{l}{\commenta alert for {\it vsnet-alert}, chat for {\it vsnet-chat},} \\
 \multicolumn{5}{l}{\phantom{\commenta} camp-sn2002ap for {\it vsnet-campaign-sn2002ap},} \\
 \multicolumn{5}{l}{\phantom{\commenta} camp-sn for {\it vsnet-campaign-sn},} \\
\end{tabular}
\end{center}
\end{table*}

\begin{table*}
\caption{Identification, astrometry and follow-up of supernovae.}\label{tab:SNid}
\begin{center}
\begin{tabular}{cccl}
\hline\hline
SN name  & VSNET articles\commenta & Relevant IAUCs & Remark \\
\hline
1998T  & chat 775, 779, 780, 786, 787 & 6859 \citep{yam98sn1998Tid}
        & see text. \\
1999et & $\cdots$ & 7344 \citep{yam00sn1999etid} & \\
2000M  & alert 4320 & 7373 \citep{dim00sn2000M} & \\
2000P  & alert 4363, 4365, 4366, 4368, 4369
                               & 7378 \citep{col00sn2000P}, & \\
       &                       & 7379 (corrigendum) & \\
2000ch & chat 2908, 2941, 2944 & 7415 \citep{KAI00sn2000chdisc}, & see text. \\
       &                       & 7417 \citep{wag00sn2000chSpec}, & \\
       &                       & 7419 \citep{yam00sn2000ch}, & \\
       &                       & 7421 \citep{fil00sn2000ch} & \\
2000cm & alert 4944, chat 3035 & 7436 \citep{mue00sn2000cm}, & \\
       &                       & 7437 \citep{jha00sn2000cm1}, & \\
       &                       & 7438 \citep{jha00sn2000cm} & \\
2001dp & alert 6317 & 7683 \citep{mig01sn2001dp} & \\
2002ao & alert 7190 & 7836 \citep{yam02sn2002aodecline} & see text. \\
2002ap & camp-sn2002ap 154 & 7836 \citep{yam02sn2002appredisc}
                            & prediscovery \\
2002dm & camp-sn 440 & 7921 \citep{san02sn2002dm}, & \\
       &             & 7923 \citep{yam02sn2002dm} & \\
2002ed & alert 7441, camp-sn 454 & 7940 \citep{mon02sn2002ed}, & \\
       &                         & 7943 \citep{yam02sn2002ed} & \\
2003ez & $\cdots$ & 8141 \citep{pap03sn2003ez}, & \\
       &          & 8142 \citep{yam03sn2003ez} & \\
\hline
  \multicolumn{4}{l}{\commenta Abbreviations same as in table \ref{tab:SNspec}.}
\end{tabular}
\end{center}
\end{table*}

\subsection{Symbiotic Variables}
\label{sec:sci:symbio}

\begin{figure*}
  \begin{center}
    \FigureFile(160mm,80mm){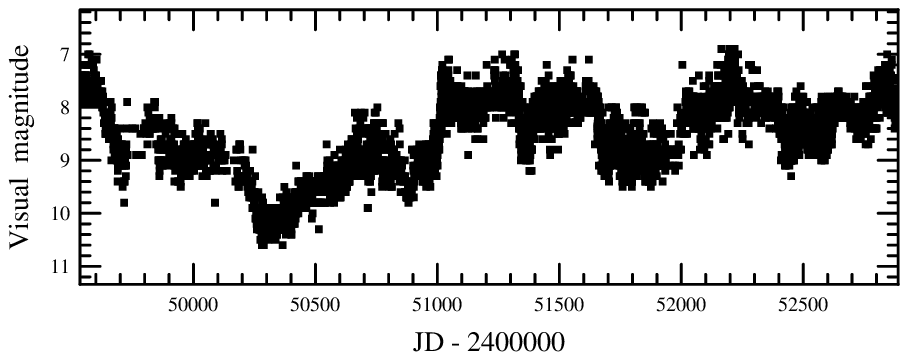}
  \end{center}
  \caption{Light curve of CH Cyg from VSNET observations.}
  \label{fig:chcyg}
\end{figure*}

   Bright symbiotic variables (cf. \cite{all84symbcatalog};
\cite{ken86book}; \cite{bel00symbiocatalog}; \cite{mun02symbiospec})
have been well-observed by many members of
the VSNET Collaboration (see figure \ref{fig:chcyg}).
These observations provided a number of
detections of outbursts and eclipses, which were immediately relayed
to more specialized researchers for detailed study.

   The eclipse phenomenon in the outbursting object FN Sgr was discovered
through the VSNET regular activity.\footnote{
$\langle$http://www.kusastro.kyoto-u.ac.jp/vsnet/Symbio/fnsgr.html$\rangle$.
} This work was summarized by \citet{wat00fnsgr}.

   The outburst and possible eclipse phenomenon in V343 Ser = AS 289
was discovered by Kesao Takamizawa (= Tmz~V17, vsnet-obs 8957)\footnote{
$\langle$http://www.kusastro.kyoto-u.ac.jp/vsnet/Mail/obs8000/\\msg00957.html$\rangle$.
} and Minoru Wakuda.  The final publication is by \citet{tak98v343ser}.

\begin{figure*}
  \begin{center}
    \FigureFile(160mm,80mm){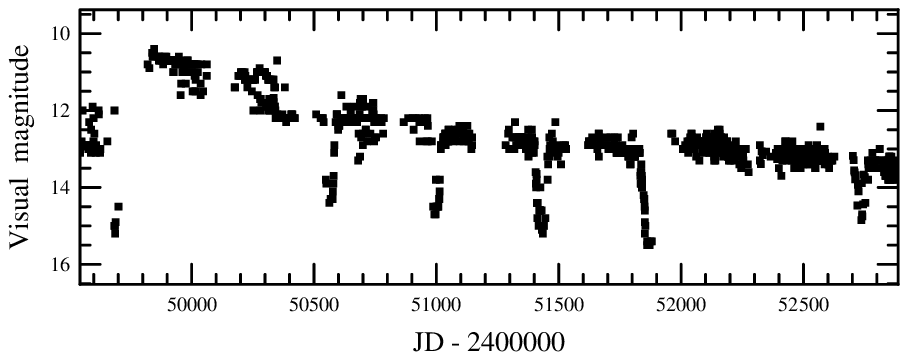}
  \end{center}
  \caption{Light curve of V1413 Aql from VSNET observations.  Deep eclipses
  are superimposed on a gradual fading from the 1995 outburst
  (\cite{eli95v1413aqliauc6165}; \cite{mun95v1413aqliauc6151}.}
  \label{fig:v1413}
\end{figure*}

   V1413 Aql = AS 338 is another object whose eclipsing symbiotic nature
(cf. figure \ref{fig:v1413})
was revealed by amateur astronomers \citep{wak88v1413aql}.\footnote{
See $\langle$http://www.kusastro.kyoto-u.ac.jp/vsnet/docs/\\v1413aql.html$\rangle$
for a full story.  See also \citet{sch85v1413aql}, \citet{mun92v1413aql},
\citet{mun93v1413aqliauc5884}.
}  The outbursts
and eclipses were regularly announced in VSNET, which have been followed
by a number of researchers (e.g. \cite{esi00v1413aql}; \cite{kol01v1413aql}).

   \citet{kat00chcyg} reported the similarity of light curves between
CH Cyg and the supersoft X-ray sources (V Sge and RX~J0513.9$-$6951)
based on VSNET observations.

   The 1997 and 2000 outbursts of Z And were reported in
vsnet-alert 938,\footnote{
$\langle$http://www.kusastro.kyoto-u.ac.jp/vsnet/Mail/\\vsnet-alert/msg00938.html$\rangle$.
} and vsnet-alert 5232, 5233,\footnote{
$\langle$http://www.kusastro.kyoto-u.ac.jp/vsnet/Mail/\\alert5000/msg00232.html$\rangle$ and
$\langle$http://www.kusastro.kyoto-u.ac.jp/vsnet/Mail/alert5000/msg00233.html$\rangle$.
} respectively.  These outbursts enabled, more or less owing to the VSNET
alerts, modern observations of this classical symbiotic binary
(\cite{sok99zand}; \cite{sko00zand}; \cite{sko03zandeclipse};
\cite{tom03zand}).

   \citet{ish01v694mon} systematically studied short-term variations
of V694 Mon = MWC 560.  During the entire period of observations,
the object showed pronounced flickering activity.  This work has been
referred to as one of the most intensive photometric observations of
ths peculiar symbiotic variable (\cite{sch01v694mon}; see also
\cite{sok01symbiorapidvar} for a recent survey work).  We now have
a dedicated mailing list for symbiotic stars {\it vsnet-symbio}\footnote{
$\langle$http://www.kusastro.kyoto-u.ac.jp/vsnet/Mail/\\vsnet-symbio/maillist.html$\rangle$.
} for informing about recent activities, particularly announcements of
outbursts and eclipses, of symbiotic variables.

\subsection{R Coronae Borealis Stars and Related Objects}
\label{sec:sci:rcb}

\begin{figure*}
  \begin{center}
    \FigureFile(160mm,80mm){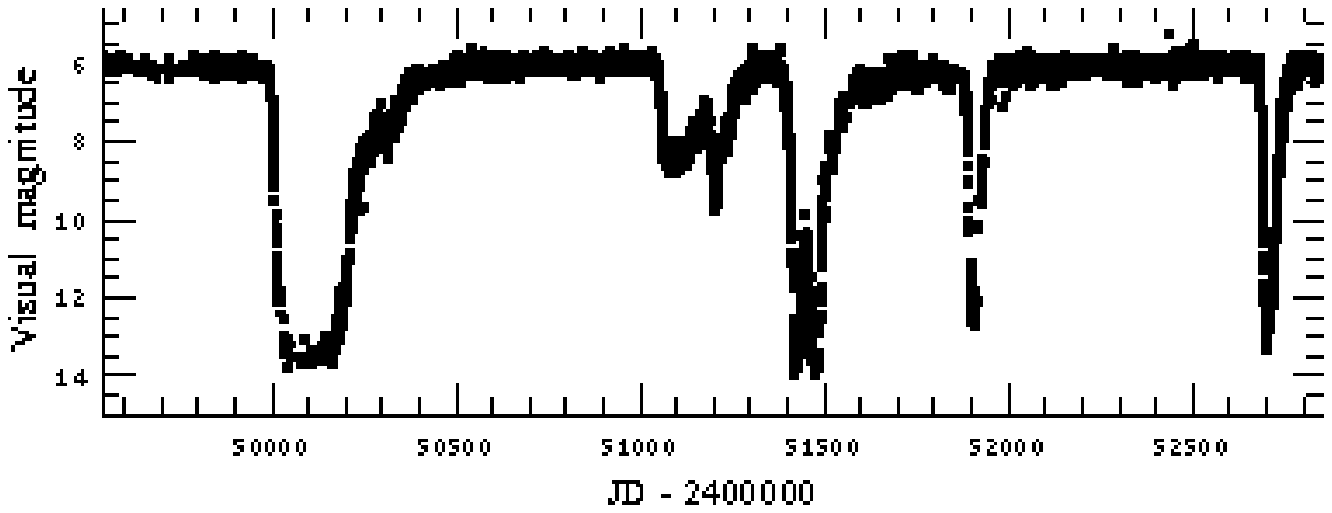}
  \end{center}
  \caption{Light curve of R CrB from VSNET observations.}
  \label{fig:rcrb}
\end{figure*}

   R CrB stars are hydrogen-deficient carbon stars which show occasional
fadings caused by dust formation \citep{cla96rcrbreview}.  A representative
light curve of R CrB from VSNET observations is shown in figure
\ref{fig:rcrb}.  Early announcements of the fadings of R CrB stars can
provide the best opportunities to study the formation mechanism of dust
in these stars.  Before the VSNET era, these fadings had been only
widely announced only when the
objects had unmistakably faded (typically below $\geq$ 1 mag below
their usual maximum); the early decline stage had been usually overlooked.
The real-time communication via the VSNET public lists broke this
historical limitation.  The most dramatic instance was with the first-ever
fading of FG Sge (also known as a final helium flash object)
in 1992 late August--September (cf. history partly recorded in
vsnet-history 200, 202).  It was only when the object had already faded
by 1 mag when IAUCs were issued (\cite{pap92fgsgeiauc5604};
\cite{waa92fgsgeiauc5609}; \cite{woo92fgsgeiauc5619}).
This detection of a possible fading (1992 August 30) reported by
Nobuhiro Makiguchi (VSOLJ, see subsection \ref{sec:vsoljcolab})
was immediately attended by a number of world-wide observers.
This fading was the first one of the series of fadings successively
occurring up tp now
(cf. \cite{iij93fgsge}; \cite{iij96fgsge}; \cite{ibe93fgsge};
\cite{jur93fgsge}; \cite{sto93fgsge}; \cite{woo93fgsgeIR};
\cite{ark94fgsge1}; \cite{sim94fgsge}; \cite{vangen94fgsge};
\cite{kip95fgsgespec}; \cite{vangen95fgsge};
\cite{ark96fgsgepuls}; \cite{ark96fgsgespec};
\cite{ark94fgsge2}; \cite{ark96fgsge}; \cite{ark98fgsge1};
\cite{ark98fgsge2}; \cite{ark99fgsge}; \cite{kip99fgsge};
\cite{gon98fgsge}; \cite{jor99fgsge}; \cite{kip01fgsge}).
The discovery of the fading in FG Sge brought a breakthrough in
the understanding of late-time low-mass stellar evolution:
our understanding of FG Sge had been slow and limited before this
phenomenon (cf. \cite{cow85fgsge}; \cite{aik85fgsge}; \cite{ark91fgsge};
although the unusual nature of this object was recognized more than 30 yr
ago and had long been discussed from different standpoints,
including binary hypothesis and a thermal pulse in stellar evolution
(e.g. \cite{kuk60fgsge}; \cite{fur62fgsge}; \cite{her68fgsge};
\cite{wen71fgsge}; \cite{pac70fgsge}; \cite{pac71fgsge}; \cite{spa73fgsge};
\cite{chr74fgsge}; \cite{lan74fgsge}; \cite{coh80fgsgebinary};
\cite{kur86fgsge}; \cite{fei90fgsge}).
It is now widely believed that FG Sge, V605 Aql, and V4334 Sgr
(cf. subsection \ref{sec:sci:CN}), as well as some unusual R CrB stars
(cf. V348 Sgr, e.g. \cite{cla02v4334sgr}) comprise a sequence of final
helium-flash objects (cf. \cite{law03v4334sgrv605aqlfgsge}).

   The dramatic fading of ES Aql, which had been suspected to be an
R CrB-type star, was first announced through the VSNET alert network;
Peter Williams was the first to detect this object getting fainter than
14.0 on 2001 March 22.  Based on this information, \citet{cla02esaql}
succeeded in clarifying the R CrB-type nature of this object.

   The other R CrB-type star recognized through the VSNET activity is
V2552 Oph = Had~V98 (\cite{kat03hadv98}, \cite{rao03hadv98};
see subsection \ref{sec:novaconfirm}).

   We now have a dedicated mailing list for R CrB stars and related
objects {\it vsnet-rcb},\footnote{
$\langle$http://www.kusastro.kyoto-u.ac.jp/vsnet/Mail/vsnet-rcb/\\maillist.html$\rangle$.
} which is best employed by the researchers of this field.
More recent announcements of rare R CrB-type fadings include the
2003 fading of V3795 Sgr (vsnet-rcb 585).\footnote{
$\langle$http://www.kusastro.kyoto-u.ac.jp/vsnet/Mail/vsnet-rcb/\\msg00585.html$\rangle$.
}
Although the attempt was not very successful, \citet{ros01rcrb} tried
to predict future light curve of R CrB using VSNET observations.

\subsection{Be Stars}
\label{sec:sci:be}

   Among recent phenomena in Be stars (B-type emission-line stars),
the case of $\delta$ Scorpii is still fresh in our memory.  The star,
which had been considered as a non-variable B-type star, underwent a
dramatic change in 2000 July.  Sebastian Otero, a VSNET member, visually
noticed a 0.1 mag brightening in $\delta$ Scorpii and issued an alert
through the VSNET (vsnet-be 2).\footnote{
$\langle$http://www.kusastro.kyoto-u.ac.jp/vsnet/Mail/vsnet-be/\\msg00002.html$\rangle$.
}  The supposed transition to a Be star was subsequently
confirmed by spectroscopic observations \citep{fab00deltascoiauc}.
The star further brightened
to a maximum of $V\sim$1.8 around 2000 July 30.  Such a dramatic change
in a bright naked-eye star is extremely rare.  The only comparable
precedent phenomenon was $\gamma$ Cassiopeiae in 1937, which brightened
to $V\sim$1.6.
This news was widely distributed through public news media,
as originating from the VSNET, and became the one of the most popular
astronomical phenomena in that year.  This object, after reaching
a temporal minimum just following the initial peak, continues to show
remarkable activities up to 2003 (\cite{ote01deltasco};
\cite{gan02deltasco}).  On several occasions in 2002 and 2003,
the star even brightened close to $V\sim$1.5, even slightly surpassing
in brightness the historical event of $\gamma$ Cassiopeiae.

   In the aftermath of this event, visual monitoring of bright Be stars
has been conducted by a number of VSNET members, most intensively by Otero.
These observations have detected of number of outbursts, e.g. in
$\mu$ Centauri (cf. \cite{baa01mucen}) and
$\omega$ Canis Majoris (cf. \cite{ote02omegacma}).
The VSNET runs dedicated lists on the Be-star
phenomenon, {\it vsnet-be}\footnote{
$\langle$http://www.kusastro.kyoto-u.ac.jp/vsnet/Mail/vsnet-be/\\maillist.html$\rangle$.
} and
{\it vsnet-campaign-be}.\footnote{
$\langle$http://www.kusastro.kyoto-u.ac.jp/vsnet/Mail/\\vsnet-campaign-be/maillist.html$\rangle$.
}

\subsection{Wolf--Rayet Stars}
\label{sec:sci:wr}

   Wolf--Rayet (WR) stars are massive, luminous, and hot stars which have
lost their hydrogen envelope, and are considered to be immediate precursors
of some kinds of supernovae, and likely GRBs.
In spite of their astrophysical importance,
WR stars were less conspicuous objects in terms of optical variability.
The only known categories of variability in WR stars had been occasional
short-period obscuration by dust production, instabilities in the wind,
or possible pulsation (see \cite{wil90}; \cite{wil95}; \cite{veen97}
for recent references).

   This situation dramatically changed by the discoveries of the two
most actively variable WR stars by the VSNET team
(WR 104 = Had~V82 = V5097 Sgr: \cite{kat02wr104},
WR 106 = Had~V84 = V5101 Sgr: \cite{kat02wr106}).
Both objects were initially reported as variable stars by Katsumi
Haseda, detected during his search for novae.  The variables soon turned
out to be identical with known WR stars.  In particular, WR 104 is
a well-known binary consisting of a late-type WR star and an OB star.
The most remarkable feature in this object is the presence of
a ``dusty pinwheel nebula" (\cite{tut99}; \cite{tut02})
co-rotating with the interferometric binary.
The importance of the detection of large-amplitude optical variation
was immediately recognized and communicated via VSNET alert network.
Such a large variation in WR 104 required a non-classical interpretation
\citep{kat02wr104}.  With these conspicuous discoveries, the VSNET
significantly broadened the scope of variability studies of WR stars,
and this field is now becoming one of the contemporary topics in
studying the WR-type activity.  Combined with the recent advance in
GRB astronomy, the importance of fundamental understanding of various
phenomena in WR stars will be a matter of key importance.

\subsection{Pre-main Sequence Stars}
\label{sec:sci:prems}

   Some pre-main sequence stars show dramatic outbursts (FU Ori stars:
FUORs) and smaller outbursts (EX Lup stars: EXORs), probably originating
from some kind of instabilities in the circumstellar disk
(e.g. \cite{kaw93fuori}).
Outbursts of these objects were occasionally followed by the VSNET
Collaboration.  The best example was the 1995 outburst of V1143 Ori
(Baba et al. in preparation).  Outbursts of V1118 Ori have also been
occasionally reported (e.g. \cite{hay98v1118ori}).

\begin{figure*}
  \begin{center}
    \FigureFile(160mm,80mm){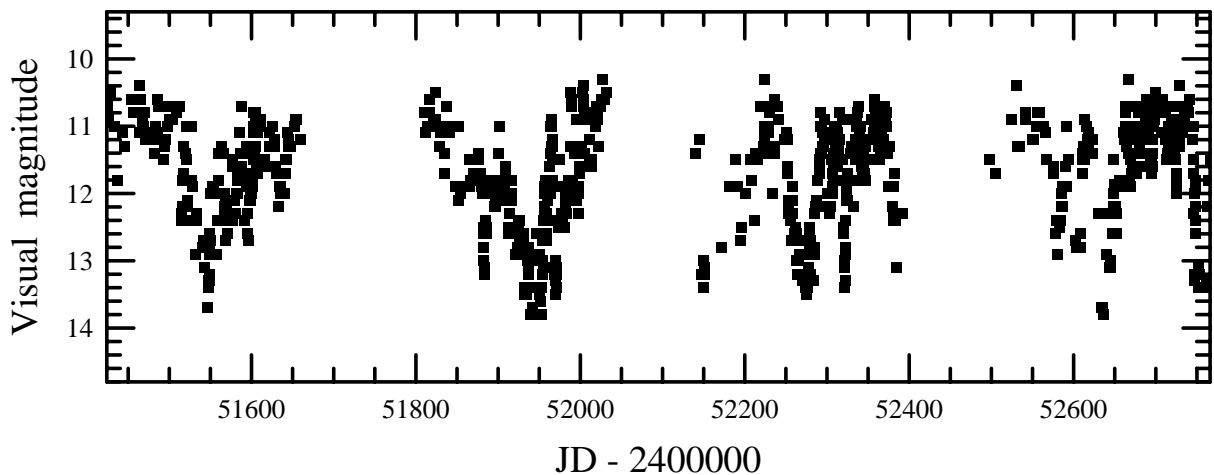}
  \end{center}
  \caption{Complex light curve of the Herbig Ae/Be star RR Tau
  from VSNET observations.}
  \label{fig:rrtau}
\end{figure*}

   Among pre-main sequence stars, Herbig Ae/Be stars have been one of
the best observed objects by VSNET members.  The recent discovery
of a likely Herbig Ae/Be star with large-amplitude variation, MisV1147,
has been widely studied through the VSNET (M. Uemura et al. in preparation).
Other objects of this class, well observed and timely informed through
the VSNET, include UX Ori, RR Tau, CQ Tau, AB Aur, RZ Psc and WW Vul
(see figure \ref{fig:rrtau}).
Remarkable variations in these objects have been relayed through
a dedicated list of pre-main sequence variables,
{\it vsnet-orion},\footnote{
$\langle$http://www.kusastro.kyoto-u.ac.jp/vsnet/Mail/vsnet-orion/\\maillist.html$\rangle$.
} and utilized for detailed follow-up studies.

\subsection{Peculiar Variable Stars}
\label{sec:sci:pec}

\begin{figure*}
  \begin{center}
    \FigureFile(160mm,80mm){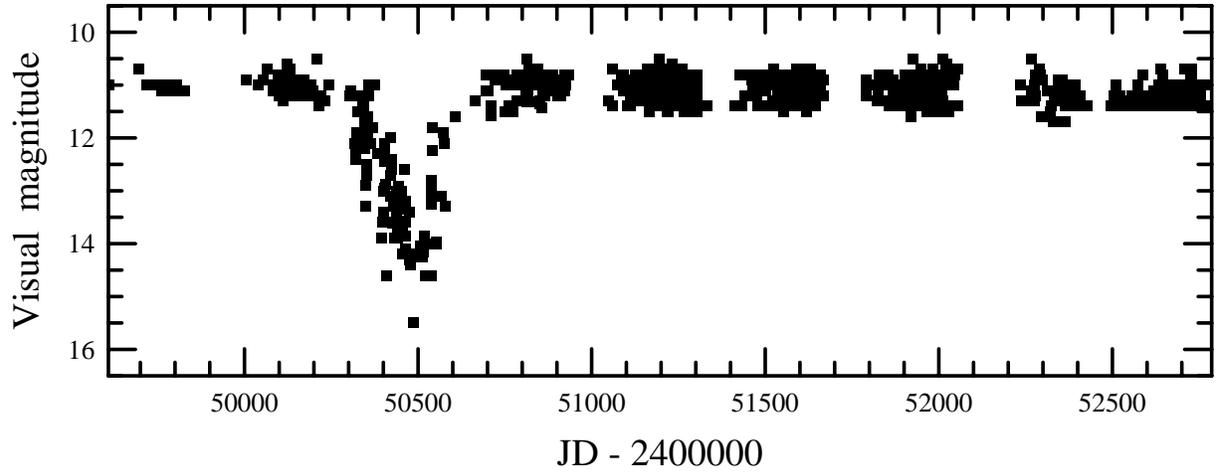}
  \end{center}
  \caption{Light curve of V651 Mon from VSNET observations.}
  \label{fig:v651mon}
\end{figure*}

   Some peculiar (or unique) variable stars were also studied by the
VSNET Collaboration.  The most striking objects include V651 Mon
(the central star of the planetary nebula NGC 2346).
This binary system containing a B-type subdwarf underwent a totally
unexpected fading in 1981--1985 (for a summary of this event, see
\cite{cos86v651mon}).  Although no similar phenomenon was
recorded in the century-long past photographic records, the object
underwent another unexpected fading episode in 1996--1997.  This
phenomenon was detected by Danie Overbeek, and immediately reported
to the VSNET (vsnet-alert 548).\footnote{
$\langle$http://www.kusastro.kyoto-u.ac.jp/vsnet/Mail/vsnet-alert/\\msg00548.html$\rangle$.
}  Following this announcement, \citet{kat01v651mon} succeeded in
recording the phenomenon in detail.  There was a sharply defined
transient clearing (brightening) even during this fading, which was
ascribed to a sharply defined, small (several times $\sim$10$^{11}$ cm)
lucent structure within the obscuring body.  See figure \ref{fig:v651mon}
for the recent light curve.

\subsection{Pulsating Variables}
\label{sec:sci:puls}

\begin{figure*}
  \begin{center}
    \FigureFile(160mm,80mm){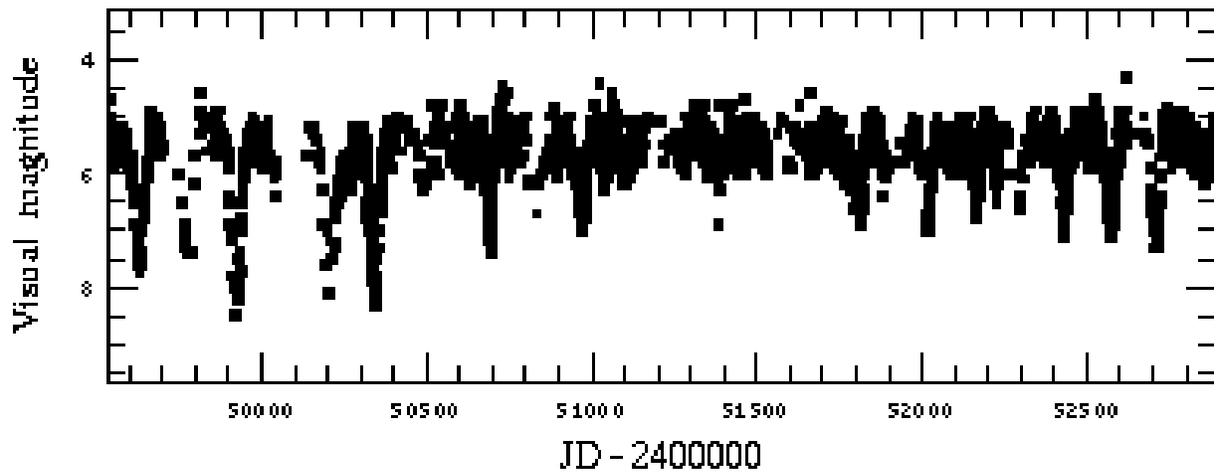}
  \end{center}
  \caption{Light curve of the RV Tau-type star R Sct from VSNET observations.}
  \label{fig:rsct}
\end{figure*}

   Although pulsating variables were only occasionally selected as
intensive targets of the VSNET Collaboration, the VSNET public data
archive (cf. figure \ref{fig:rsct}),
as well as the VSOLJ database, have been frequently used in
period analysis, and correlation studies with other observational
modalities.  These references to the VSNET/VSOLJ data have been so
numerous that we only list most recent ones: historical archive for
$\mu$ Cephei (\cite{ski01mucep}), evolution of R Hya
(\cite{zij02rhya}), dust formation in L$_2$ Pup (\cite{bed02l2pup}),
possible chaotic behavior in R Cyg (\cite{kis02rcyg}),
period determination of V648 Oph (\cite{kaz02v648oph}),
non-variability of EK And (\cite{yos03ekand}),
and period variation in T UMi (\cite{sza03tumi}).

   Numerous new variable stars reported in {\it vsnet-newvar}
(subsection \ref{sec:newvar}) have been studied in detail, and have
been reported in a numerous number of papers published in Information
Bulletin on Variable Stars (IBVS).

\begin{figure*}
  \begin{center}
    \FigureFile(160mm,80mm){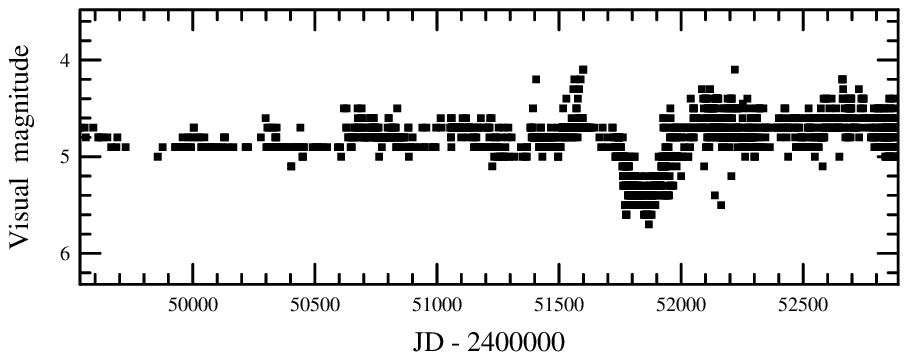}
  \end{center}
  \caption{Light curve of $\rho$ Cas from VSNET observations.}
  \label{fig:rhocas}
\end{figure*}

   $\rho$ Cas, originally classified as a yellow semiregular
variable, is now considered to be a low-temperature counterpart
(yellow hypergiants: \cite{dejag98rhocas}; \cite{isl99rhocas};
\cite{dejag01rhocas}; \cite{sto01rhocas}) of the extremely luminous
hot hypergiants (luminous blue variables: LBVs).  This object
occasionally undergoes temporary optical fadings caused by
huge mass-loss events, usually once in decade(s)
(figure \ref{fig:rhocas}).
The most spectacular recent event occurred in 2000 (\cite{lob03rhocas}).
Independent detections, including that by one of the authors (TK),
of this phenomenon were circulated through
the VSNET (vsnet-alert 5186, 5187),\footnote{
$\langle$http://www.kusastro.kyoto-u.ac.jp/vsnet/Mail/alert5000/\\msg00186.html$\rangle$ and
$\langle$http://www.kusastro.kyoto-u.ac.jp/\\vsnet/Mail/alert5000/msg00187.html$\rangle$.
} which enabled first dense optical coverage of this kind
of phenomenon.  Upon recognition of the astrophysical significance
of this event, we set up a dedicated mailing list
{\it vsnet-campaign-rhocas} in 2000 August.\footnote{
$\langle$http://www.kusastro.kyoto-u.ac.jp/vsnet/Mail/\\vsnet-campaign-rhocas/maillist.html$\rangle$.
}  These observations have provided the primary resource on this rare
event, and are referenced on Lobel's dedicated page on
$\rho$ Cassiopeiae.\footnote{
$\langle$http://cfa-www.harvard.edu/$^{\sim}$alobel/$\rangle$.
}
In 2003 there was a small signature of a line variation similar to
the precursor event in 2000.  This news was widely announced through
{\it vsnet-campaign-rhocas} and the object has been intensively observed.

\subsection{Eclipsing Binaries}
\label{sec:sci:ecl}

   Although classical (other than CV-type or symbiotic type) binaries are
usually not intensive targets for the VSNET Collaboration, there have
been several calls for observations, particularly for long-period
eclipsing binaries (e.g. OW Gem, EE Cep).  Visual and CCD/photoelectric
observations, as well as comparison star sequence information,
have been exchanged on VSNET lists.
The discovery of eclipses in the bright naked-eye star
$\delta$ Velorum was one of the breaking news mediated through the VSNET
(see also \cite{ote00deltavel}).  A dedicated list for $\delta$ Velorum
observations has been set up, {\it vsnet-campaign-deltavel}.\footnote{
$\langle$http://www.kusastro.kyoto-u.ac.jp/vsnet/Mail/\\vsnet-campaign-deltavel/maillist.html$\rangle$.
}

   The VSNET runs dedicated lists on eclipsing binaries,
{\it vsnet-ecl}\footnote{
$\langle$http://www.kusastro.kyoto-u.ac.jp/vsnet/Mail/vsnet-ecl/\\maillist.html$\rangle$.
} which are now recognized as one of world-wide networks exchanging
information in eclipsing binaries.  Kazuo Nagai (VSOLJ) has been summarizing
the times of minimum of eclipsing binaries reported to {\it vsnet-ecl}.

\subsection{Blazars}
\label{sec:sci:blazar}

   Blazars (BL Lac objects and optically violently variable quasars) are
also one of the targets of long-term and intensive observing campaign.
We run a dedicated list {\it vsnet-campaign-blazar}\footnote{
$\langle$http://www.kusastro.kyoto-u.ac.jp/vsnet/Mail/\\vsnet-campaign-blazar/maillist.html$\rangle$.
} for exchanging information on blazar activities and upcoming collaboration
with other world-wide blazar groups.  From the very establishment of
the Whole Earth Blazar Telescope (WEBT: \cite{WEBT}),\footnote{
$\langle$http://www.to.astro.it/blazars/webt/$\rangle$.
} we have been continuously
in collaboration with this group.

   In addition to continuous visual monitoring campaigns on strongly
active blazars, such as OJ 287, Markarian 421, 3C 279, quick electronic
circulation and prompt feedback via the VSNET discussion group led to the
notable discovery of unexpectedly large, short-term intranight variation
of BL Lac in 1997.\footnote{
$\langle$http://www.kusastro.kyoto-u.ac.jp/vsnet/BLLAC/bllac.html$\rangle$.
}  This intensive observation was initiated by a report of a bright state
of BL Lac in IAUC \citep{mat99bllaciauc7189}, which initially
revealed ``outbursts" every five to ten days.
Following this stage, the object entered a more active phase
in mid-July to early August.  During this stage, real-time comparisons
of visual observations through the VSNET discussion group revealed
substantial discrepancies depending on observers' longitudes.
These discrepancies, which were much larger than what had been recognized
as blazar microvariability, were initially considered as a result of
an inhomogeneous magnitude system.  This possibility was soon disproved
by the real-time distribution of the modern photoelectric comparison
star magnitudes.  By the end of 1997 July, the incredible short-term
variation (0.8 mag in four hours) in BL Lac had become a doubtless
phenomenon.

   In response to these visual observations, the Kyoto University team
obtained long time-resolved CCD photometry at Ouda Station, revealing
the unprecedented complexity and fast variation in the light curve.
A similar conclusion was reached with time-resolved CCD photometry
by Tonny Vanmunster, which was also rapidly communicated via VSNET.
The variation has a power-law temporal properties, analogous to those
of AGN variability (although the variation in BL Lac was much more
violent and rapid), which can be tracked down to five minutes.  This
is one of the shortest time scales ever recorded in blazar optical
variation \citep{mat99bllac}.  The early results by the VSNET team
were presented at the 23rd IAU General Assembly held in Kyoto
held in 1997 August.

   Since then, several studies have been performed in collaboration with
the WEBT (\cite{rav02bllac}; \cite{vil02bllac}).
Long-term light curves of selected blazars from VSNET observations have
been widely used for correlation with multiwavelength data.

\subsection{Gamma-Ray Bursts}
\label{sec:sci:grb}

   The VSNET collaboration team has tried to search optical afterglows of
GRBs and reported their detection or upper-limits of magnitudes.
Earlier announcements of Beppo Satellite per Astronomia X (BeppoSAX)
GRB detections were reported to
{\it vsnet-alert}.  Since 2000 June, we set up a dedicated list
for GRB announcements ({\it vsnet-grb}).\footnote{
$\langle$http://www.kusastro.kyoto-u.ac.jp/vsnet/Mail/vsnet-grb/\\maillist.html$\rangle$.
}  This mailing list is currently employed to distribute satellite-borne
alerts, as a secondary distribution node of
the GRB Coordinates Network (GCN).\footnote{
$\langle$http://gcn.gsfc.nasa.gov/$\rangle$.
}  A public mailing list {\it vsnet-grb-info},\footnote{
$\langle$http://www.kusastro.kyoto-u.ac.jp/vsnet/Mail/\\vsnet-grb-info/maillist.html$\rangle$.
} originally extended from {\it vsnet-grb}, currently provides
GCN Circular information as well as our own observations by the
VSNET Collaboration.  Table \ref{tab:GRBobs} summarizes our own GRB
afterglow observations (not all observations are listed),
which is an extension of table 1 in \citet{yam03GRBsmalltel}.
See the corresponding GCN Circulars for more details.

\begin{table*}
\caption{GRB Afterglows observed by the Kyoto team.}
\label{tab:GRBobs}
\begin{center}
\begin{tabular}{cccccc}
\hline\hline
GRB name  & Alert type & Epoch (d)\commenta & Telescope (cm)
          & Magnitude\commentb & GCN Circular \\
\hline
001025A & IPN  & 1.7   & 60    & $>$18.0 & 866 \citep{uem00grb001025agcn866} \\
001025B & IPN  & 1.6   & 60    & $>$19.0 & unpublished \\
001212  & IPN  & 2.0   & 25    & $>$17.5 & 902 \citep{ish00grb001212gcn902} \\
010214  & SAX  & 0.29  & 30    & $>$18.0 & 948 \citep{uem01grb010214gcn948} \\
010220  & SAX  & 1.5   & 30    & $>$17.5 & unpublished \\
010222  & SAX  & 0.23  & 30    & 17.8 & 984 \citep{uem01grb010222gcn984} \\
        &      & 0.55  & 30    & 19.3 & \\
011030  & SAX  & 0.28  & 25    & $>$16.0 & unpublished \\
011130  & HETE & 0.47  & 25    & OF & unpublished \\
020124  & HETE & 0.079 & 25,30 & OF & unpublished \\
020331  & HETE & 0.033 & 25,30 & 17.9 & 1363 \citep{kat02grb020331} \\
020812  & HETE & 0.018 & 60    & $>$21.2
                & 1515 \citep{oha02grb020812gcn1515} \\
        &      & 0.038 & 25,30 & $>$19.1
                & 1521 \citep{uem02grb020812gcn1521} \\
020813  & HETE & 0.36  & 60    & $\sim$19 & unpublished \\
020819  & HETE & 0.14  & 60    & $>$15.0 & unpublished \\
020903  & HETE & 0.18  & 25    & OF or $>$17.5
                & 1537 \citep{uem02xrf020903gcn1537} \\
021004  & HETE & 0.036 & 25,30 & 16.3 & 1566 \citep{uem02grb021004gcn1566} \\
030227  & HETE & 0.046 & 25    & $>$18.0
                & 1899 \citep{uem03grb030227gcn} \\
030329  & HETE & 0.053 & 25,30 & 12.6
                & 1989 \citep{uem03grb030329gcn1989}, \\
        &      &       &       &
                & 1994 \citep{uem03grb030329gcn1994}, \\
        &      &       &       &
                & 2147 \citep{yam03grb030329gcn} \\
030528  & HETE & 0.004 & 25    & $>$16.0
                & 2252 \citep{uem03grb030528gcn} \\
030823  & HETE & 1.24  & 60    & $>$20.5
                & 2370 \citep{ima03grb030823gcn2370} \\
\hline
 \multicolumn{6}{l}{\commenta Day after the burst detection.} \\
 \multicolumn{6}{l}{\commentb Unfiltered CCD or $R_{\rm c}$ magnitude.
                    OF = optical counterpart out of the field of view.} \\
\end{tabular}
\end{center}
\end{table*}

   Here we summarize our significant detections and their importance in GRB
astronomy:

\vskip 2mm
\noindent
{\it GRB010222}: Our observation of the afterglow of GRB010222 covered a
period around the jet break (\cite{uem01grb010222gcn984}).
This is the first detection of the GRB afterglow in Japan.\footnote{
  The first detection with a Japanese telescope was the infrared afterglow
  of GRB000926 observed with the Subaru telescope in Hawaii
  (\cite{kob00grb000926}).
} This observation encouraged observers having small telescopes in Japan
to try observations of GRB afterglows.

\vskip 2mm
\noindent
{\it GRB020331}: We succeeded in observing an early afterglow with
$R_{\rm c}=17.9$ (17.5--18.7; 1-$\sigma$ limits) 64 min after
the burst (\cite{kat02grb020331}).  This is the earliest observation
for this afterglow.  Our observation revealed that the light curve can be
described with a single power-law from the early phase ($\sim 1$ hr) of
the afterglow.  

\vskip 2mm
\noindent
{\it GRB021004}: Owing to the prompt identification by the HETE-2
satellite (\cite{shi02gcn1565}), we first revealed the continuous
behavior of the GRB afterglow around 1 hr after the burst
(\cite{uem02grb021004gcn1566}; \cite{uem03grb021004pasj}).  An early 
afterglow was also observed in GRB990123 (\cite{sar99grb990123}),
however, our observation covered a period corresponding to an observing
gap of the early afterglow light curve of GRB990123.  In the light
curve of GRB021004, the initial fading phase was terminated by a short
plateau phase that lasted for about 2 hours from 0.024 to 0.10 d after the
burst (\cite{uem03grb021004pasj}).  The object then entered an ordinary
power-law fading phase.  We propose that the plateau phase is
evidence that the maximum of the synchrotron emission from a forward
shock region appears around $\sim 0.1$ d after the burst, as expected
from theoretical calculations (\cite{sar99prediction};
\cite{kob03grb021004}).  The initial fading phase can be interpreted as
part of the optical flash, which was recorded in GRB990123.  If this is
the case, the color of the afterglow would have dramatically changed
from blue to red around the maximum, while no color information is
available in this early phase (\cite{sar98grb}).
Detections the color change in the
early phase will be an important future step for GRB astronomy.
On the other hand, the feature around 0.1 d after the burst can be
one of a series of bumps observed 1 d after the burst
(\cite{nak03grb021004}; \cite{laz02grb021004density};
\cite{fox03grb021004nature}; \cite{sch03grb021004progenitor};
\cite{li03grb021004wind}).

\begin{figure*}
  \begin{center}
    \FigureFile(150mm,100mm){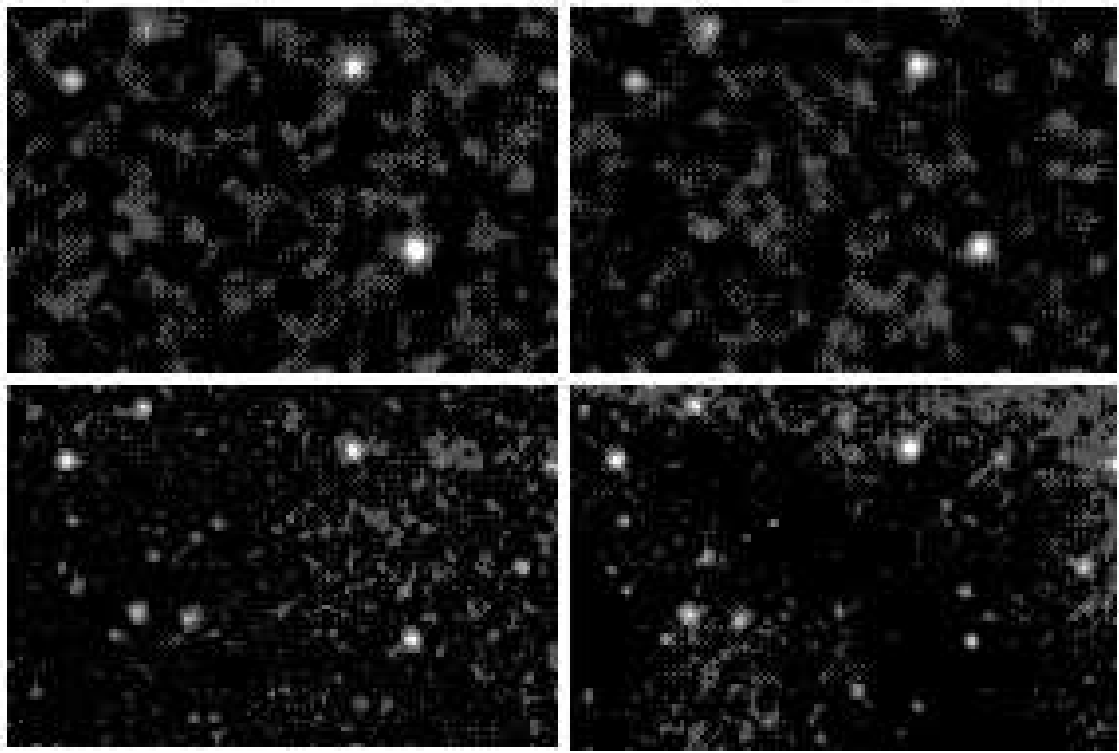}
  \end{center}
  \caption{Fading afterglow of GRB030329.  The images were taken by
  the VSNET Collaboration.  The individual images (from left top to
  right bottom) correspond to 1.27, 3.37, 8.18 and 28.65 hr
  after the gamma-ray burst.}
  \label{fig:g0329}
\end{figure*}

\vskip 2mm
\noindent
{\it GRB030329}: The world-famous ``monster GRB'' GRB030329
(figure \ref{fig:g0329}) occurred closest to us
($z=0.168$), which gave a chance for us to study the detailed structure
of the GRB afterglow (\cite{van03grb030329}; \cite{gre03grb030329};
\cite{cal03grb030329}).  We detected a bright, 12th-mag optical
afterglow 76 min after the burst (\cite{pet03grb030329};
\cite{tor03grb030329}; \cite{uem03grb030329gcn1994}).
With the international
collaboration through the VSNET, we obtained an $\sim 11$ hr continuous
light curve of this afterglow (\cite{uem03grb030329nature}).  Our
observation revealed that the afterglow experienced repetitive
modulations even in the early phase of our observation.  In conjunction
with public data reported to the GCN Circulars,
the light curve was rich in
unexpectedly complicated structures throughout $\sim 6$ days after the
burst.  It is a surprise that the amplitude of modulations was almost
constant with time.  This feature of modulations is difficult to be
understood with density variations of the interstellar medium
(\cite{nak03grb021004}).  The energy in the shock region must have
changed with time, while the mechanism to generate additional energy is
an open issue (\cite{uem04grb030329pasjvar}).  The VSNET team also
contributed to a another collaborative work on the earliest stage
afterglow \citep{tor03grb030329apjl}.

\section{Conclusion}

   Variable Star Network (VSNET) is a global professional-amateur network
of researchers in variable stars and related objects, particularly in
transient objects, such as cataclysmic variables, black hole binaries,
supernovae and gamma-ray bursts.  The VSNET has been playing a pioneering
role in establishing the field of {\it transient object astronomy},
by effectively incorporating modern advance in observational astronomy
and global electronic network, as well as collaborative progress
in theoretical astronomy and astronomical computing.
The VSNET is now one of the best-featured global networks in this field
of astronomy.  We review on the historical progress, design concept,
associated technology.  We also review on the breathtaking scientific
achievements, as well as regular variable star works, particularly
focusing on dwarf novae (discovery of ER UMa stars, works in WZ Sge-type
dwarf novae, more usual SU UMa-type dwarf novae, eclipsing dwarf novae),
black hole X-ray transients (discoveries of an unexpected
violent outburst of V4641 Sgr, rapid optical variations from the
same object), and recent achievements in gamma-ray bursts.

\vskip 3mm

We are grateful to Seiji Masuda and Katsura Matsumoto, who greatly
contributed to the activities of the VSNET administrator group.
We are grateful to many VSNET members who have been
continuously supporting our activity.
We are grateful to Emile Schweitzer (AFOEV), Keiichi Saijo
and Makoto Watanabe (VSOLJ) for kindly allowing us to use
AFOEV and VSOLJ public database for drawing light curves.
We are also grateful to Dave Monet for making USNO A1 CD-ROMs readily
available for us.
This work is partly supported by a grant-in-aid [13640239, 15037205 (TK),
14740131 (HY)] from the Japanese Ministry of Education, Culture, Sports,
Science and Technology.
Part of this work is supported by a Research Fellowship of the
Japan Society for the Promotion of Science for Young Scientists (MU, RI).
This research has made use of the astronomical catalogs at Astronomical
Data Centers operated by National Astronomical Observatory, Japan,
and NASA.  This research has also made use of the Digitized Sky Survey
producted by STScI, the ESO Skycat tool, the VizieR catalogue
access tool, and the Electronic Edition of the GCVS.

\appendix
\section{Poster Presented at the 23rd IAU General Assembly}
\label{sec:app:iaupos}

\begin{center}
{\bf Development of a Mailing List on\\ Variable Stars, VSNET}

(Daisaku Nogami, Taichi Kato,\\ Hajime Baba, Chatief Kunjaya)
\vskip 3mm

{\bf Abstract}
\end{center}

As the computer environment has been developing drastically
these years, the style of astronomical study has been changed.
The key words of these changes are thought to be ``real-time'' and
``interactivity''.  Suspecting that they have potential to made an
great effect on the study of transient objects (cataclysmic
variables, X-ray binary, super novae, and so on), we set up the
mailing list, VSNET, in 1994.  Our policy on VSNET since the start
is that VSNET is world-widely opened to any kinds of researchers
including professionals and amateurs, observers and theorists.
Subscribers have increased as time goes and now are over 400 from
over 40 countries.  Although VSNET started as a mailing list, it
at present consists of five sub-maling lists, vsnet: vsnet-alert,
vsnet-obs, vsnet-chat, and vsnet-chart, and each of these lists
works independently for different purposes.  Using the different
characteristics of these sub-lists to advantage, various types of
study have been proposed and carried out.  VSNET will be developed
further with cooperation of subscribers.  If you have any comments
or questions, please feel free to contact with VSNET
administrators (vsnet-adm@kusastro.kyoto-u.ac.jp).
\vskip 3mm

\subsection*{Introduction}

The computer environment is making remarkable progress with
development of the infrastructure and the machine power.  The
World-Wide Web (WWW, Berners-Lee et al. 1992) is explosively
coming into wide as well, redrawing the common sense on the
information distribution.

These have an great affect on the astronomical study, for example,
1) the complicated rapid control of telescopes (adaptive optics,
etc.) becomes possible, 2) the scale of simulations are getting
much larger, 3) editors of various journals encourage the
electronic submission, 4) the electronic publication of papers on
WWW or by CD-ROM is discussed and partly realized , 5) IAU
circulars are distributed by e-mail, 6) preprints are usually
distributed by e-mail or on web pages (e.g. http://www.lanl.gov,
Ginsparg 1996), 7) the search of papers has become quite easy by
the foundation of the Astrophysics Data System
(NASA, Accomazzi et al. 1995), and so on.

The key words of this revolution are thought to be ``real-time''
and ``interactivity''.

Is there a possibility for these characteristics to be quite
useful in the study of transient objects?  We had an eye on the
mailing list since the system makes it possible to share much
information among much people quite quickly and can be used
interactively by much people.  Then, in July 1994, we started a
mailing list, VSNET (vsnet@kusastro.kyoto-u.ac.jp), mainly on
cataclysmic variables, supernovae, X-ray binaries, and so on, with
a policy that we aim to make a contribute to the astronomical
community by providing a mechanism to make information shared
and a room to discuss all aspects of the astronomy concerning
those transient objects among subscribers from all over the world
irrespective of the status (an amateur or a professional), the
style of astronomical study (an observer or a theorist), and other 
properties.

\subsection*{History of VSNET}

The first member of VSNET is several tens of amateurs belongs to
amateur associations (The American Association of Variable Stars
(AAVSO, see http://www.aavso.org/), Association Fran\c{c}aise
des Observateurs d'Etoiles Variables (AFOEV), Variable Star Observer
League in Japan (VSOLJ), and so on) and a few tens of
professionals.  Subscribers, then, have increased by
mouse-to-mouse advertising and our invitation to authors of papers 
in astronomical journals.  Now the number of subscribers from
about 50 countries exceeds 400, which means that major part of
researchers in this field has already subscribed.

At first, all e-mails other than administrative ones for
subscription and unsubscription were distributed to all subscribers.
However, accepting a request mainly from theorists that they do
not need daily data but want to know final results deduced from
observations, we modified the system of VSNET to divide VSNET to
three sub-mailing lists in October 1994, 1) vsnet-obs
(vsnet-obs@kusastro.kyoto-u.ac.jp) for reporting daily
observations, 2) vsnet-alert (vsnet-alert@kusastro.kyoto-u.ac.jp)
for alert on the discovery of supernovae, novae, rare outbursts,
discovery of new variable star, dramatic change of known variable
stars, and so on, and 3) vsnet (vsnet@kusastro.kyoto-u.ac.jp) for
the general information not suited for the former two lists, for
example, complied data, finding charts, preprints, calls for
observations for international co-observation campaigns, and so on.

In order to refer to an old log correctly, serial numbers in the
subject were added as [vsnet-obs 1997] since January 1995.

In June 1995, the VSNET Web pages
(http://www.kusastro.kyoto-u.ac.jp/vsnet/) were opened.  You can
see all articles ever posted to VSNET on WWW.  You can see light
curves drawn from observations distributed via vsnet-obs in the
recommended format, too.  The most important results among
vsnet-alert logs and public information like conference
announcement are gathered on the top page almost daily up-dated.
Many useful tools developed and provided by various groups are
available on the pages as well.  At the same time we started VSNET
anonymous ftp service (ftp.kusastro.kyoto-u.ac.jp/vsnet/) where
you can get objects almost same as on WWW.

In January 1997, we made an addition of a new sub-mailing-list of
vsnet-chat (vsnet-chat@kusastro.kyoto-u.ac.jp) in order to discuss 
various subjects.  Though VSNET was used nearly only for
distribution of information, this modification widened the usage
of VSNET.

Having the background as mentioned here, VSNET at present consists
of four sub-mailing-lists (vsnet, vsnet-alert, vsnet-obs, and
vsnet-chat) as well as VSNET Web pages and anonymous ftp service.
However, one of the sub-lists, vsnet, is now closed, since
commercial informations never related to astronomy, so-called
SPAM, were posted again and again.  This type of problem, although 
common to all mailing lists, is hard to completely solve.

The number of e-mail posted to vsnet-obs, vsnet-alert, and
vsnet-chat is 20-40, a few, $\sim$5 per day, although changing
from day to day.

Tables show several e-mails posted to VSNET andf igures
show light curves available on WWW.

\subsection*{VSNET: Now and Future}

The major usages of VSNET are 1) to forecast the behavior of
variable stars from daily observations available via vsnet-obs and 
plot a strategy of the observation, 2) to check the status of a
variable star in optical on vsnet-obs at the time of the
observations in UV or X-ray, 3) to publish new results on
vsnet-alert (discoveries of supernovae, determination of superhump 
period in SU UMa-type dwarf novae, and so on), 4) to call the
follow-up observations of transient objects detected in UV or
X-ray, 5) to call the co-operation for international
co-observation campaigns, 6) to notify conferences newly held, 7)
to discuss various subjects ranging from the first step of the
observation to theoretical interpretations of new interesting
phenomena, and so on.  For reference, tables list a part of
subjects of e-mails posted to VSNET.  These usages would be born
in the active environment under the complex of ``real-time'',
``interactivity'', and the VSNET policy.

In addition, new original applications of VSNET are appreciated
all time, and, actually, will be proposed.  We, VSNET
administrators, would be much glad if studies completed on use of
VSNET contribute to the progress of astrophysics.  If you have any 
questions, requests, and suggestions, please do not hesitate to
communicate with VSNET administrators (vsnet-adm@kusastro.kyoto-u.ac.jp).

\vskip 3mm

Part of this work was supported by a Research Fellowship of the
Japan Society for the Promotion of Science for Young Scientists
(D.N.).

\subsection*{\bf References}

Accomazzi, A., Eichhorn, G., Grant, C. S., Murray,
S. S., and Kurtz, M. J. 1995, Vistas in Astronomy 39, 63

Berners-Lee, T. J., et al. 1992, in ``Electronic
Networking: Research, Applications and Policy'' (Meckler
Publishing, Westport), Vol. 2, No.1, p52

Ginsparg, P. 1996,
http://xxx.lanl.gov/blurb/\\pg96unesco.html

\section{VSNET Data Reporting System}
\label{sec:app:report}

   As described in subsection \ref{sec:rephist}, the VSNET data handling,
database managing, and data analysis tools are mostly common to the ones
developed for the VSOLJ database project \citep{kat90VSOLJreport},
and later adapted for
wider range of observations.  The programs were originally written for
microcomputers running on MS-DOS$^{\rm TM}$, and later ported to Linux.
The following information is mainly an excerpt from
\citet{kat90VSOLJreport}, partly rewritten for the recent changes for the
VSNET management.

\subsection{Electronic Format of Observations}
\label{app:format}

   The VSNET reporting format is an UT extension of
the VSOLJ electronic reporting format, whose design was established in 1987.
The standard format VSNET observations are composed of lines separated by
new-line characters.  Since a full description of the format used in these
data or files may not be necessary to all readers, the minimal requirements
to interpret the data which are made available through the VSNET
World-Wide Web service.

   Each line contains the following items.

\begin{enumerate}
\item[(a)] Name of the object
\item[(b)] Time of the observation in decimals of UT
\item[(c)] Observed magnitude
\item[(c')] Magnitude system (CCD/photoelectric) or film and filter (photographic)
\item[(d)] Observer's code
\end{enumerate}

   These items were designed to express original observations as exactly
as possible (mainly in terms of significant digits).  Comparison stars,
charts used, or any other text information can be written as a comment
following the item (d); they are usually used for future reference or
examination, and not directly used at present in the VSNET regular
database management.

   Each item is separated by one or more space characters (ASCII code 20 in
hexadecimal) and does not contain spaces within the item.

\subsubsection*{(a) Name of the Object}
\vskip 2mm

   This is an identifier for the object.  If the name is unique enough
to discriminate the object from other celestial objects, any expression
is basically allowed; the use of the names listed in the General Catalogue
of Variable Stars (GCVS), Beyer names, numbers in the New catalogue of
Suspected Variable Stars (NSV), and Durchmusterung numbers (BD, CD, CPD),
however, is strongly recommended.  When there is no specific relevant
catalog, the use of GSC and USNO identifier can be used.

   For GCVS and Beyer names, three-letter IAU code of constellation
(in upper-case letters) precedes the name of the star in the
constellation.  Greek letters are written in the standard English
expressions.

\vskip 3mm
\begin{tabular}{ll}
Examples: & ANDR for R And, \\
          & ORIalpha for Betelgeuse ($\alpha$ Ori)
\end{tabular}
\vskip 3mm

   Translation of reported names into standard expressions, when necessary,
is performed either automatically or manually using the
{\sl ALIAS} database.
This process particularly becomes necessary when new GCVS names are
released in the form of regular name-list updates, or when a new GCVS
designation is given for a new nova.  All the name translation rules are
centralized in the {\sl ALIAS} database and reflected on the entire VSNET
system, eliminating additional efforts to modify individual observation
reports or expressions in VSNET circulars.  This is an great advantage
of the VSNET database managing process, and a newly released name-list
update (containing some hundreds of newly designated variable stars)
can be usually reflected within a day of the release.  Observers are
allowed to continue using the old expressions as long as the expressions
can be uniquely and automatically translated into the new expressions.

\subsubsection*{(b) Time of the Observation}
\vskip 2mm

   In order to facilitate detecting errors by eye, times are expressed
in the following decimal format using UTC.

\begin{center}
Example: 20030701.123 (2003 July 1.123 UTC).
\end{center}

   No heliocentric or barycentric corrections are introduced at this stage.
The conversion to TAI or TD (in any expression including Julian Date),
or helio-(or bary-)centric corrections are left to data analysis software
and users.  These measures are partly because widely used software packages
at the observer's end are known to frequently contain problems.
The other reason is that leap seconds do not allow to distribute fixed
tables for conversion beforehand.

\subsubsection*{(c) Observed Magnitude}
\vskip 2mm

   Decimal points are explicitly used to show the significant digits.
Upper limit observation (non-detection) is expressed by a prefix `$<$'.
Expression may be followed by one of `:' or `?' to show uncertainty.

\subsubsection*{(c') Magnitude System, Film and Filter}
\vskip 2mm

   If the observation is visually performed, this item is not necessary.
Otherwise, the code of the magnitude system (or the film and the filter
if the observation is done photographically) follows the item (c)
without placing a space.  Well-defined standard photoelectric systems
(e.g. Johnson-Cousins $U$, $B$, $V$, $R_{\rm c}$, $I_{\rm c}$, $R_{\rm j}$,
$I_{\rm j}$) systems are used as in usual sense.  Other VSNET-specific
codes include ``C" (unspecified unfiltered CCD magnitude), ``CR"
(unfiltered CCD magnitude calibrated on $R_{\rm c}$-band) and
``p" (unspecified photographic magnitude).
The complete list of the codes representing the
films and filters at present is available at the VSNET website.\footnote{
$\langle$http://www.kusastro.kyoto-u.ac.jp/vsnet/etc/format.html$\rangle$.
}

\begin{center}
Example: $<$15.5Rc  (fainter than 15.5 in $R_{\rm c}$-band)
\end{center}

\subsubsection*{(d) Observer's code}
\vskip 2mm

   The code is usually a three-letter code, as canonically used by
the American Association of Variable Star Observers (AAVSO)\footnote{
$\langle$http://www.aavso.org$\rangle$.
} and the VSOLJ.  The code may be immediately followed by a
period mark (`.') and the organization code.  When there is no fixed
affiliation for the observer, the VSNET manager group issues a code,
which scheme also enables archiving historical or literature
observations.

   With this format, each line of observation has all information in
itself; any operation of moving lines in a data file (such as sorting)
does not affect the properties of the data.  The above design of
electronic data has greatly facilitated overall aspects of the following
electronic data management, and this pioneering concept has been taken
over by a number of world-wide variable star organizations.

\section{Design of the VSNET Database}

   Most of programs are written in the language C, and were originally
compiled with Turbo-C (Borland International).  Many of machine-independent
source codes (handling text-based data) can be compiled by ANSI C
(ANSI X3.159) compilers (such as gcc) without much correction.
In the actual VSNET database management, these codes are either compiled
on Windows$^{\rm TM}$/DOS personal computers, or Linux workstations.
Most of the text-based works are currently done on Linux workstations.

   Besides storing original data as text files (as reported in vsnet-obs),
they are also incorporated into an observational database in rewritable
media allowing random access.  Several interesting graphic programs and
data analysis programs run in this mode.  The maximum number of
observations handled at one time depends only on capacity of the media
and the addressing capacity of the operating system.  For example,
we now handle dynamic and random access to the entire VSNET observations,
consisting of more than 1.2 million visual observations ($\sim$130 MB)
of randomly accessible data, and we have registered more than 1.3 million
CCD observations submitted to the VSNET Collaboration.  We have also
confirmed that a combination test of the VSNET, VSOLJ and AFOEV public
data (about 3 million observation) has yielded satisfactory efficiency
of handling a huge volume of randomly accessible data.

   In addition to the observational database, there also exist other
system databases including {\sl VARTYPE}
(database on individual variable star
types), and {\sl ALIAS} (name resolver database).
A programmer on this system
can inquire these databases for the type of variability or the standard
expression of the name for the object specified by a given identifier.

\subsection{Structure of the Database}
\label{sec:app:database}

   Because the number of the data or the objects is very large, and both
random and sequential accesses are necessary for easy operation on the
light curves, we adopted the combination of B+tree and bidirectional
linear list.  This basic structure of the database was originally
established in 1989 to fully incorporate the VSOLJ database.

   Complete data of observations (other than remarks) are stored in
a file equivalent to an index file of an relational database, and there
is no need for reading an additional data file.  The size of a storage
block is 16 kilobytes, but this can be modified upon different compilation.
The main memory is dynamically allocated to simulate a virtual memory
in order to minimize access to the storage media.  With this feature,
the main database module only requires an order of a few hundreds of
megabytes of actual memory.  The kernel of the database module is written
transparently to the upper modules, so upper modules have only to pass
the key or the virtual address to the kernel to inquire
the next or the previous data.  The modules are written to enable
simultaneous handling of multiple databases (e.g. observational database
and {\sl VARTYPE}) without interference.  The database functions are
prepared as a form of C-language based application interface (API),
but we skipped the details of individual APIs because they are too
technical to be presented in this paper.

\subsection{Basic Command Operations}

   Basic user operations to the database can be performed by command-line
tasks on MS-DOS$^{\rm TM}$ or Linux.
The most frequently used and basic commands
include:

\begin{enumerate}
\item[{\bf inittree}] creates a new database.
\item[{\bf mrg}] merges a text file to a database.
\item[{\bf vdel}] deletes data specified by a text file from a database.
\item[{\bf vmrg}] merges an observational data file in the standard format
to the observational database, and lists potential
errors by referring to the previously registered data and
{\sl VARTYPE} database.
\item[{\bf vcut}] lists data of a given object for a given chronological
period.
\item[{\bf vren}] replaces an identifier for an object by another.
\item[{\bf vlist}] lists all data in a database.
\item[{\bf liststar}] lists all stars in a database.
\item[{\bf seltype}] selects data from a text file by specifying types of
variability, by referring to the {\sl VARTYPE} database.
\item[{\bf chknew}] checks observation data file if they contain newly
reported objects by comparison with the existing observational
database..
\item[{\bf chkstd}] checks observation data file for grammar, and lists
discordant data and potential errors by referring to the previously
registered data and {\sl VARTYPE} database.
\item[{\bf looktype}] returns the variable star type by referring to the
{\sl VARTYPE} database.
\item[{\bf settype}] sets or modifies the variable star type in the
{\sl VARTYPE} database.
\item[{\bf calias}] converts a variable star name to the standard name
by referring to {\sl ALIAS} database.
\item[{\bf alias}] sets or modifies the entry in the {\sl ALIAS}
database (this command has a different name on Linux because of
the collision with the shell built-in command).
\end{enumerate}

   Most basic operations on the VSNET data are done with these basic
commands, and are frequently used as a combination regulated by a shell
script.

\subsection{Data Analysis Application Programs}

\begin{figure*}[t]
  \begin{center}
    \FigureFile(170mm,90mm){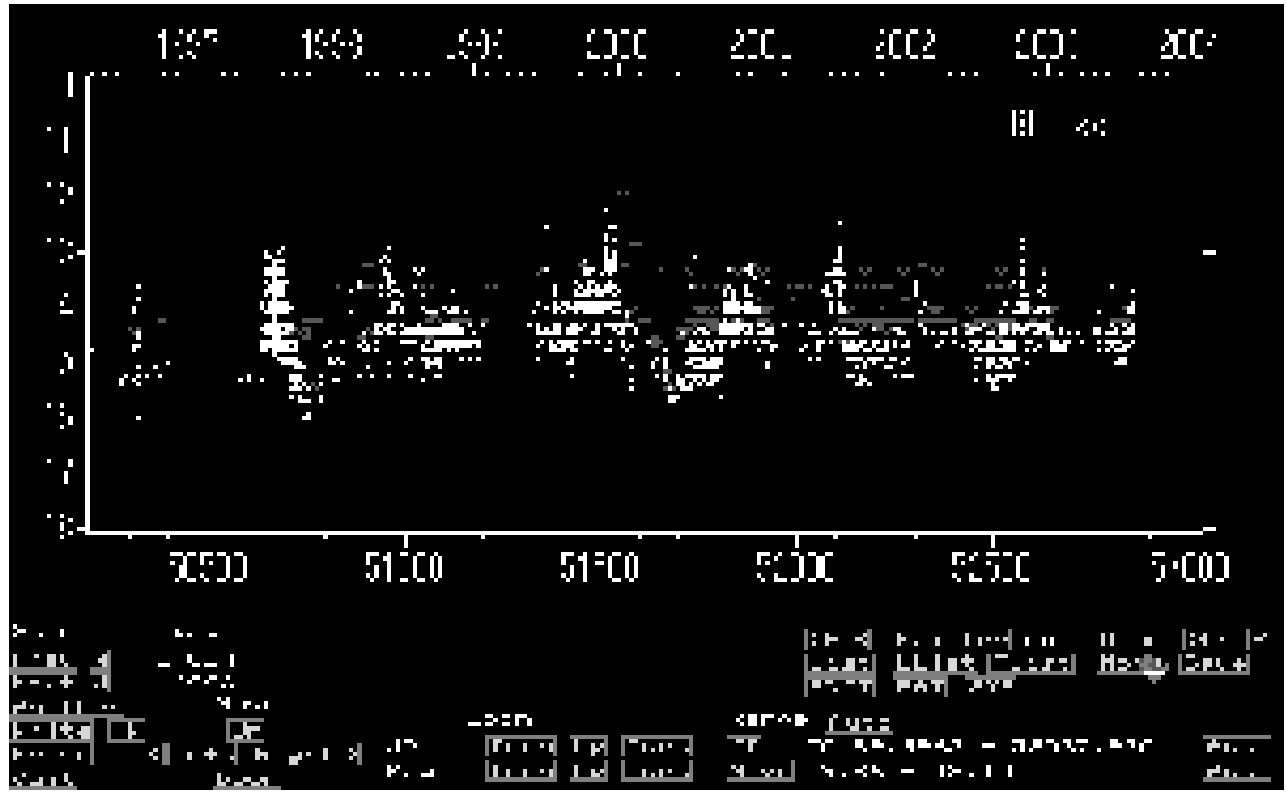}
  \end{center}
  \caption{Sample image by {\bf grp} interactive light curve viewer
  program.  The light curve is BL Lac from VSNET observations.
  White and red marks represent positive and upper limit observations,
  respectively.
  The program enables interactive GUI-based operation on photometric
  database constructed on the database engine described in Appendix
  \ref{sec:app:database}.}
  \label{fig:bl}
\end{figure*}

   Most of these programs use a GUI.
These packages are architecture-specific, and presently run in limited
environments (e.g. NEC PC 9801 machines), although efforts have been
taken to port these applications to Windows$^{\rm TM}$ operating system,
or to write an equivalent wrapper GUI applications written
in Java$^{\rm TM}$.  Figure \ref{fig:bl} shows a sample image
by {\bf grp} interactive light curve viewer program.

\begin{enumerate}
\item[{\bf grp}] displays light curves, and enables interactive zooming, data
selection, data editing.
\item[{\bf autogrp}] displays and prints light curves automatically.
An equivalent package written in Java$^{\rm TM}$ is presently used to
produce VSNET online light curves.
\item[{\bf circ}] automatically produces VSNET CV Circulars.
\end{enumerate}

\subsection{Data Processing}

   Besides drawing light curves, there also exist several kinds of scientific
data-analysis programs.  Some require databases and others require only
data files in the text format.  A few samples are shown here.  Many of
the general-purpose programs (mostly with source codes) are available
from the VSNET website ``Tools and Programs" section.\footnote{
$\langle$http://www.kusastro.kyoto-u.ac.jp/vsnet/etc/prog.html$\rangle$.
}

\subsubsection*{Heliocentric and Barycentric Corrections}
\vskip 2mm

   We have implemented both conventional heliocentric corrections for
the observed JDs, by using the well-known Newcomb's expansion of the
planets.  We have also implemented barycentric corrections by numerically
integrating the DE 200 ephemeris produced by NASA.  The original source
code of the DE 200 ephemeris was imported from a software package
written by Mr. Novas.

\subsubsection*{Period Analysis Tools}
\vskip 2mm

   We have our own implementations of Discrete Fourier Transform (DFT),
Phase Dispersion Minimization (PDM: \cite{PDM}), Discrete Wavelet
Transform (DWT: \cite{DWT}) and other period analysis tools.
The PDM package was ported to Windows$^{\rm TM}$ by Andreas Wijaya,
and has been conveniently used by many users.

\subsubsection*{CV Circulars}
\vskip 2mm

   CV Circulars were originally prepared monthly and issued by the VSOLJ,
from the observations reported by the end of the next month.  In the modern
VSNET service, these circulars are issued almost daily to fully
incorporate the daily changes in the rapidly varying CVs.
The corresponding list is
{\it vsnet-cvcirc}.\footnote{
$\langle$http://www.kusastro.kyoto-u.ac.jp/vsnet/Mail/vsnet-cvcirc/maillist.html$\rangle$.
}

   CV Circulars contain information about all reported outbursts and
standstills of dwarf novae in the form of nightly averages, and nightly
averages of other peculiar objects (CVs other than dwarf novae,
X-ray binaries, symbiotic variables, eruptive variable such as
R CrB stars, S Dor stars, FU Ori stars, and supernovae, active galactic
nuclei and other objects of special interest).

   Unpredictable variations of brightness of these objects make it
extremely difficult to check observations as automatically done for
pulsating variables.  Together with faintness of these objects (which
implies existence of large number of negative observations), cross checks
between different observers are indispensable.

   The present program {\bf circ}, referring to a list of more than 1000
objects containing their properties (type of variability, normal range of
variation, ephemerides of eclipses etc.), checks all observations using
databases.  The list is updated by the editor whenever new information
becomes available in order to maintain circulars up-to-date.

   The results are then listed as a file in the prototypical form of
CV Circulars, containing the reports of potentially discordant data
(or rapid intrinsic changes) with special marks.  The present version
can handle more than tens of thousands of observations per month, and
produces a circular in a minute.

   The same program, slightly modified to produce long-term averages,
is used to produce VSNET Mira Circulars {\it vsnet-miracirc}.\footnote{
$\langle$http://www.kusastro.kyoto-u.ac.jp/vsnet/Mail/vsnet-miracirc/maillist.html$\rangle$.
}

\section{Chart-Plotting and Star Identification Program}
\label{sec:app:chart}

   When identifying variable stars and locating the CCD field of view,
chart-plotting and star identification programs are very useful,
particularly when the network resource was unavailable.  For this
purpose, one of the authors (TK) developed in 1990 a chart-plotting
computer program, when machine-readable Guide Star Catalog (GSC) 1.0
\citep{GSCI1} was released.  The program was designed to run on a
stand-alone personal computer running on MS-DOS$^{\rm TM}$
non-extended (less than 640 KB) memory.

   Since the distribution form of GSC 1.0 was composed of huge ASCII
tables on two CD-ROMs (1.2 GB), and since the objects were randomly
arranged within individual files covering 2--3 degrees square,
it was an absolute requirement to compress these data and make them
quickly and randomly accessible.  In this software, we subdivided
the entire GSC data into 0.5 degree bins in declination, and sorted
the objects within the same bin according to right ascension.
In order to compress the data, we used binary files which can be
directly mapped in C structure.  In order to avoid redundancy in
right ascensions and enable quick random access, we used a separate
jump table which records the file positions of given coordinate
meshes.  Individual entries in the 0.5-deg bins only contain
residuals to the mesh coordinates.  In original compression of
GSC 1.0, each catalog entry corresponded to 5 bytes.
By combining the information of coordinate mesh in the jump table
and the individual entries, one can obtain the fully decoded
coordinates.  These functions (including sequential reading functions
of a given box) were implemented as transparent APIs, many of which
were designed to take a pointer to the display function as a call-back
function.  With this compression, the entire GSC 1.0 (disregarding
the object names, plate numbers) can be compressed into 100 MB.
This program was one of the earliest chart-plotting software packages
that used GSC as source catalogs and implement encoded compressed
catalogs, and the design was taken over in various successive
third-party applications.  The program is also able to display
GCVS and NSV variable star catalogs, and IRAS PSC objects, which
were compressed in a similar manner employed in GSC compression.

\begin{figure*}[t]
  \begin{center}
    \FigureFile(160mm,90mm){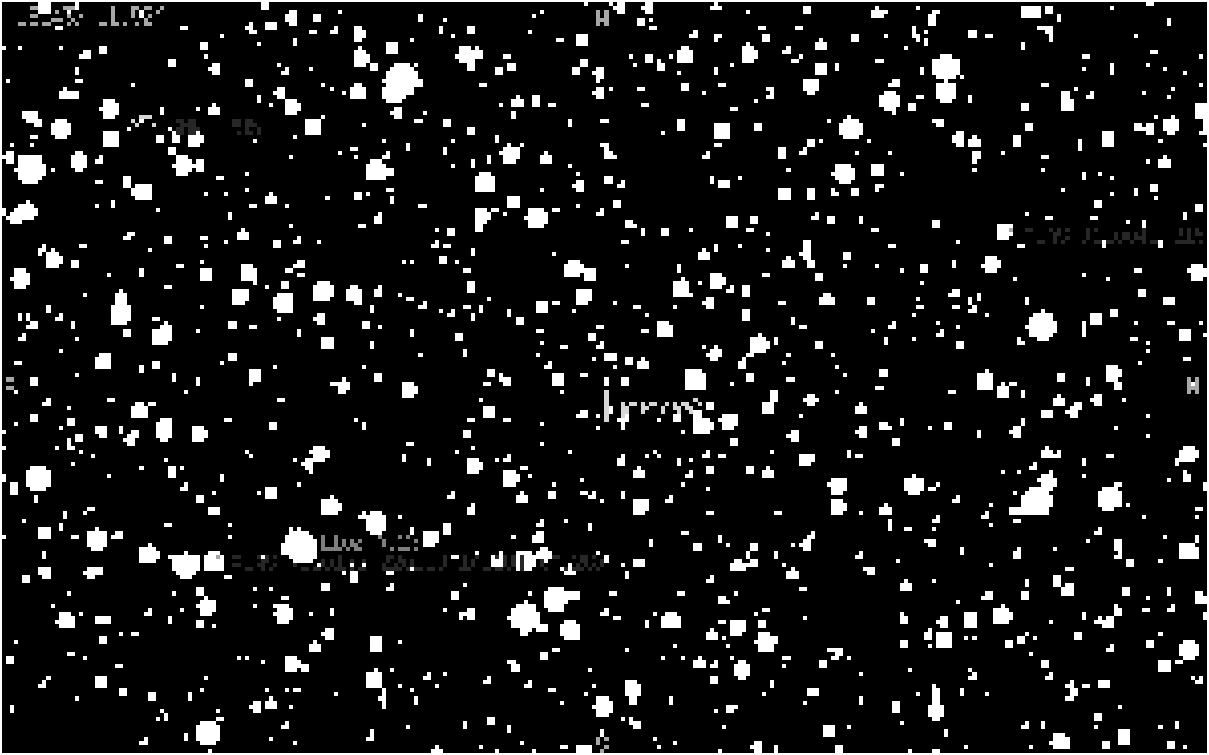}
  \end{center}
  \caption{Sample chart of V4743 Sgr (Nova Sgr 2002 No. 3).  The nova
  progenitor is marked.}
  \label{fig:v4743}
\end{figure*}

   The program was later updated to accommodate
GSC 1.1 \citep{GSC11} and USNO A1.0 \citep{USNOA10}
catalogs in 1998, and other catalogs (with variable-length object
labels).
With the increase of capacity of storage media, the present
version uses 12 bytes for one GSC 1.1 full entry and 8 bytes for
one USNO A1.0 entry.  With this compression method, the entire
USNO A1.0 data (more than 4.8$\times$10$^8$ objects)
can be stored in 4 GB storage and can be quickly and randomly accessed.
The present version is equipped with a function to handle the
2MASS point source catalog in the same manner.  The program also
handles a name resolver by using the same database engine described
in Appendix \ref{sec:app:database}.  The entire program is
presently ported to Windows$^{\rm TM}$ and
Linux using {\it xlib} graphic library (see figure \ref{fig:v4743}
for an example).  With this program and APIs,
one can very quickly identify new and known variable stars either
interactively or in a batch.  Online VSNET charts with Hipparcos
and Tycho magnitudes (subsection \ref{sec:standard}) have been prepared
with this software operated in a batch mode.  The object databases
have been regularly updated, especially when a name lists for
new variable stars is released.  The source codes are available
upon request to the author.

\end{document}